\documentclass[a4paper,11pt]{article}
\pdfoutput=1 

\usepackage{jheppub} 
                              
\usepackage[T1]{fontenc} 
\usepackage{xcolor}
\usepackage{multirow}
\usepackage{tikz}
\usepackage{adjustbox}
\usepackage{ulem}

\def\gev{{\rm GeV}}

\allowdisplaybreaks

\title{\bf \boldmath CP asymmetries of $t \to c \gamma$ and $t \to cg$ decays in the aligned two-Higgs-doublet model}

\author[a]{Fang-Min Cai,}
\author[b]{Rui-Lin Fan,}
\author[b,c,1]{Xin-Qiang Li,\note{Corresponding author.}}
\author[a,b]{and Ya-Dong Yang}

\affiliation[a]{Institute of Particle and Nuclear Physics, Henan Normal University, Xinxiang 453007, China}
\affiliation[b]{Institute of Particle Physics and Key Laboratory of Quark and Lepton Physics~(MOE), Central China Normal University, Wuhan, Hubei 430079, China}
\affiliation[c]{Center for High Energy Physics, Peking University, Beijing 100871, China}

\emailAdd{caifangmin@htu.edu.cn}
\emailAdd{fanrl@mails.ccnu.edu.cn}
\emailAdd{xqli@mail.ccnu.edu.cn}
\emailAdd{yangyd@mail.ccnu.edu.cn}

\abstract{We study the CP asymmetries of the rare top-quark decays $t \to c \gamma$ and $t \to cg$ in the aligned two-Higgs-doublet model (A2HDM), which is generically characterized by new sources of CP violation beyond the Standard Model (SM). Specifically, the branching ratios and CP asymmetries of these rare top-quark decays are explicitly formulated, with an emphasis on the origins of weak and strong phases in the A2HDM. Taking into account the most relevant constraints on this model, we evaluate the variations of these observables with respect to the model parameters. It is found that the branching ratios of $t \to c \gamma$ and $t \to cg$ decays can maximally reach up to $1.47\times10^{-10}$ and $4.86\times10^{-9}$ respectively, which are about four and three orders of magnitude higher than the corresponding SM predictions. While the branching ratios are almost independent of the relative phase $\varphi$ between the two alignment parameters $\varsigma_u$ and $\varsigma_d$ within the allowed parameter space, the CP asymmetries are found to be very sensitive to $\varphi$. When the two alignment parameters are complex with a non-zero $\varphi$ varied within the range $[50^\circ,150^\circ]$, the magnitudes of the CP asymmetries can be significantly enhanced relative to both the SM and the real case. In particular, the maximum absolute values of the CP asymmetries can even reach up to $\mathcal{O}(1)$ for these two decay modes, in the range $\varphi \in [70^\circ,100^\circ]$. These interesting observations could be utilized to discriminate the SM and the different scenarios of the A2HDM.}

\begin{document} 
\maketitle
\flushbottom

\section{Introduction}
\label{sec:introduction}

As the heaviest elementary particle known to date, the top quark plays a special role in validating the Standard Model (SM) of particle physics and probing new physics (NP) beyond it~\cite{Beneke:2000hk,TopQuarkWorkingGroup:2013hxj,Schwienhorst:2022yqu}. Especially, being the only quark with a coupling to the Higgs boson of order unity, the top quark provides a unique laboratory to test our understanding of matter and fundamental interactions at the electroweak symmetry-breaking scale and beyond. Another significant feature of the top quark is that, due to its short lifetime, the top quark is expected to decay before top-flavoured hadrons or $t\bar{t}$-quarkonium bound states can form~\cite{Bigi:1986jk}. Since its discovery at the Tevatron in 1995~\cite{CDF:1995wbb,D0:1995jca}, the top-quark properties have been investigated in great detail both in production and in decay~\cite{ParticleDataGroup:2024cfk}, and will also be among the core physical programs at future high-energy colliders~\cite{Azzi:2019yne,Cerri:2018ypt,LHeCStudyGroup:2012zhm,ILC:2013jhg,CLICdp:2018esa,FCC:2018byv,CEPCStudyGroup:2018ghi}. 

Among the various top-quark decay modes, the rare flavour-changing neutral-current (FCNC) decays $t \to q X$, with $q=c, u$ and $X=\gamma$ (the photon), $g$ (the gluon), $Z$ (the electroweak gauge boson), $h$ (the SM Higgs), are of particular interest, because their rates are exactly zero at tree level and are severely suppressed by the Glashow-Iliopoulos-Maiani (GIM) mechanism~\cite{Glashow:1970gm} at the loop level within the SM. Explicitly, due to the unitarity of the Cabibbo-Kobayashi-Maskawa (CKM)~\cite{Cabibbo:1963yz,Kobayashi:1973fv} matrix and the smallness of the mass splittings among the down-type quarks running in the loop, these rare FCNC decays are predicted to have branching ratios ranging from $\mathcal{O}(10^{-17})$ to $\mathcal{O}(10^{-12})$ at the one-loop order within the SM~\cite{Eilam:1990zc,Mele:1998ag,Aguilar-Saavedra:2002lwv,Aguilar-Saavedra:2004mfd,Abbas:2015cua}, which are far below the current experimental upper limits of $\mathcal{O}(10^{-4}-10^{-5})$~\cite{ParticleDataGroup:2024cfk,LHCTop}. As a result, any observation of the top-quark FCNC phenomena at the LHC and the future colliders would be a clear signal of NP beyond the SM. There are, indeed, a number of NP models that can increase the branching ratios of these rare FCNC decays by several orders of magnitude, such as the two-Higgs-doublet model (2HDM)~\cite{Eilam:1990zc,Abbas:2015cua,Diaz-Cruz:1989tem,Grzadkowski:1990sm,Deshpande:1991pn,Hou:1991un,Luke:1993cy,Atwood:1996vj,Bejar:2000ub,Diaz:2001vj,Iltan:2001yt,Arhrib:2005nx,Baum:2008qm,Chen:2013qta,Botella:2015hoa,Altmannshofer:2019ogm,Hou:2020ciy,Cai:2022xha}, the supersymmetric model~\cite{Li:1993mg,Lopez:1997xv,deDivitiis:1997sh,Yang:1997dk,Guasch:1999jp,Eilam:2001dh,Cao:2007dk,Dedes:2014asa,Bardhan:2016txk,Yang:2018utw,Sun:2023zeb,Liu:2023vlm}, the extra dimensional model~\cite{Agashe:2006wa,Azatov:2009na,Gao:2013fxa,Dey:2016cve,Diaz-Furlong:2016ril,Chiang:2018oyd}, the littlest Higgs model~\cite{Hong-Sheng:2007acy,Han:2009zm,Aranda:2021kza}, the left-right symmetric model~\cite{Gaitan:2004by,Frank:2005vd,Frank:2023fkc}, and other specific NP scenarios~\cite{Jueid:2024cge,Chen:2023eof,Chen:2022dzc,Crivellin:2022fdf,Liu:2021crr,Balaji:2020qjg,Bie:2020sro,Bolanos:2019dso,Banerjee:2018fsx,Aguilar-Saavedra:2002lwv,Hung:2017tts}. For a recent review, we refer the readers to refs.~\cite{Castro:2022qkg,dEnterria:2023wjq}. These processes have also been studied~\cite{Zhang:2013xya,Durieux:2014xla,Aguilar-Saavedra:2018ksv,Bhattacharya:2023beo} in the framework of the SM effective field theory~\cite{Buchmuller:1985jz,Grzadkowski:2010es,Brivio:2017vri}. It is expected that, with the increase of both the center-of-mass energy and the accumulated luminosity, as well as the improvements on the signal-to-background optimization techniques, the high-luminosity LHC (HL-LHC)~\cite{Cerri:2018ypt,Azzi:2019yne} and the future colliders~\cite{LHeCStudyGroup:2012zhm,ILC:2013jhg,CLICdp:2018esa,FCC:2018byv,CEPCStudyGroup:2018ghi} will enable much more stringent bounds on the top-quark FCNC decays, which can therefore put even stronger constraints on NP beyond the SM. For example, making full use of the potential of the future circular hadron-hadron collider (FCC-hh), the $95\%$ confidence-level limits on the branching ratios are estimated to be $\mathcal{O}(10^{-7})$ for $t \to c\gamma$ with an integrated luminosity of $30~\text{ab}^{-1}$~\cite{FCC:2018byv}, and $\mathcal{O}(10^{-7}-10^{-8})$ for $t \to cg$ with an integrated luminosity of $10~\text{ab}^{-1}$~\cite{Oyulmaz:2019jqr,Khanpour:2019qnw}.

Among the various NP scenarios, the 2HDM is one of the simplest extensions of the SM by adding a second scalar doublet with the same quantum numbers as the SM Higgs doublet~\cite{Gunion:1989we,Branco:2011iw,Ivanov:2017dad}. The model is characterized by a rich and flexible scalar spectrum with three neutral and one pair of charged scalars, and hence opens many interesting possibilities like new sources of CP violation~\cite{Lee:1973iz,Wu:1994ja,Ginzburg:2004vp,Gunion:2005ja,Maniatis:2007vn,Grzadkowski:2013rza,Cao:2022rgh,Darvishi:2023fjh}, the neutrino mass generation~\cite{Ma:2006km,Gabriel:2006ns,Wang:2006jy,Davidson:2009ha,Kanemura:2013qva,Li:2022yna}, the electroweak baryogenesis~\cite{Turok:1990zg,Cline:1995dg,Fromme:2006cm,Cline:2011mm,Fuyuto:2015jha,Chiang:2016vgf,Dorsch:2016nrg,Basler:2021kgq,Enomoto:2021dkl,Enomoto:2022rrl}, as well as the dark matter~\cite{LopezHonorez:2006gr,Goudelis:2013uca,Belyaev:2016lok} and axion-like~\cite{Celis:2014zaa,Espriu:2015mfa} phenomenologies. The 2HDM is also realized as the low-energy effective theory of some more elaborate NP scenarios~\cite{Mohapatra:1974hk,Haber:1984rc,Kim:1986ax}. However, for a most generic 2HDM, there will be unavoidably tree-level FCNCs that must be small enough to avoid conflict with the present experimental data~\cite{ParticleDataGroup:2024cfk}, because we cannot diagonalize all the Yukawa matrices involved simultaneously. To guarantee the absence of these potentially dangerous interactions, one usually imposes discrete $Z_{2}$ symmetries in the Lagrangian, such that only one of the two scalar doublets couples to a given type of right-handed fermion fields~\cite{Glashow:1976nt,Paschos:1976ay}. This results in four different types of $Z_{2}$-symmetric models, corresponding respectively to the so-called type-I, type-II, type-X (lepton specific) and type-Y (flipped) 2HDMs~\cite{Branco:2011iw,Haber:1978jt,Donoghue:1978cj,Hall:1981bc,Barger:1989fj,Grossman:1994jb,Aoki:2009ha}. 

A more generic theoretical framework to avoid the appearance of tree-level FCNCs in the 2HDM is to assume that the Yukawa matrices associated with the same type of right-handed fermions are aligned in the flavour space, so that the resulting interactions of the two scalar doublets with fermions share the same flavour structure~\cite{Pich:2009sp,Penuelas:2017ikk,Manohar:2006ga,Lee:2021oaj}. This is known as the aligned two-Higgs-doublet model (A2HDM)~\cite{Pich:2009sp}, in which the flavour violation is minimal~\cite{DAmbrosio:2002vsn,Altmannshofer:2012ar,Dery:2013aba} and the highly-suppressed FCNCs appear only at higher perturbative orders~\cite{Pich:2009sp,Jung:2010ik,Penuelas:2017ikk,Ferreira:2010xe,Braeuninger:2010td,Bijnens:2011gd,Botella:2015yfa,Gori:2017qwg}. Interestingly, all the four 2HDMs based on $Z_{2}$ symmetries can be recovered by adjusting the alignment parameters of the A2HDM~\cite{Pich:2009sp}. The model is also featured by new sources of CP violation beyond the single complex phase of the CKM matrix within the SM, present in both the scalar and Yukawa sectors~\cite{Pich:2009sp}. As a consequence, the A2HDM has led to many compelling phenomenologies in the electric dipole moments of the leptons, the neutron and various atoms~\cite{Jung:2013hka,Dekens:2014jka}, the muon anomalous magnetic moment~\cite{Han:2015yys,Ilisie:2015tra,Cherchiglia:2016eui,Cherchiglia:2017uwv,DelleRose:2020oaa}, the low-energy flavour physics~\cite{Jung:2010ik,Jung:2010ab,Jung:2012vu,Celis:2012dk,Li:2013vlx,Li:2014fea,Enomoto:2015wbn,Chang:2015rva,Hu:2016gpe,Cho:2017jym,Hu:2017qxj,DelleRose:2019ukt}, the rare top-quark decays~\cite{Abbas:2015cua,Cai:2022xha}, the high-energy colliders~\cite{Celis:2013rcs,Duarte:2013zfa,Celis:2013ixa,Wang:2013sha,Ilisie:2014hea,Ayala:2016djv,Abbas:2018pfp,Kanemura:2022cth,Connell:2023jqq} and so on. Global fits of the parameter space of the A2HDM by incorporating the theoretical constraints required by perturbative unitarity and boundedness of the scalar potential from below, the low-energy flavour bounds, as well as the LHC and LEP data have been performed in refs.~\cite{Eberhardt:2020dat,Karan:2023kyj}. 

In this paper, following our previous studies~\cite{Abbas:2015cua,Cai:2022xha}, we shall proceed to investigate the CP asymmetries of the rare top-quark decays $t \to c \gamma$ and $t \to c g$ in the A2HDM. As these CP asymmetries are predicted to be very tiny within the SM~\cite{Deshpande:1990ua,Aguilar-Saavedra:2002lwv,Balaji:2020qjg,Deshpande:1991pn}, any significant observation of these signals would be a clear indication of NP beyond the SM. In some extensions of the SM, such as the 2HDM and the supersymmetric model, CP asymmetries of the top-quark decays can be significantly enhanced. Experimentally, the CP violation effects in top-quark productions and decays have been investigated in high energy $e^+ e^-$, $\gamma \gamma$, $\mu^+\mu^-$, $p p$ and $p\bar{p}$ colliders. For more details, the readers are referred to Ref.~\cite{Atwood:2000tu}. Although CP violation beyond the SM has yet to be observed, we suspect that it must be there in order to explain the baryon asymmetry of the Universe. This gives paramount importance to new search strategies for CP violation in the top-quark sector in the context of NP scenarios~\cite{Atwood:2000tu}. Specific to the A2HDM, the sources of CP violation for these rare processes arise from the imaginary parts of the loop integrals, the single complex phase of the CKM matrix, and the relative phase $\varphi$ between the two alignment parameters $\varsigma_u$ and $\varsigma_d$ (see section~\ref{sec:decay process} for their definitions). Here we shall focus primarily on how large the CP asymmetries of these rare FCNC decays are possible in the A2HDM, after taking into account the most relevant constraints on the model parameters~\cite{Eberhardt:2020dat,Karan:2023kyj}. It is found that the branching ratios of $t \to c \gamma$ and $t \to cg$ decays can maximally reach up to $1.47\times10^{-10}$ and $4.86\times10^{-9}$ respectively, which are about four and three orders of magnitude higher than the corresponding SM predictions~\cite{Aguilar-Saavedra:2002lwv,Balaji:2020qjg}, but are still below the current experimental upper limits of $\mathcal{O}(10^{-5})$ and $\mathcal{O}(10^{-4})$~\cite{ParticleDataGroup:2024cfk,LHCTop}. While the branching ratios are almost independent of the relative phase $\varphi$ within the allowed parameter space, the CP asymmetries are found to be very sensitive to the parameter $\varphi$: when $\varphi$ varies within the range $[50^\circ,150^\circ]$, the magnitudes of the CP asymmetries can be significantly enhanced relative to both the SM and the real case, with their maximum absolute values even reaching up to $\mathcal{O}(1)$ in the range $\varphi \in [70^\circ,100^\circ]$. As a consequence, we could make full use of these observations to discriminate the SM and the different scenarios of the A2HDM. 

This paper is organized as follows. In section~\ref{sec:A2HDM}, we introduce the A2HDM in the Higgs basis, as well as its scalar and Yukawa sectors that are most relevant to our study. In section~\ref{sec:decay process}, our procedure for calculating and the generic form of the decay amplitudes, the branching ratios, and the CP asymmetries of $t \to c \gamma (g)$ decays at the one-loop order in the A2HDM are presented. The numerical results and discussions are then given in section~\ref{sec:numerical results}, where we show the maximum values of these observables that can be achieved in the A2HDM, together with their variations with respect to the model parameters. Our conclusion is finally made in section~\ref{sec:conclusion}. For convenience, several appendices are also provided. 

\section{Aligned two-Higgs-doublet model}
\label{sec:A2HDM}

\subsection{Higgs basis}
\label{sec:higgsbasis}

The 2HDM extends the SM with a second complex scalar $SU(2)_{L}$ doublet that has the same weak hypercharge as the SM one, \textit{i.e.} $Y=\frac{1}{2}$. Making use of the freedom in defining the two scalar doublets and assuming that the vacua of the theory respect the $U(1)_{em}$ symmetry, we can, without loss of generality, parametrize the two scalar doublets as~\cite{Branco:2011iw,Penuelas:2017ikk} 
\begin{equation}
	\phi_{a}=\mathrm{e}^{i \theta_{a}}\left[\begin{array}{c}
		\phi_{a}^{+} \\[0.1cm]
		\frac{1}{\sqrt{2}}\left(v_{a}+\rho_{a}+i \eta_{a}\right)
	\end{array}\right]\,, \qquad (a=1,2)\,,
\end{equation}
where each vacuum expectation value (vev) could be complex, $\langle 0\vert \phi_a^T(x)\vert 0 \rangle=(0,v_a e^{i \theta_a}/\sqrt{2})$, with $v_a \geq 0$. As only the relative phase $\theta \equiv \theta_2-\theta_1$ is relevant, we can further enforce $\theta_1 = 0$ through an appropriate $U(1)_{Y}$ transformation. 

To make the separation between the physical scalars and the Goldstone modes clear, it is more convenient to perform a suitable $SU(2)$ global transformation that rotates the original scalar basis $(\phi_1,\phi_2)$ to the so-called Higgs basis $(\Phi_1,\Phi_2)$~\cite{Davidson:2005cw,Haber:2006ue,Haber:2010bw}, 
\begin{equation}
	\left( \begin{array}{c} \Phi_1 \\[0.1cm] -\Phi_2 \end{array} \right) \equiv
	\left( \begin{array}{cc} \cos\beta & \sin\beta \\[0.1cm] \sin\beta & -\cos\beta \end{array} \right)\,
	\left( \begin{array}{c} \phi_1 \\[0.1cm]  \mathrm{e}^{-i\theta}\phi_2 \end{array} \right)\,,
\end{equation}
such that only the first scalar doublet $\Phi_1$ acquires a non-zero and real vev, $\langle 0\vert \Phi_1^T(x)\vert 0 \rangle=(0,v/\sqrt{2})$, with $v=\sqrt{v_{1}^{2}+v_{2}^{2}}=246~\gev$. The rotation angle $\beta$ is defined by $\tan\beta \equiv e^{-i\theta} \langle \phi_{2}^{0} \rangle / \langle \phi_{1}^{0} \rangle = v_2/v_1$ and, by convention, its value is limited to the first quadrant due to $v_{1,2}\geq0$. In the Higgs basis, the two scalar doublets can then be parametrized as~\cite{Pich:2009sp}
\begin{equation} \label{eq:Higgsbasis}
	\Phi_1=\left[ \begin{array}{c} G^+ \\[0.1cm] \frac{1}{\sqrt{2}}\, (v+S_1+iG^0) \end{array} \right] \,,
	\qquad
	\Phi_2 = \left[ \begin{array}{c} H^+ \\[0.1cm] \frac{1}{\sqrt{2}}\, (S_2+iS_3) \end{array} \right] \,.
\end{equation}
It can be seen that the Goldstone fields $G^{\pm}$ and $G^{0}$ now fill only the first doublet, just as in the SM. The physical scalar spectrum contains a pair of charged fields $H^{\pm}(x)$ and three neutral ones $h(x)$, $H(x)$, $A(x)$, which are linear combinations of the neutral components $S_{1,2,3}(x)$ through an orthogonal transformation, $\varphi^0_i(x) =\{h(x), H(x), A(x)\}=\mathcal{R}_{ij}S_j$. Here the orthogonal matrix $\mathcal{R}$ determines the mass eigenstates of the three neutral scalars, and its explicit form is fixed by the scalar potential of the theory~\cite{Branco:2011iw,Celis:2013rcs}.

\subsection{Scalar sector}
\label{sec:scalar sector}

In the Higgs basis, the most general scalar potential allowed by the SM gauge symmetry is given by~\cite{Branco:2011iw,Celis:2013rcs}
\begin{align} \label{eq:potential}
	V &= \mu_{1}\,\left(\Phi_{1}^{\dagger}\Phi_{1}\right)\,+\,\mu_{2}\,\left(\Phi_{2}^{\dagger}\Phi_{2}\right)\,
	+\,\left[\mu_{3}\,\left(\Phi_{1}^{\dagger}\Phi_{2}\right)\,+\,\mu_{3}^{\ast}\,\left(\Phi_{2}^{\dagger}\Phi_{1}
	\right)\right]\nonumber\\[0.13cm]
	& +\lambda_{1}\,\left(\Phi_{1}^{\dagger}\Phi_{1}\right)^{2}\,+\,\lambda_{2}\,\left(\Phi_{2}^{\dagger}\Phi_{2}\right)^{2}\,
	+\,\lambda_{3}\,\left(\Phi_{1}^{\dagger}\Phi_{1}\right)\left(\Phi_{2}^{\dagger}\Phi_{2}\right)\,
	+\,\lambda_{4}\,\left(\Phi_{1}^{\dagger}\Phi_{2}\right)\left(\Phi_{2}^{\dagger}\Phi_{1}\right)\nonumber\\[0.13cm]
	& +\left[\left(\lambda_{5}\,\Phi_{1}^{\dagger}\Phi_{2}\,+\,\lambda_{6}\,\Phi_{1}^{\dagger}\Phi_{1}\,+\,\lambda_{7}\,
	\Phi_{2}^{\dagger}\Phi_{2}\right)\left(\Phi_{1}^{\dagger}\Phi_{2}\right)\,+\,\mathrm{h.c.}\right]\,,
\end{align}
where, due to the hermiticity of the scalar potential, all the parameters should be real except $\mu_{3}$ and $\lambda_{5,6,7}$. Using the minimization conditions of the scalar potential, $\langle 0\vert \Phi_1^T(x)\vert 0 \rangle=(0,v/\sqrt{2})$ and $\langle 0\vert \Phi_2^T(x)\vert 0 \rangle=(0,0)$, we have the relations
\begin{align} \label{eq:minimizationcondition}
        \mu_1= -\lambda_1\,v^2,\qquad \mu_3 = -\frac{1}{2}\,\lambda_6\,v^2\,,
\end{align}
which allow us to trade the parameters $\mu_1$ and $\mu_3$ by $\lambda_1$ and $\lambda_6$, respectively.

Plugging eq.~\eqref{eq:Higgsbasis} into eq.~\eqref{eq:potential}, we can obtain the mass terms for the scalars from the part of the scalar potential that is bilinear in the fields,
\begin{eqnarray}
	V_{2} &=& m_{H^{\pm}}^{2}H^{+}H^{-} + \frac{1}{2}\begin{pmatrix}
	S_{1} & S_{2} & S_{3}	
	\end{pmatrix} \mathcal{R}^T\,\mathcal{R}\, \mathcal{M}\, \mathcal{R}^T\,\mathcal{R} \begin{pmatrix}
	S_{1} \\
	S_{2} \\
	S_{3}	
	\end{pmatrix} \nonumber \\
    &=& m_{H^{\pm}}^{2}H^{+}H^{-} + \frac{1}{2} m_{h}^{2}\,h^{2} + \frac{1}{2} m_{H}^{2}\,H^{2} + \frac{1}{2} m_{A}^{2}\,A^{2} \,,
\end{eqnarray}
where the charged-Higgs mass squared is given by
\begin{eqnarray}   
  m_{H^{\pm}}^{2}=\mu_{2}+\frac{1}{2} \lambda_{3} v^{2} \,.
\end{eqnarray}
The masses of the three physical neutral scalars are obtained by diagonalizing the symmetric mass-squared matrix $\mathcal{M}$ with the orthogonal matrix $\mathcal{R}$, \textit{i.e.}, $\mathrm{diag}\left( m_h^2, m_H^2,m_A^2\right)=\mathcal{R}\,\mathcal{M}\,\mathcal{R}^T$~\cite{Branco:2011iw,Celis:2013rcs}. In the CP-conserving limit of the scalar potential, $\lambda_{5,6,7}$ are all real, and the neutral field $S_3(x)$ does not mix with the other two. In this case, the CP-odd field $A(x)$ corresponds to the field $S_{3}(x)$, while the two CP-even scalars $h(x)$ and $H(x)$ are orthogonal combinations of $S_1(x)$ and $S_2(x)$. The orthogonal matrix $\mathcal{R}$ is now simplified as  
\begin{eqnarray} \label{eq: R matrix}
	\mathcal{R} = \begin{pmatrix}
		\cos{\tilde\alpha} & \sin{\tilde\alpha} & 0\\[0.1cm]
		-\sin{\tilde\alpha} & \cos{\tilde\alpha} & 0\\[0.1cm]
		0 & 0 & 1
	\end{pmatrix}\,,
\end{eqnarray}
where the mixing angle $\tilde\alpha$ is determined by
\begin{equation}\label{eq:tanalphatilde}
	\sin2{\tilde\alpha} = \frac{-2\lambda_6 v^2 }{m_H^2-m_{h}^{2}}\,,\qquad
	\cos2{\tilde\alpha} = \frac{m_A^2+2(\lambda_{5}-\lambda_{1}) v^2 }{m_H^2-m_{h}^{2}} \,.
\end{equation}
In such a CP-symmetric limit, the squared masses of the three neutral scalars are given, respectively, by~\cite{Celis:2013rcs}
\begin{eqnarray}
       m_{h}^2= \frac{1}{2}(\Sigma-\Delta)\,,\qquad m_{H}^2= \frac{1}{2}(\Sigma+\Delta)\,,\qquad m_{A}^2= m_{H^{\pm}}^2+v^2 \left(\frac{\lambda_4}{2}-\lambda_5\right)\,,
\end{eqnarray}
with 
\begin{align}
	\Sigma &= m_{H^{\pm}}^2+v^2\,\left(2\lambda_1+\frac{\lambda_4}{2}+\lambda_5\right),\\[0.15cm]
	\Delta &= \sqrt{\left[m_{A}^2+2v^2\,(\lambda_5-\lambda_1)\right]^2+4v^4\,\lambda_6^{2}}.
\end{align}
Here we have adopted the convention $m_{H} \ge m_{h}$, and used the freedom of re-phasing the second scalar doublet $\Phi_2 \to \mathrm{e}^{i \eta} \Phi_2$ in the Higgs basis to fix our choices of $\lambda_{6} \le 0$ and hence $0 \le \tilde\alpha \le \frac{\pi}{2}$~\cite{ONeil:2009fty,Cai:2022xha}. In particular, the SM limit is recovered when $\tilde\alpha=0$.

\subsection{Yukawa sector}
\label{sec:yukawa sector}

Specific to the Higgs basis, the Yukawa interactions of the two scalar doublets $\Phi_{1,2}(x)$ with the SM fermions are described by the following most general Lagrangian~\cite{Branco:2011iw,Pich:2009sp}:
\begin{align}\label{eq:Yukawa_interaction}
	\mathcal{L}_{Y,\text{weak}} = & -\frac{\sqrt{2}}{v}\,\Big[\bar{Q}^\prime_L (M^\prime_d \Phi_1 + Y^\prime_d \Phi_2) d^\prime_R + \bar{Q}^\prime_L (M^\prime_u \tilde{\Phi}_1 + Y^\prime_u \tilde{\Phi}_2) u^\prime_R \nonumber \\[0.1cm]
	& \quad \quad\,\, + \bar{L}^\prime_L (M^\prime_\ell \Phi_1 + Y^\prime_\ell \Phi_2) \ell^\prime_R \Big] + \mathrm{h.c.}\,,
\end{align}
where $Q_L^\prime$ and $L_L^\prime$ are the left-handed quark and lepton doublets, and $u^\prime_R$, $d^\prime_R$, $\ell^\prime_R$ denote the right-handed fermion singlets, all being written as three-dimensional flavour vectors; for instance, $u^\prime_R=(u^\prime_R,c^\prime_R,t^\prime_R)^T$, and similarly for $d^\prime_R$, $\ell^\prime_R$, $Q_L^\prime$ and $L_L^\prime$. The superscript `$^\prime$' indicates that the fermion fields are all given in the weak-interaction basis. The charge-conjugated fields $\tilde{\Phi}_a(x) \equiv i\sigma_2\Phi_a^{\ast}(x)$, with $\sigma_2$ the second Pauli matrix, have a weak hypercharge $Y=-1/2$. The non-diagonal matrices $M^{\prime}_f$ ($f=u,d,\ell$) encode both the fermion masses and the Yukawa couplings of the first scalar doublet $\Phi_{1}$ to fermions, while $Y^{\prime}_f$ characterize only the Yukawa interactions of the second scalar doublet $\Phi_{2}$ with fermions. 

It should be noted that, for a given type of the right-handed fermion field $u^\prime_R$, $d^\prime_R$, or $\ell^\prime_R$, the corresponding matrices $M^{\prime}_f$ and $Y^{\prime}_f$ in eq.~\eqref{eq:Yukawa_interaction} cannot be simultaneously diagonalized in the flavour space. When the theory is expressed in terms of the fermion mass-eigenstate basis, $f_{L,R}=U_{L,R}^{f\dagger}f^\prime_{L,R}$, the mass matrices $M_f=U_L^{f\dagger}M^{\prime}_fU_R^f$ will be diagonal, while the Yukawa matrices $Y_f=U_L^{f\dagger}Y^{\prime}_fU_R^f$ remain still non-diagonal and hence give rise to non-vanishing tree-level FCNC interactions in the neutral scalar sector. These unwanted tree-level FCNC vertices can, however, be eliminated by requiring that the two matrices $M^{\prime}_f$ and $Y^{\prime}_f$ are aligned in the flavour space~\cite{Pich:2009sp,Penuelas:2017ikk}. This results in the following alignment relations between the mass and the Yukawa matrices~\cite{Pich:2009sp}:
\begin{equation}\label{eq:alignment}
	Y_{d,\ell}=\varsigma_{d,\ell} \, M_{d,\ell}, \qquad  Y_u=\varsigma_u^* \, M_u,
\end{equation}
where $\varsigma_{f}$ are the alignment parameters that can be, in general, arbitrary complex and hence bring about new sources of CP violation beyond the SM. These parameters also satisfy universality among the three different generations and are scalar-basis independent~\cite{Pich:2009sp}. When  $\varsigma_{f}$ take the particular values as shown in table~\ref{tab:2HDM}, the four conventional 2HDMs based on discrete $Z_{2}$ symmetries can be recovered.

\begin{table}[t]
	\begin{center}	
		\renewcommand{\arraystretch}{1.3}
		\tabcolsep=0.315cm
        \begin{adjustbox}{width=0.98\textwidth,center}
        \begin{tabular}{|c||c|c|c|c|c|c||c|c|c|}
			\hline\hline
                 & $\phi_{1}$ & $\phi_{2}$ & $u_{R}$ & $d_{R}$ & $e_{R}$ & $Q_{L}$, $L_{L}$ & $\varsigma_{u}$ & $\varsigma_{d}$ & $\varsigma_{\ell}$ \\
			\hline 
                \text{Type-I } & $+$ & $-$ & $-$ & $-$ & $-$ & $+$ & $\cot\beta$ & $ \cot\beta$ & $ \cot\beta$ \\
			\text{Type-II} & $+$ & $-$ & $-$ & $+$ & $+$ & $+$ & $\cot\beta$ & $-\tan\beta$ & $-\tan\beta$ \\
			\text{Type-X } & $+$ & $-$ & $-$ & $-$ & $+$ & $+$ & $\cot\beta$ & $ \cot\beta$ & $-\tan\beta$ \\
			\text{Type-Y } & $+$ & $-$ & $-$ & $+$ & $-$ & $+$ & $\cot\beta$ & $-\tan\beta$ & $ \cot\beta$ \\
			\hline\hline
		\end{tabular}
        \end{adjustbox}
		\caption{$Z_{2}$-charge assignments on the scalar and fermion fields in the four conventional 2HDMs and the corresponding values of the alignment parameters $\varsigma_{f}$ in the A2HDM.} \label{tab:2HDM} 
	\end{center}
\end{table}

In terms of the fermion mass-eigenstate fields $f_{L,R}$ and with the alignment conditions specified by eq.~\eqref{eq:alignment} taken into account, the Yukawa Lagrangian of the A2HDM can be finally rewritten as~\cite{Pich:2009sp,Cai:2022xha}
\begin{align} \label{eq:yukawa in mass basis}
	\mathcal{L}_{Y,\text{mass}} =&-i \frac{G^0}{v} \Big\{\bar{d}_{L} M_{d} d_{R}+\bar{u}_{R} M_{u}^{\dagger} u_{L}+\bar{\ell}_{L} M_{\ell} \ell_{R}\Big\} \nonumber \\[0.1cm]
	& -\left(1+\frac{S_{1}}{v}\right) \Big\{\bar{u}_{L} M_{u} u_{R}+\bar{d}_{L} M_{d} d_{R}+\bar{\ell}_{L} M_{\ell} \ell_{R}\Big\} \nonumber \\[0.1cm]
	& -\frac{1}{v} \left(S_2+i S_3\right) \Big\{\varsigma_d\,\bar{d}_{L} M_{d} d_{R}+\varsigma_u\,\bar{u}_{R} M_{u}^{\dagger} u_{L}+\varsigma_\ell\,\bar{\ell}_{L} M_{\ell} \ell_{R}\Big\}\nonumber \\[0.1cm]
	& - \frac{\sqrt{2}}{v} G^{+} \Big\{\bar{u}_{L} V_{\mathrm{CKM}} M_{d} d_{R} -\bar{u}_{R} M_{u}^{\dagger} V_{\mathrm{CKM}} d_{L} 
	+\bar{\nu}_{L} M_{\ell} \ell_{R} \Big\}\nonumber \\[0.1cm]
	& - \frac{\sqrt{2}}{v} H^{+} \Big\{\varsigma_d\,\bar{u}_{L} V_{\mathrm{CKM}} M_{d} d_{R} - \varsigma_u\,\bar{u}_{R} M_{u}^{\dagger} V_{\mathrm{CKM}} d_{L} + \varsigma_\ell\,\bar{\nu}_{L} M_{\ell} \ell_{R}\Big\}+\mathrm{h.c.}\,,
\end{align}
where $V_{\text{CKM}}=U_L^{u\dagger}U_L^d$ is the CKM matrix~\cite{Cabibbo:1963yz,Kobayashi:1973fv}, and the diagonal fermion mass matrices $M_f$ are given, respectively, by
\begin{align}
    M_{u}=\mathrm{diag}\left(m_{u}, m_{c}, m_{t}\right), \quad
    M_{d}=\mathrm{diag}\left(m_{d}, m_{s}, m_{b}\right), \quad M_{\ell}=\mathrm{diag}\left(m_{e}, m_{\mu}, m_{\tau}\right)\,.
\end{align}
It can be seen from eq.~\eqref{eq:yukawa in mass basis} that all fermion-scalar interactions are proportional to the corresponding fermion mass matrices, as in the SM. The tree-level FCNC vertices are automatically absent in the A2HDM, and the only source of flavour-changing interactions is the CKM matrix $V_{\text{CKM}}$, which appears in the couplings of $H^{\pm}$ and $W^{\pm}$ with fermions.

For the rare top-quark decays $t\to c \gamma$ and $t\to c g$ considered here, the NP contribution at the one-loop order involves only the interactions of charged scalars $H^{\pm}$ with quarks. This means that only the two alignment parameters $\varsigma_{u,d}$ and the charged-Higgs mass $m_{H^{\pm}}$ are involved throughout this work. For convenience, the relevant Feynman rules for the charged-Higgs contributions to the decays are collected in appendix~\ref{sec:Feynman rules}.

\section{Calculation of the decay processes}
\label{sec:decay process}

In this section, let us establish the theoretical framework for calculating the rare top-quark decays $t \to q \gamma$ and $t \to q g$, with the particle polarization information kept, at the one-loop order in the A2HDM. In addition to the branching ratios~\cite{Cai:2022xha,Abbas:2015cua}, we shall pay particular attention to the CP asymmetries of these rare FCNC decays, because the complex alignment parameters $\varsigma_{u,d}$ in the model can provide new sources of CP violation beyond the SM, and thus potentially large deviations from the corresponding SM predictions~\cite{Deshpande:1990ua,Aguilar-Saavedra:2002lwv,Balaji:2020qjg} could be expected in these observables. As the decay rates with an up quark in the final state are suppressed by the ratio $|V_{ub}/V_{cb}|^{2}\simeq 0.0088$ compared to that with a charm both within the SM and in the A2HDM, we shall focus on the $t \to c \gamma$ and $t \to c g$ decays. 

\subsection{Procedure for calculating and generic form of the decay amplitudes}  
\label{sec:decay amplitude}

\begin{figure}[t]
  \centering
  \includegraphics[width=0.98\textwidth]{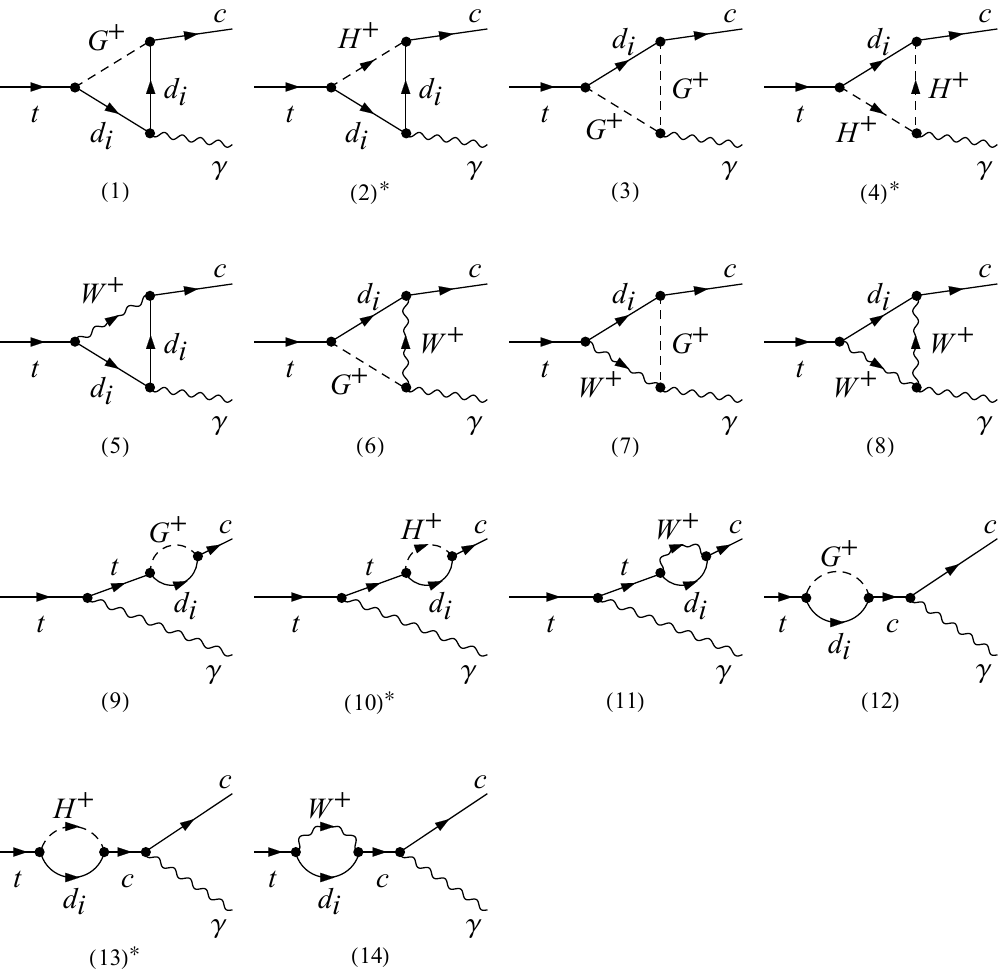}
  \caption{One-loop Feynman diagrams contributing to the $t \to c\gamma$ decay in the 't Hooft-Feynman gauge, with $d_i=d,s,b$, in the A2HDM, where the first eight correspond to the vertex diagrams, while the remaining ones to the flavour-changing $t-c$ fermion self-energy diagrams. The diagrams indexed with a superscript `$^\ast$' result from the charged-Higgs contributions.} \label{Fig:tcg}
\end{figure}

In the A2HDM, due to the absence of tree-level FCNC interactions, both the $t \to c \gamma$ and $t \to c g$ decays occur firstly at the one-loop order~\cite{Cai:2022xha,Abbas:2015cua}, as within the SM~\cite{Eilam:1990zc,Aguilar-Saavedra:2004mfd}. The corresponding one-loop Feynman diagrams in the 't Hooft-Feynman gauge are depicted in Fig.~\ref{Fig:tcg} for the $t \to c \gamma$ decay, where the first eight correspond to the vertex diagrams, while the remaining ones to the flavour-changing $t-c$ fermion self-energy diagrams.\footnote{The one-loop Feynman diagrams contributing to the $t \to c g$ decay in the 't Hooft-Feynman gauge in the A2HDM have already been given in Fig.~1 of ref.~\cite{Cai:2022xha}, which can also be obtained from Fig.~\ref{Fig:tcg} shown here by replacing each external photon by a gluon line and omitting the diagrams involving the triple-boson vertices (\textit{i.e.}, the diagrams indexed by $(3)$, $(4)^\ast$, and $(6)$--$(8)$ in Fig.~\ref{Fig:tcg}).}  The diagrams labelled with a superscript `$^\ast$' represent the additional charged-Higgs contributions to the decay, which are of the same order as the $W$-boson contributions within the SM. 

For the calculation of these one-loop diagrams, we have followed the same procedure as in our previous studies~\cite{Cai:2022xha,Abbas:2015cua}: Firstly, to generate the model file together with a complete set of Feynman rules, we have implemented the A2HDM into the Mathematica-based package \texttt{FeynRules}~\cite{Christensen:2008py,Alloul:2013bka}. The generated model file is then exported into the package \texttt{FeynArts}~\cite{Kublbeck:1990xc,Hahn:2000kx}, based on which we can obtain the one-loop Feynman diagrams as shown in Fig.~\ref{Fig:tcg}, as well as the corresponding amplitudes for the $t \to c \gamma$ and $t \to c g$ decays. To manipulate the decay amplitudes generated and numerically evaluate the one-loop Feynman integrals, we have used the packages \texttt{FeynCalc}~\cite{Mertig:1990an,Shtabovenko:2016sxi,Shtabovenko:2020gxv,Shtabovenko:2023idz} and \texttt{LoopTools}~\cite{Hahn:1998yk}, respectively. In the meantime, some partial cross-checks have been made with the help of the \texttt{Package-X}~\cite{Patel:2015tea,Patel:2016fam}. Finally, we can get the branching ratios and CP asymmetries from the squared amplitudes by performing the necessary phase-space integration. Throughout this work, the one-loop calculation is performed in $D=4-2\epsilon$ space-time dimensions, and the dimensional regularization scheme~\cite{tHooft:1972tcz,Bollini:1972ui} is used to regulate the ultraviolet divergence present in the loop integrals. We have also made the GIM mechanism manifest by dropping the terms that are independent of the internal down-type quark masses $m_{d_i}$, after summing over the three internal quark flavours $d_i=d,s,b$ and imposing the unitarity relation $\sum_{d_i=d,s,b}V_{td_i}^{\ast}V_{cd_i}=0$. The sum of the amplitudes resulting from all the diagrams shown in Fig.~\ref{Fig:tcg} is then found to be ultraviolet finite. Although the whole calculation is carried out in the 't Hooft-Feynman gauge, in which the divergences are more manageable but at the cost of having extra diagrams with un-physical scalars, we have checked explicitly the gauge independence of our final results by performing the same calculation in an arbitrary $R_{\xi}$ gauge.

Let us now detail the generic form of the polarized $t\to c \gamma_{\pm}$ amplitudes in the A2HDM, where $\gamma_{+}$ and $\gamma_{-}$ represent the positively and negatively polarized photons, respectively. Here we follow the same conventions as used in refs.~\cite{Balaji:2019fxd,Balaji:2020oig,Balaji:2020qjg,Balaji:2021lpr} to specify the polarization information of the initial- and final-state particles. Explicitly, the transition amplitudes of the $t(p_{\text{i}}) \to c(p_{\text{f}}) \gamma_{\pm}(q)$ decays can be, in full generality, written as 
\begin{equation} \label{eq:decay_amplitude}
	i \mathcal{M} (t \to c + \gamma_{\pm})  = i \bar{u}(p_{\text{f}}) \,\Gamma_{\text{fi}, \gamma}^{\mu}(q^2) \,u(p_{\text{i}}) \,\varepsilon^{\ast}_{\pm,\mu}(q)\,, 
\end{equation}
where $u(p_{\text{i}})$ and $u(p_{\text{f}})$ are the Dirac spinors of the initial top and the final charm state respectively, and $\varepsilon_{\pm}^{\mu}(q)$ denote the photon polarization vectors, with its momentum given by $q=p_{\text{i}}-p_{\text{f}}$ from energy-momentum conservation. Their explicit expressions in the initial top-quark rest frame can be found in appendix~\ref{sec:Polarized amplitudes}. As $q^2$ is the only available Lorentz-invariant kinematic quantity for an on-shell top quark ($p_{\text{i}}^2=m_t^2$) decaying into an on-shell charm quark ($p_{\text{f}}^2=m_c^2$), the one-loop effective vertex $\Gamma_{\text{fi}, \gamma}^\mu(q^2)$ depends only on $q^2$. Making use of the Gordon identities and the unitarity of the CKM matrix $\sum_{d_i=d,s,b}V_{td_i}^{\ast}V_{cd_i}=0$, we can, in the most general case, decompose $\Gamma_{\text{fi}, \gamma}^\mu(q^2)$ as~\cite{Deshpande:1981zq,Deshpande:1991pn}
\begin{equation} \label{eq:vertex_general}
       \Gamma_{\text{fi}, \gamma}^\mu(q^2) = \left(q^2 \gamma^{\mu}-q^{\mu}q\!\!\!/\right) \left(F_{1,\text{fi}}^{L}P_{L}+F_{1,\text{fi}}^{R}P_{R}\right) +i\sigma^{\mu\nu}q_{\nu} \left(F_{2,\text{fi}}^{L}m_{c}P_{L}+F_{2,\text{fi}}^{R}m_{t}P_{R}\right)\,,
\end{equation}
as required by Lorentz and electromagnetic gauge invariance. Here $\sigma^{\mu\nu}=\frac{i}{2}[\gamma^{\mu},\gamma^{\nu}]$, and the chiral projection operators are defined as $P_{L,R}= \frac{1}{2}(1 \mp \gamma_{5})$. When the emitted photons are taken to be on-shell, we have $q^2=0$ and $q \cdot \varepsilon_{\pm}(q)=0$, which dictate that the first term in eq.~\eqref{eq:vertex_general} has a vanishing contribution to the decays we are considering. The effective vertex can then be rewritten as
\begin{equation} \label{eq:vertex function1}
      \Gamma_{\text{fi}, \gamma}^\mu(q^2) = i\sigma^{\mu\nu}q_{\nu} \left(f_{\text{fi},\gamma}^{L}P_{L}+f_{\text{fi},\gamma}^{R}P_{R}\right)\,,
\end{equation}
where the form factors $f_{\text{fi},\gamma}^{L}$ and $f_{\text{fi},\gamma}^{R}$ contain information about the couplings, the CKM matrix elements, and the one-loop integrals. For convenience, detailed derivations of the polarized amplitudes of $t\to c \gamma_{\pm}$ and their CP-conjugated modes $\bar{t}\to \bar{c} \gamma_{\mp}$ are presented in appendix~\ref{sec:Polarized amplitudes}. A similar analysis can also be applied to the $t\to cg_{\pm}$ decays. 

To be more specific, we can parametrize the one-loop effective vertex for the processes $t \to c \gamma_{\pm}$ in the A2HDM in terms of the product of the alignment parameters $\varsigma_{u} \varsigma_{d}^{\ast}$ as
\begin{equation} \label{eq:effective vertex2}
	 \Gamma_{\text{fi}, \gamma}^\mu(q^2)  =  i\sigma^{\mu\nu}q_{\nu} V_{t\alpha}^{\ast} V_{c\alpha} \left[  \left(\mathcal{F}_{\alpha}^{L}+ \varsigma_{u} \varsigma_{d}^{\ast} \mathcal{N}^{L}_{\alpha} \right)m_{c} P_{L} + \left(\mathcal{F}_{\alpha}^{R}+ \varsigma_{d} \varsigma_{u}^{\ast} \mathcal{N}^{R}_{\alpha} \right) m_{t} P_{R} \right]\,,  
\end{equation}
where $\alpha$ must be summed over all the three contributing down-type quarks, $\alpha=d,s,b$, and $V_{t\alpha}$ and $V_{c\alpha}$ are the CKM matrix elements. Among the pure loop kinetic terms, $\mathcal{F}_{\alpha}^{L}$ and $\mathcal{F}_{\alpha}^{R}$ result from both the SM contribution and the NP terms associated with $|\varsigma_{u}|^2$ and $|\varsigma_{d}|^2$, while $\mathcal{N}_{\alpha}^{L}$ and $\mathcal{N}_{\alpha}^{R}$ from the NP contributions associated with $\varsigma_{u}\varsigma_{d}^{\ast}$ and $\varsigma_{d}\varsigma_{u}^{\ast}$, respectively. Defining the relative phase $\varphi$ between the two alignment parameters $\varsigma_{u}$ and $\varsigma_{d}$ as $\varsigma_{u}^{\ast} \varsigma_{d}=|\varsigma_{u}||\varsigma_{d}|e^{-i\varphi}$, we can see that the two terms in eq.~\eqref{eq:effective vertex2} involve different weak phases, which is crucial for CP studies. It should be noted that, due to the hermiticity of the effective electromagnetic current $\bar{u}(p_{\text{f}}) \,\Gamma_{\text{fi}, \gamma}^{\mu}(q^2) \,u(p_{\text{i}})$ introduced in eq.~\eqref{eq:decay_amplitude}, the loop kinetic terms $\mathcal{F}_{\alpha}^{L}$ ($\mathcal{N}^{L}_{\alpha}$) and $\mathcal{F}_{\alpha}^{R}$ ($\mathcal{N}^{R}_{\alpha}$) are symmetric about the initial top and final charm masses; \textit{i.e.}, $\mathcal{F}_{\alpha}^{R}$ and $\mathcal{N}^{R}_{\alpha}$ are obtained from $\mathcal{F}_{\alpha}^{L}$ and $\mathcal{N}^{L}_{\alpha}$ by interchanging $m_{c}$ and $m_{t}$, respectively. These loop kinetic terms can be expressed in terms of the one-loop Feynman integrals, and their explicit results for the $t\to c\gamma$ and $t\to cg$ decays are collected in appendix~\ref{sec:loop kinetic terms}. Combining eqs.~\eqref{eq:vertex function1} with \eqref{eq:effective vertex2}, we have the following relations between the form factors and the loop kinetic terms:
\begin{equation} \label{eq:ffs1}
\begin{aligned}
	f_{\text{fi},\gamma}^{L} = V_{t \alpha}^{\ast} V_{c \alpha} \left(\mathcal{F}_{\alpha}^{L}+\varsigma_{u} \varsigma_{d}^{\ast} \mathcal{N}^{L}_{\alpha} \right)m_{c}\,, \\[0.2cm]
	f_{\text{fi},\gamma}^{R} = V_{t \alpha}^{\ast} V_{c \alpha} \left(\mathcal{F}_{\alpha}^{R}+ \varsigma_{d} \varsigma_{u}^{\ast} \mathcal{N}^{R}_{\alpha} \right)m_{t}\,.
\end{aligned} 
\end{equation}
As the charm-quark mass is much smaller than that of the top quark, the effective couplings are predominantly right-handed. Especially in the limit $m_{c}=0$, the form factor $f_{\text{fi},\gamma}^{L}$ will vanish. During our numerical evaluations, however, we have kept all the particle masses.

\subsection{Derivation of the branching ratios}  
\label{sec: branching ratio formulas}

With the polarized amplitudes at hand (see appendix~\ref{sec:Polarized amplitudes} for details), it is now easy to obtain the polarized decay widths $\Gamma(t\to c+\gamma_{\pm})$ by integrating over the two-body phase space, 
\begin{eqnarray} \label{eq:decay width}
	\Gamma(t\to c+\gamma_{\pm} ) = \frac{m_{t}^{2}-m_{c}^{2}}{16 \pi m_{t}^{3}}|\mathcal{M}(t\to c+\gamma_{\pm} )|^{2}\,. 
\end{eqnarray}
As there are only two non-vanishing polarized amplitudes for $t \to c\gamma_{\pm}$ decays, with
\begin{equation} \label{eq:amplitude1}
\begin{aligned}
	\mathcal{M}(t \to c + \gamma_+) = +\sqrt{2} f_{\text{fi}, \gamma}^{L} 
        \left(m_t^2 - m_c^2\right)\,, \\[0.15cm]
	\mathcal{M}(t \to c + \gamma_-) = - \sqrt{2} f_{\text{fi}, \gamma}^{R} 
        \left(m_t^2 - m_c^2\right)\,,
\end{aligned}
\end{equation}
the corresponding polarized decay widths can be written explicitly as
\begin{eqnarray} 
	\Gamma(t\to c+\gamma_{+} ) &=& \frac{(m_{t}^{2}-m_{c}^{2})^3}{8 \pi m_{t}^{3}} |f_{\text{fi}, \gamma}^{L}|^{2}\,, \label{eq:decaywidthtcr1} \\[0.15cm]
	\Gamma(t\to c+\gamma_{-} ) &=& \frac{(m_{t}^{2}-m_{c}^{2})^3}{8 \pi m_{t}^{3}}|f_{\text{fi}, \gamma}^{R}|^{2}\,. \label{eq:decaywidthtcr2}
\end{eqnarray}
The total un-polarized decay width of $t \to c \gamma$ decay can be obtained by summing over the two polarized decay widths and averaging over the initial-state spins, $\Gamma(t\to c+\gamma )=\frac{1}{2}[\Gamma(t\to c+\gamma_{+})+\Gamma(t\to c+\gamma_{-})]$, which results in
\begin{eqnarray} \label{eq:decaywidthtcr_total}
	\Gamma(t\to c+\gamma ) = \frac{(m_{t}^{2}-m_{c}^{2})^3}{16 \pi m_{t}^{3}}\left(|f_{\text{fi}, \gamma}^{L}|^{2}+|f_{\text{fi}, \gamma}^{R}|^{2}\right)\,.
\end{eqnarray}
Similarly, we can obtain the polarized and un-polarized $t \to cg$ decay widths, 
\begin{eqnarray}
	\Gamma(t\to c + g_{+}) &=& \frac{C_{F}(m_{t}^{2}-m_{c}^{2})^3}{8 \pi m_{t}^{3}}|f_{\text{fi}, g}^{L}|^{2}\,, \\[0.15cm]
 	\Gamma(t\to c + g_{-}) &=& \frac{C_{F}(m_{t}^{2}-m_{c}^{2})^3}{8 \pi m_{t}^{3}}|f_{\text{fi}, g}^{R}|^{2}\,, \\[0.15cm]
	\Gamma(t\to c + g) &=& \frac{C_{F}(m_{t}^{2}-m_{c}^{2})^3}{16 \pi m_{t}^{3}}\left(|f_{\text{fi}, g}^{L}|^{2}+|f_{\text{fi}, g}^{R}|^{2}\right)\,,
\end{eqnarray}
where $C_{F}=\frac{4}{3}$ is the $SU(3)_C$ colour factor, and the form factors $f_{\text{fi}, g}^{L,R}$ can be calculated from the one-loop Feynman diagrams shown in Fig.~\ref{Fig:tcg} by replacing each external photon by a gluon line and omitting the diagrams indexed by $(3)$, $(4)^\ast$ and $(6)$--$(8)$, together with the replacement $iQe\gamma^{\mu}$ for the quark-photon by $i g_{s} \gamma^{\mu} T^{a}$ for the quark-gluon vertices.  

In the SM, the dominant decay channel of the top quark is the tree-level two-body decay $t \to b W^{+}$, which can be approximated as the total width of the top quark, $\Gamma_\mathrm{tot}(t) \simeq \Gamma(t \to b W^+)$. Here we shall use as input the $t \to b W^{+}$ decay width predicted at the next-to-leading order (NLO) in QCD~\cite{Li:1990qf,Eilam:1991iz},\footnote{For the state-of-the-art calculation of the decay width within the SM, we refer the readers to refs.~\cite{Yan:2024hbz,Chen:2023osm,Chen:2022wit,Chen:2023dsi} and references therein.}
\begin{eqnarray} \label{eq:t2bWNLO}
	\Gamma(t \to b W^+) & = & \Gamma_0(t \to b W^+) \Biggl\{ 1+\frac{C_{F}\alpha_s}{2\pi}\left[\,2\frac{(1-\beta_W^2)(2\beta_W^2-1)(\beta_W^2-2)}
	{\beta_W^4(3-2\beta_W^2)}\ln(1-\beta_W^2)\right.\notag\\[0.15cm]
	&& \hspace{-1.5cm} \left. - \frac{9-4\beta_W^2}{3-2\beta_W^2}\ln\beta_W^2+2\mathrm{Li}_2(\beta_W^2) -2\mathrm{Li}_2(1-\beta_W^2)-\frac{6\beta_W^4-3\beta_W^2-8}{2\beta_W^2(3-2\beta_W^2)}-\pi^2\,\right]\Biggr\}\,,
\end{eqnarray}
where $\alpha_{s}=g_{s}^{2}/(4\pi)$ is the strong coupling constant, and $\beta_W = \sqrt{1-m_W^2/m_t^2}$ the velocity of the $W^{+}$ boson in the top-quark rest frame, with $m_{W}$ and $m_{t}$ being the $W$-boson and the top-quark mass respectively. The dilogarithm function is defined by $\mbox{Li}_2(x)=-\int_0^x \text{ln}(1-t)/t\,dt$. The leading-order (LO) decay width $\Gamma_0(t \to b W^+)$ in eq.~\eqref{eq:t2bWNLO} is given by~\cite{Denner:1990ns}
\begin{equation}\label{eq:t2bWLO}
	\Gamma_0(t \to b W^+) =\frac{G_F |V_{tb}|^2 \sqrt{\lambda(m_t,m_b,m_W)}}{8\pi\sqrt{2}\, m_t^3}\Big[(m_t^2-m_b^2)^2+(m_t^2+m_b^2)m_W^2-2m_W^4\Big]\,,
\end{equation}
where $G_{F}$ is the Fermi constant, and $V_{tb}$ the CKM matrix element. $\lambda(x,y,z)=(x^2-y^2-z^2)^2-4y^2z^2$ denotes the usual triangle (or K\"{a}ll\'{e}n) function for a two-body decay, with $m_{b}$ being the bottom-quark pole mass.

In the A2HDM, the charged-Higgs $m_{H^{\pm}}$ can also induce flavour-changing charged-current interactions between an up- and a down-type quark. In the case of $m_{H^{\pm}} < m_t - m_b$, the additional decay mode $t\to b H^{+}$ must be considered to account for the total top-quark width, $\Gamma_\mathrm{tot}(t) \simeq \Gamma(t \to b W^+) + \Gamma(t \to b H^+)$, with the LO partial width $\Gamma(t \to bH^{+})$ in the A2HDM given by~\cite{Abbas:2015cua}
\begin{eqnarray} \label{eq:t2bHLO}
	\Gamma(t \to bH^{+}) &=& \frac{G_F|V_{tb}|^2\sqrt{\lambda(m_t,m_b,m_{H^{\pm}})}}{8\pi\sqrt{2}\, m_t^3} \Big[\left(m_t^2 + m_b^2 - m_{H^{\pm}}^2\right)\left(m_b^2|\varsigma_d|^2 + m_t^2|\varsigma_u|^2\right)\notag\\
	&& \hspace{5.0cm}  
	- 4 m_b^2 m_t^2 \,\mathrm{Re}(\varsigma_d \varsigma_u^{\ast})\,\Big]\,.
\end{eqnarray}
To be consistent with the experimental data~\cite{ParticleDataGroup:2024cfk}, however, the approximation $\Gamma_\mathrm{tot}(t) \simeq \Gamma(t \to b W^+)$ is also expected to hold in the A2HDM. This means that the extra charged-Higgs contribution to the total top-quark width should be negligibly small. For this purpose, we shall assume that the charged-Higgs mass $m_{H^{\pm}}$ is larger than the top-quark mass $m_{t}$, and the $t\to bH^{+}$ process will be, therefore, kinematically forbidden in the A2HDM. In this case, the branching ratios of $t\to c \gamma$ and $t \to c g$ decays can be defined as
\begin{eqnarray}
	\mathcal{B}(t\to c V)=\frac{\Gamma (t\to c V)}{\Gamma(t \to bW^{+})}\,,
\end{eqnarray}
where, depending on the decays we are considering, $V$ can be an un-polarized or a polarized final-state photon (gluon).

\subsection{CP transformation properties of the polarized amplitudes}
\label{sec:CP transformantion}

Now we discuss the CP transformation properties of the polarized amplitudes of $t \to c \gamma_{\pm}$ decays. Under the CP transformation, each particle is replaced by its antiparticle, with all charges and other additive quantum numbers reversed, and its momentum $p=(\,p^{0}, \vec{p}\,)$ is changed to $\tilde{p}=(\,p^{0}, -\vec{p}\,)$, while its spin remains unchanged. Therefore, the CP-conjugated processes of $t(p_{\text{i}})\to c(p_{\text{f}}) \gamma_{\pm} (q)$ are represented by $\bar{t} (\tilde{p}_{\text{i}})\to \bar{c} (\tilde{p}_{\text{f}}) \gamma_{\mp} (\tilde{q})$. As the decay amplitudes are invariant under the spatial rotations and Lorentz boosts, we can always choose a proper reference frame for the decays $\bar{t} (\tilde{p}_{\text{i}})\to \bar{c} (\tilde{p}_{\text{f}}) \gamma_{\mp} (\tilde{q})$, in which $\tilde{p}_{i,f}=p_{i,f}$ and $\tilde{q}=q$. Consequently, the polarized amplitudes of the CP-conjugated modes can be written as
\begin{eqnarray} \label{eq:CP amplitude}
	i\mathcal{M}(\bar{t}\to \bar{c} + \gamma_{\mp}) = i\mathcal{M}^{CP}(t\to c + \gamma_{\pm}) = i \bar{v}(p_{\text{i}}) \bar{\Gamma}_{\text{if}, \gamma}^\mu(q^2) v(p_{\text{f}}) \varepsilon^{\ast}_{\mp,\mu}(q) \,,
\end{eqnarray}
where $v(p_{\text{i}})$ and $v(p_{\text{f}})$ are the Dirac spinors for the initial and final anti-fermion states. The one-loop vertex function $\bar{\Gamma}_{\text{if},\gamma}^{\mu}$ can be written in a similar form as of eq.~\eqref{eq:vertex function1},
\begin{eqnarray}
	\bar{\Gamma}_{\text{if}, \gamma}^\mu(q^2) = i\sigma^{\mu\nu}q_{\nu} \left(\bar{f}_{\text{if}, \gamma}^{L}P_{L}+\bar{f}_{\text{if}, \gamma}^{R}P_{R} \right) \,. \label{eq:vertex function2}
\end{eqnarray}
As detailed in appendix~\ref{sec:Polarized amplitudes}, the non-vanishing polarized amplitudes are determined to be
\begin{equation} 
\begin{aligned}
	\mathcal{M}(\bar{t} \to \bar{c} + \gamma_-) = +\sqrt{2} \bar{f}_{\text{if}, \gamma}^{R} 
        \left(m_t^2 - m_c^2\right)\,, \\[0.15cm]
	\mathcal{M}(\bar{t} \to \bar{c} + \gamma_+) = -\sqrt{2} \bar{f}_{\text{if}, \gamma}^{L} 
        \left(m_t^2 - m_c^2\right) \,, \label{eq:amplitude2}
\end{aligned}
\end{equation}
from which the corresponding polarized and un-polarized decay widths can be obtained, respectively, as
\begin{equation} \label{eq:decaywidth2}
\begin{aligned}
	\Gamma(\bar{t}\to \bar{c} + \gamma_{-}) &= \frac{(m_{t}^{2}-m_{c}^{2})^3}{8 \pi m_{t}^{3}}|\bar{f}_{\text{if}, \gamma}^{R}|^{2}\,, \\[0.15cm]
	\Gamma(\bar{t}\to \bar{c} + \gamma_{+}) &= \frac{(m_{t}^{2}-m_{c}^{2})^3}{8 \pi m_{t}^{3}} |\bar{f}_{\text{if}, \gamma}^{L}|^{2}\,, \\[0.15cm]
	\Gamma(\bar{t}\to \bar{c} + \gamma) &= \frac{(m_{t}^{2}-m_{c}^{2})^3}{16 \pi m_{t}^{3}} \left(|\bar{f}_{\text{if}, \gamma}^{R}|^{2} + |\bar{f}_{\text{if}, \gamma}^{L}|^{2}\right)\,.
\end{aligned} 
\end{equation}

Let us now find out the relations between the form factors $f_{\text{fi}}^{L}$, $f_{\text{fi}}^{R}$ and $\bar{f}_{\text{if}}^{R}$, $\bar{f}_{\text{if}}^{L}$ under the CP transformation. Firstly, using the CP transformation properties of the quark bilinears and the electromagnetic field,
\begin{equation} \label{eq:transformation properties}
\begin{aligned}
	(CP)\,\bar{\psi} \sigma^{\mu \nu} \chi\, (CP)^{\dagger} &= - \bar{\chi} \sigma_{\mu \nu} \psi \,, \\[0.15cm]
	(CP) \bar{\psi} \sigma^{\mu \nu} \gamma_{5} \chi (CP)^{\dagger} &= \bar{\chi} \sigma_{\mu \nu} \gamma_{5} \psi \,, \\[0.15cm]
	(CP) A^{\mu} (CP)^{\dagger} &= -A_{\mu} \,,
\end{aligned}
\end{equation}
one can easily find that the CP transformation flips the chirality in eq.~\eqref{eq:effective vertex2}, \textit{i.e.} $P_{L}\leftrightarrow P_{R}$. Furthermore, due to the hemiticity of the interaction Lagrangian, the complex parameters $V_{t\alpha}$, $V_{c\alpha}$, $\varsigma_{u}$ and $\varsigma_{d}$ will transform to their complex conjugates under the CP transformation. Therefore, the one-loop effective vertex for the $\bar{t}\to \bar{c} \gamma_{\pm}$ decays can be determined as
\begin{eqnarray}
	\bar{\Gamma}_{\text{if}, \gamma}^\mu(q^2)  = i\sigma^{\mu\nu}q_{\nu} V_{t\alpha} V_{c\alpha}^{\ast} \left[ \left(\mathcal{F}_{\alpha}^{R}+ \varsigma_{u} \varsigma_{d}^{\ast} \mathcal{N}^{R}_{\alpha} \right) m_{t} P_{L} + \left(\mathcal{F}_{\alpha}^{L}+ \varsigma_{d} \varsigma_{u}^{\ast} \mathcal{N}^{L}_{\alpha} \right) m_{c} P_{R} \right]\,.  \label{eq:effective vertex3}
\end{eqnarray}
Then, the two form factors $\bar{f}_{\text{if}, \gamma}^{L}$ and $\bar{f}_{\text{if}, \gamma}^{R}$ can be written, respectively, as
\begin{equation} \label{eq:ffs2}
\begin{aligned}
	\bar{f}_{\text{if}, \gamma}^{L} = V_{t \alpha} V_{c \alpha}^{\ast} \left(\mathcal{F}_{\alpha}^{R}+\varsigma_{u} \varsigma_{d}^{\ast} \mathcal{N}^{R}_{\alpha} \right)m_{t}\,, \\[0.15cm]
	\bar{f}_{\text{if}, \gamma}^{R} = V_{t \alpha} V_{c \alpha}^{\ast} \left(\mathcal{F}_{\alpha}^{L}+ \varsigma_{d} \varsigma_{u}^{\ast} \mathcal{N}^{L}_{\alpha} \right)m_{c}\,, 
\end{aligned}
\end{equation} 
where explicit expressions of the loop kinetic terms $\mathcal{F}_{\alpha}^{L,R}$ and $\mathcal{N}^{L,R}_{\alpha}$ are given in appendix~\ref{sec:loop kinetic terms}.

\subsection{Derivation of the CP asymmetries}
\label{sec:CP asymmetries}

The polarized CP asymmetries $\Delta_{\text{CP},+}$ between $t \to c \gamma_{+}$ and its CP-conjugated process $\bar{t} \to \bar{c} \gamma_{-}$ as well as $\Delta_{\text{CP},-}$ between $t \to c \gamma_{-}$ and its CP-conjugated process $\bar{t} \to \bar{c} \gamma_{+}$ are defined, respectively, as~\cite{Balaji:2019fxd}
\begin{eqnarray} \label{eq: polarized CP0}
	\Delta_{\text{CP},+} &=& \frac{\Gamma(t\to c \gamma_{+})-\Gamma(\bar{t} \to \bar{c} \gamma_{-})}{\Gamma(t\to c \gamma)+\Gamma(\bar{t} \to \bar{c} \gamma)} \,, \\[0.15cm]
	\Delta_{\text{CP},-} &=& \frac{\Gamma(t\to c \gamma_{-})-\Gamma(\bar{t} \to \bar{c} \gamma_{+})}{\Gamma(t\to c \gamma)+\Gamma(\bar{t} \to \bar{c} \gamma)} \,.
\end{eqnarray}
The photon polarization independent CP asymmetry of the process can then be expressed as the sum of $\Delta_{\text{CP},+}$ and $\Delta_{\text{CP},-}$, which gives
\begin{eqnarray} \label{eq:Delta_CP_unpolarized}
	\Delta_{\text{CP}} 
	&=& \frac{\Gamma(t\to c \gamma_{+})-\Gamma(\bar{t} \to \bar{c} \gamma_{-})+\Gamma(t\to c \gamma_{-})-\Gamma(\bar{t} \to \bar{c} \gamma_{+})}{\Gamma(t\to c \gamma)+\Gamma(\bar{t} \to \bar{c} \gamma)} \nonumber\\[0.15cm]
	&=& \frac{\Gamma(t\to c \gamma)-\Gamma(\bar{t} \to \bar{c} \gamma)}{\Gamma(t\to c \gamma)+\Gamma(\bar{t} \to \bar{c} \gamma)} \,.	
\end{eqnarray}
Analogous expressions can be obtained for the $t \to cg_{\pm}$ decays. Using the decay widths given by eqs.~\eqref{eq:decaywidthtcr1}--\eqref{eq:decaywidthtcr_total} and \eqref{eq:decaywidth2}, we can further write the three CP asymmetries as
\begin{eqnarray} \label{eq: polarized CP}
	\Delta_{\text{CP},+} &=& \frac{|f_{\text{fi}}^{L}|^{2}-|\bar{f}_{\text{if}}^{R}|^{2} }{|f_{\text{fi}}^{L}|^{2}+|f_{\text{fi}}^{R}|^{2}+|\bar{f}_{\text{if}}^{L}|^{2}+|\bar{f}_{\text{if}}^{R}|^{2}} \,, \\[0.15cm]
	\Delta_{\text{CP},-} &=& \frac{|f_{\text{fi}}^{R}|^{2}-|\bar{f}_{\text{if}}^{L}|^{2}}{|f_{\text{fi}}^{L}|^{2}+|f_{\text{fi}}^{R}|^{2}+|\bar{f}_{\text{if}}^{L}|^{2}+|\bar{f}_{\text{if}}^{R}|^{2}} \,, \\[0.15cm]
	\Delta_{\text{CP}} &=& \frac{|f_{\text{fi}}^{L}|^{2}+|f_{\text{fi}}^{R}|^{2}-|\bar{f}_{\text{if}}^{R}|^{2}-|\bar{f}_{\text{if}}^{L}|^{2} }{|f_{\text{fi}}^{L}|^{2}+|f_{\text{fi}}^{R}|^{2}+|\bar{f}_{\text{if}}^{L}|^{2}+|\bar{f}_{\text{if}}^{R}|^{2}} \,.
\end{eqnarray}

Making use of the parametrizations of the form factors $f_{\text{fi}}^{L,R}$ and $\bar{f}_{\text{if}}^{L,R}$ as given by eqs.~\eqref{eq:ffs1} and \eqref{eq:ffs2}, the two polarized CP asymmetries can be finally expressed in terms of the particle masses, the loop functions, as well as the CKM matrix elements and the alignment parameters as
\begin{align} 
        \Delta_{\text{CP},+} 
	&= -\frac{\sum_{\alpha, \beta} \left[ \mathcal{J}_{\alpha \beta} \operatorname{Im}\left(\mathcal{F}_{\alpha}^{L} 
        \mathcal{F}_{\beta}^{L\ast}\right) +\mathcal{J}_{\alpha \beta}^{\mathbf{N}} \operatorname{Im}\left(\mathcal{N}_{\alpha}^{L} \mathcal{N}_{\beta}^{L\ast}\right)+2\mathcal{J}_{\alpha \beta}^{\mathbf{SN,L}} \operatorname{Im}\left(\mathcal{F}_{\alpha}^{L} \mathcal{N}_{\beta}^{L\ast}\right)  \right ]m_{c}^{2}  }{\mathcal{D} }\,, \label{eq: polarized CPp} \\[0.15cm]
	\Delta_{\text{CP},-}
        &= -\frac{\sum_{\alpha, \beta} \left[ \mathcal{J}_{\alpha \beta} \operatorname{Im}\left(\mathcal{F}_{\alpha}^{R} \mathcal{F}_{\beta}^{R\ast}\right) +\mathcal{J}_{\alpha \beta}^{\mathbf{N}} \operatorname{Im}\left(\mathcal{N}_{\alpha}^{R} \mathcal{N}_{\beta}^{R\ast}\right)+2\mathcal{J}_{\alpha \beta}^{\mathbf{SN,R}} \operatorname{Im}\left(\mathcal{F}_{\alpha}^{R} \mathcal{N}_{\beta}^{R\ast}\right) \right ]m_{t}^{2}  }{\mathcal{D} }\,, \label{eq: polarized CPm}
 \end{align}
with the denominator given by
\begin{eqnarray}
        \mathcal{D} &=& \sum_{\alpha, \beta} \bigg\{ \mathcal{R}_{\alpha \beta}\Big[\operatorname{Re}\Big(\mathcal{F}_{\alpha}^{R} \mathcal{F}_{\beta}^{R\ast}\Big) m_{t}^{2}+ \operatorname{Re}\Big(\mathcal{F}_{\alpha}^{L} \mathcal{F}_{\beta}^{L\ast}\Big) m_{c}^{2}\Big]  +\mathcal{R}_{\alpha \beta}^{\mathbf{N}}\Big[\operatorname{Re}\Big(\mathcal{N}_{\alpha}^{R} \mathcal{N}_{\beta}^{R\ast}\Big) m_{t}^{2} \nonumber\\[0.15cm]
		&& \quad +\operatorname{Re}\Big(\mathcal{N}_{\alpha}^{L} \mathcal{N}_{\beta}^{L\ast}\Big) m_{c}^{2}\Big]  +2\mathcal{R}_{\alpha \beta}^{\mathbf{SN}}\Big[\operatorname{Re}\Big(\mathcal{F}_{\beta }^{R} \mathcal{N}_{\alpha }^{R\ast}\Big) m_{t}^{2}+ \operatorname{Re}\Big(\mathcal{F}_{\alpha}^{L} \mathcal{N}_{\beta}^{L\ast}\Big) m_{c}^{2}\Big]\bigg\}\,, \label{eq: Denominator} 
\end{eqnarray}
where $\alpha,\beta=\{d, s, b\}$ and
\begin{equation} \label{eq:Jarlskog_quantities}
      \begin{aligned}
	 \mathcal{J}_{\alpha \beta}&=\operatorname{Im} \left( V_{t\alpha }^{\ast } V_{c\alpha } V_{t\beta  } V_{c\beta  }^{\ast } \right) \,, \qquad &&
     \mathcal{R}_{\alpha \beta}=\operatorname{Re} \left( V_{t\alpha }^{\ast } V_{c\alpha } V_{t\beta  } V_{c\beta  }^{\ast } \right)\,, \\[0.15cm]
	 \mathcal{J}_{\alpha \beta}^{\mathbf{N}}&=\operatorname{Im} \left( V_{t\alpha }^{\ast } V_{c\alpha } V_{t\beta  } V_{c\beta  }^{\ast }\varsigma_{u}\varsigma_{u}^{\ast }\varsigma_{d}\varsigma_{d}^{\ast }  \right)\,, \qquad &&
     \mathcal{R}_{\alpha \beta}^{\mathbf{N}}=\operatorname{Re} \left( V_{t\alpha }^{\ast } V_{c\alpha } V_{t\beta  } V_{c\beta  }^{\ast } \varsigma_{u}\varsigma_{u}^{\ast }\varsigma_{d}\varsigma_{d}^{\ast } \right)\,,  \\[0.15cm]
	 \mathcal{J}_{\alpha \beta}^{\mathbf{SN,L}}&=\operatorname{Im} \left( V_{t\alpha }^{\ast } V_{c\alpha } V_{t\beta  } V_{c\beta  }^{\ast } \varsigma_{u}^{\ast }\varsigma_{d} \right) \,,  \qquad &&
	\mathcal{R}_{\alpha \beta}^{\mathbf{SN}}=\operatorname{Re} \left( V_{t\alpha }^{\ast } V_{c\alpha } V_{t\beta  } V_{c\beta  }^{\ast } \varsigma_{u}^{\ast }\varsigma_{d} \right)\,,  \\[0.15cm]
	 \mathcal{J}_{\alpha \beta}^{\mathbf{SN,R}}&=\operatorname{Im} \left( V_{t\alpha }^{\ast } V_{c\alpha } V_{t\beta  } V_{c\beta  }^{\ast } \varsigma _{d}^{\ast }\varsigma _{u} \right)\,.
    \end{aligned}
\end{equation}
It can be seen from eqs.~\eqref{eq: polarized CPp} and \eqref{eq: polarized CPm} that, to generate a non-vanishing CP asymmetry, there must exist at least two interfering amplitudes with different weak and different strong phases. The weak phases arise from the complex couplings in the Lagrangian and appear as the Jarlskog and Jarlskog-like quantities, $\mathcal{J}_{\alpha \beta}$, $\mathcal{J}_{\alpha \beta}^{\mathbf{N}}$, $\mathcal{J}_{\alpha \beta}^{\mathbf{SN,L}}$ and $\mathcal{J}_{\alpha \beta}^{\mathbf{SN,R}}$, which can be utilized to measure the strength of the CP violation. Especially, all the Jarlskog invariants $\operatorname{Im}(V_{\text{i}\alpha}^{\ast} V_{\text{f}\alpha} V_{\text{i}\beta} V_{\text{f}\beta}^{\ast })$ are equal to each other up to a difference in sign, and they are invariant under any phase transformation of the CKM matrix elements~\cite{Jarlskog:1985ht,Wu:1985ea}. Experimentally, $|\operatorname{Im}(V_{\text{i}\alpha }^{\ast} V_{\text{f}\alpha} V_{\text{i}\beta} V_{\text{f}\beta}^{\ast})|=\mathcal{O}(10^{-5})$, and thus the CP violation within the SM is predicted to be very small~\cite{Deshpande:1990ua,Aguilar-Saavedra:2002lwv,Balaji:2020qjg}. Being related to the model parameters $\varsigma _{u}$ and $\varsigma_{d}$, on the other hand, the Jarlskog-like quantities $\mathcal{J}_{\alpha \beta}^{\mathbf{N}}$, $\mathcal{J}_{\alpha \beta}^{\mathbf{SN,L}}$ and $\mathcal{J}_{\alpha \beta}^{\mathbf{SN,R}}$ could be enhanced to some extent. 

The strong phases arise from the imaginary parts of the loop integrals, which can be obtained by applying the optical theorem~\cite{Peskin:1995ev}
\begin{equation}
	-i[\mathcal{M}(a\to b)-\mathcal{M}^{\ast}(b\to a)] = 2 \operatorname{Im} \mathcal{M}(a\to b) = \sum_{n}\int d\Pi_{n}\mathcal{M}^{\ast}(b\to n) \mathcal{M}(a\to n) \,,
\end{equation}
where the sum runs over all possible sets of final-state particles that are allowed by the theory. When a non-zero imaginary part is generated in the loop integrals, the threshold condition for particle masses, $m_{a}>m_{n}$ or $m_{b}>m_{n}$, is required. In this work, $a$ denotes a top quark, $b$ is a charm quark plus a photon or a charm quark plus a gluon, and $n$ can be a $W$ boson plus a down-type quark or a charged-Higgs plus a down-type quark at the one-loop approximation. To avoid the presence of $t\to b H^{+}$ decay, we set $m_{H^{\pm}} > m_t$, which means that the imaginary terms $\operatorname{Im}(\mathcal{N}_{\alpha}^{L,R})$ are exactly zero. Therefore, the terms $\operatorname{Im}(\mathcal{N}_{\alpha}^{L,R} \mathcal{N}_{\beta}^{L,R\ast})$ will vanish in the case of $m_{H^{\pm}} > m_t$. As the particle masses satisfy $m_{t}>m_{W}+m_{\alpha}$, with $\alpha=d,s,b$, the imaginary parts $\operatorname{Im}(\mathcal{F}_{\alpha}^{L,R})$ can be non-zero and are related only to the SM loop integrals. As a summary, there could exist large deviations between the SM and the A2HDM predictions of the CP asymmetries $\Delta_{\text{CP},\pm}$, due to the extra weak phases from the alignment parameters $\varsigma_{u}$ and $\varsigma_{d}$.

\section{Numerical results and discussions}
\label{sec:numerical results}

\subsection{Input parameters}
\label{sec:input}

\begin{table}[t]
	\centering
	\let\oldarraystretch=\arraystretch
	\renewcommand*{\arraystretch}{1.3}
	{\tabcolsep=0.875cm \begin{tabular}{|cccc|}
			\hline\hline
			\multicolumn{4}{|l|}{\textbf{QCD and electroweak parameters~\cite{ParticleDataGroup:2024cfk}}}
			\\
			\hline
			$G_{F}[10^{-5}~\gev^{-2}]$
			& $m_{W}[\gev]$
			& $m_{Z}[\gev]$
			& $\alpha_s^{(5)}(m_Z)$
			\\
			$1.1663788$
			& $80.3692$
			& $91.1880$
			
			& $0.1180$
			\\
			\hline
			  
			& 
			& $\sin^2{\theta_W}$
			& $\alpha_s^{(6)}(m_t^{\rm pole})$
			\\
			
			&
			& $0.231$
			& $0.1077$
			\\
			\hline
	\end{tabular}}
	{\tabcolsep=0.549cm \begin{tabular}{|cccccc|}
			\multicolumn{6}{|l|}{\hspace{0.17cm} \textbf{Quark masses~[GeV]~\cite{ParticleDataGroup:2024cfk}}}
			\\
			\hline
			$m_t^{\rm pole}$
			& $m_c^{\rm pole}$
                & $m_b^{\rm pole}$
			& $\bar{m}_b(\bar{m}_b)$
			& $\bar{m}_s(2~\gev)$
			& $\bar{m}_d(2~\gev)$
			\\
			$172.57$
			& $1.67$
                & $4.78$
			& $4.18$
			& $0.0935$
			& $0.00470$
			\\
			\hline
			&
			&
			& $\bar{m}_b(m_t^{\rm pole})$
			& $\bar{m}_s(m_t^{\rm pole})$
			& $\bar{m}_d(m_t^{\rm pole})$
			\\
			&
			&
			& $2.72$
			& $0.0512$
			& $0.00257$
			\\
			\hline
	\end{tabular}}
	{\tabcolsep=1.40cm \begin{tabular}{|cccc|}
			\multicolumn{4}{|l|}{\hspace{-0.64cm} \textbf{CKM parameters~\cite{Karan:2023kyj}}}
			\\
			\hline
			$\lambda$
			& $A$
			& $\bar{\rho}$
			& $\bar{\eta}$
			\\
			$0.2249$
			& $0.806$
			& $0.173$
			& $0.368$
			\\
			\hline\hline
	\end{tabular}}
	\caption{Summary of the SM input parameters used throughout this work. For the external top and charm quarks their pole masses $m^{\rm pole}$ are taken as input, while for the internal down-type quarks their $\overline{\rm MS}$ running masses $\bar{m}(\mu)$ at the initial scale $\mu$ are used. The CKM parameters are obtained by fitting to the measured CKM entries with less impact from the A2HDM contributions~\cite{Karan:2023kyj}. \label{tab:inputs}}
\end{table}

The relevant input parameters for our numerical analysis are listed in table~\ref{tab:inputs}. For the external top and charm quarks we take their pole masses as input, while for the internal down-type quarks we adopt the running masses evaluated at $\mu_{t}=m_{t}^{\rm pole}$, the characteristic scale of the top-quark decays, in the modified minimal subtraction ($\overline{{\mathrm{MS}}}$) scheme. To evaluate the renormalization group running of the quark masses, we have used the \texttt{Mathematica} package \texttt{RunDec}~\cite{Chetyrkin:2000yt}, in which the QCD renormalization group equations are implemented up to the four-loop level. For the CKM matrix elements, we use as input the same values of the four Wolfenstein parameters $\lambda$, $A$, $\bar{\rho}$ and $\bar{\eta}$ as given in ref.~\cite{Karan:2023kyj}, which are extracted by fitting to the measured CKM entries that are less affected by the A2HDM contributions. Since the CP violation is highly correlated with the single complex phase of the CKM matrix, we have adopted the exact standard parametrization~\cite{Chau:1984fp} of the matrix in terms of the four independent parameters $s_{12}$, $s_{23}$, $s_{13}$ and $\delta$, which are related to the four Wolfenstein parameters to all orders in $\lambda$ through~\cite{Buras:1994ec,Buras:2020xsm}
\begin{equation}
	s_{12}=\lambda,\qquad s_{23}=A\lambda^{2},\qquad s_{13}e^{-i\delta}=A\lambda^{3}(\rho-i\eta)\,,
\end{equation}
with
\begin{equation}
       \rho=\frac{\bar{\rho}}{1-\lambda^2/2}, \qquad \eta=\frac{\bar{\eta}}{1-\lambda^2/2}\,,
\end{equation}
where $s_{ij}=\sin{\theta_{ij}}$ with $i$ and $j$ being the generation labels $(i,j=1,2,3)$. The other input parameters listed in table~\ref{tab:inputs} are taken from the latest version of \textsc{Particle Data Group}~\cite{ParticleDataGroup:2024cfk}.

In the A2HDM, the NP parameters involved in the $t \to c\gamma$ and $t \to cg$ decays are the two alignment parameters $\varsigma_{u}$, $\varsigma_{d}$ and the charged-Higgs mass $m_{H^{\pm}}$. In particular, $\varsigma_{u}$ and $\varsigma_{d}$ can be, in the most general case, complex, which can introduce new sources of CP violation beyond the SM. For convenience, we define their product as $\varsigma_{u}^{\ast} \varsigma_{d}=|\varsigma_{u}||\varsigma_{d}|e^{-i\varphi}$, where $\varphi$ is the relative phase between the two alignment parameters. The magnitudes of $\varsigma_{u}$ and $\varsigma_{d}$ are set within the limits $\sqrt{2}|\varsigma_{u,d}| m_{u,d}/v \lesssim 1$ allowed by the perturbative constraints~\cite{Karan:2023kyj,Allwicher:2021rtd}. To avoid the presence of $t \to b H^{+}$ decay, the charged-Higgs mass $m_{H^{\pm}}$ must be larger than the top-quark mass $m_t$. The upper limit on $m_{H^{\pm}}$ is deduced from the variation of the branching ratios of $t \to c\gamma(g)$ decays with respect to $m_{H^{\pm}}$ and their comparison with the corresponding SM predictions, which will be explained in the next subsection. These considerations motivate us to choose the following priors of the model parameters~\cite{Karan:2023kyj}:
\begin{equation}\label{eq:zetauzetadrange}
	|\varsigma_{u}| \in [0,\,1.5], \qquad |\varsigma_{d}| \in [0,\,50], \qquad \varphi \in [-180^\circ,\,180^\circ], \qquad m_{H^{\pm}} \in [200,\,600]~\gev.
\end{equation}

\begin{figure}[t]
    \centering 
    \includegraphics[width=0.98\textwidth]{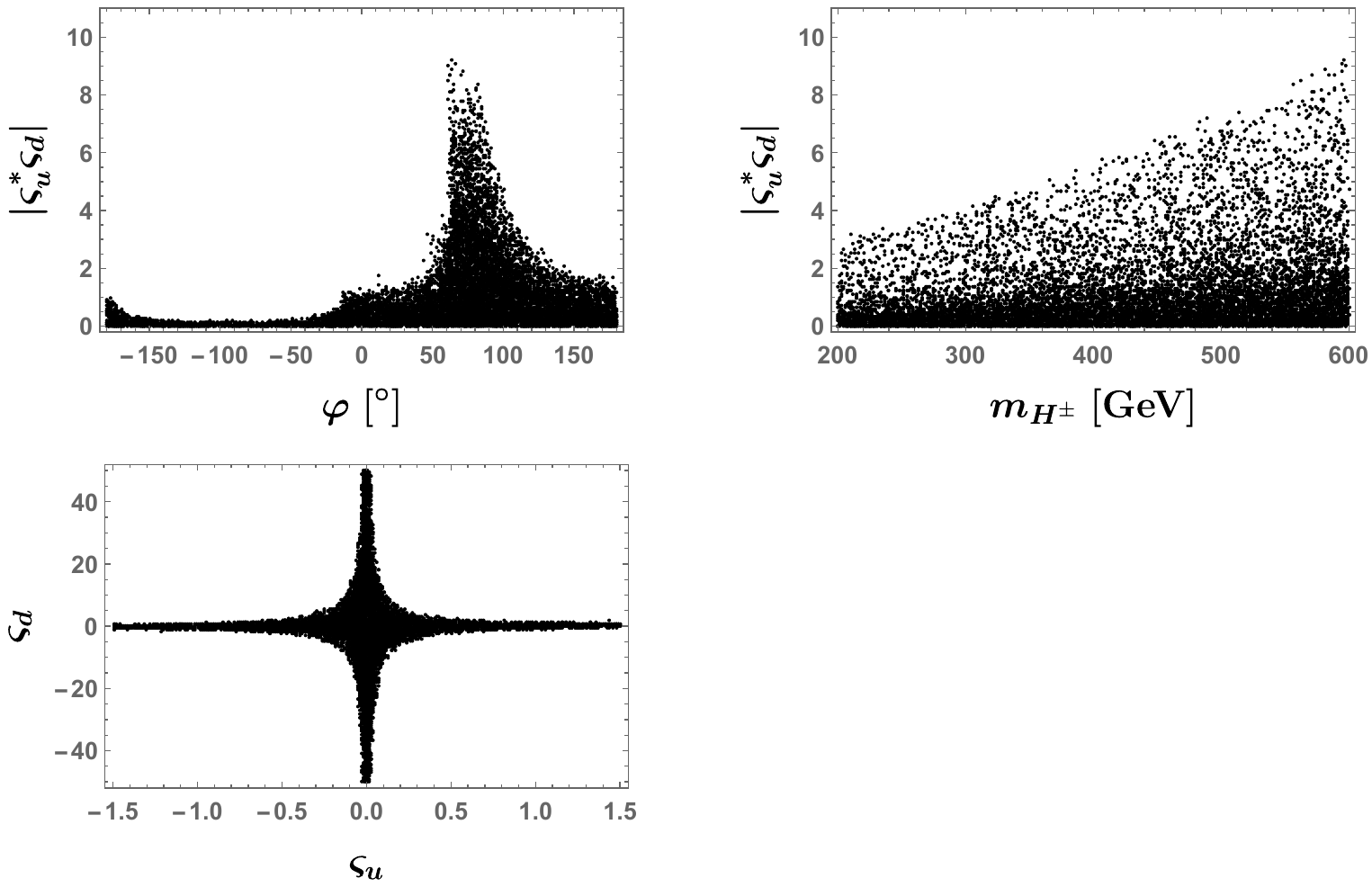}
    \caption{The parameter space resulted from the combined constraints from $\mathcal{B}(B \to X_{s} \gamma)$, $\Delta(K^{\ast}\gamma)$ and $\Delta(\rho\gamma)$. The model parameters are projected onto the $|\varsigma_{u}^{\ast}\varsigma_{d}|-\varphi $ and $|\varsigma_{u}^{\ast}\varsigma_{d}|-m_{H^{\pm}}$ planes (complex couplings) in the first two plots, while onto the $\varsigma_{u}-\varsigma_{d}$ plane (real couplings) in the third plot. \label{Fig:paraspace}}
\end{figure}

To further refine the parameter space of the A2HDM allowed by the current experimental data, we follow the same procedure as in refs.~\cite{Jung:2012vu,Li:2013vlx}, while updating the input parameters and the experimental measurements~\cite{HFLAV:2022esi,ParticleDataGroup:2024cfk}. To this end, we mainly consider the most relevant constraints from the measured branching ratio of the inclusive radiative $B \to X_{s} \gamma$ decay, $\mathcal{B}(B \to X_{s} \gamma)$, as well as the isospin asymmetries of the exclusive radiative $B \to K^{\ast} \gamma$ and $B \to \rho \gamma$ decays, $\Delta(K^{\ast}\gamma)$ and $\Delta(\rho\gamma)$. The A2HDM contributions to these observables arise from the photon penguin diagrams mediated by the charged-Higgs boson, which contributes at the same level as that of the $W$-boson within the SM. Due to the good agreement of the SM prediction~\cite{Misiak:2015xwa} with the experimental measurement~\cite{HFLAV:2022esi,ParticleDataGroup:2024cfk}, the observable $\mathcal{B}(B \to X_{s} \gamma)$ provides a stringent constraint on the model parameters~\cite{Jung:2012vu,Li:2013vlx}. The observables $\Delta(K^{\ast}\gamma)$ and $\Delta(\rho\gamma)$, while being still plagued by larger experimental uncertainties, can generate very complementary constraints on the model parameters~\cite{Li:2013vlx}. To obtain the allowed regions of the parameter space, we vary both the SM predictions and the experimental measurements of these observables within the $2\sigma$ error bars. Taking into account the priors of the model parameters specified by eq.~\eqref{eq:zetauzetadrange}, we show in Fig.~\ref{Fig:paraspace} the final combined constraints from the three observables $\mathcal{B}(B \to X_{s} \gamma)$, $\Delta(K^{\ast}\gamma)$ and $\Delta(\rho\gamma)$, where the first two plots correspond to the projections of the complex parameter space onto the $|\varsigma_{u}^{\ast}\varsigma_{d}|-\varphi$ and $|\varsigma_{u}^{\ast}\varsigma_{d}|-m_{H^{\pm}}$ planes, while the third to the projection of the real parameter space onto the $\varsigma_{u}-\varsigma_{d}$ plane. It can be seen that, in the complex parameter space, the combination $|\varsigma_{u}^{\ast}\varsigma_{d}|$ has a strong correlation with the relative phase $\varphi$, and it depends also on the charged-Higgs mass, with larger values allowed only for larger $m_{H^{\pm}}$. In the real case, $\varsigma_{u}$ and $\varsigma_{d}$ cannot be large simultaneously. In the following studies, we shall also investigate the variations of the branching ratios and the CP asymmetries of $t \to c\gamma$ and $t \to cg$ decays with respect to the two alignment parameters, while fixing the charged-Higgs mass at three benchmark values, $m_{H^{\pm}} = 200$, $400$, and $600~\gev$, respectively.

\subsection{Branching ratios of \texorpdfstring{$t \to c\gamma$}{ttocgamma} and \texorpdfstring{$t \to cg$}{ttocg} decays}

\begin{figure}[t]
    \centering 
    \includegraphics[width=0.49\textwidth]{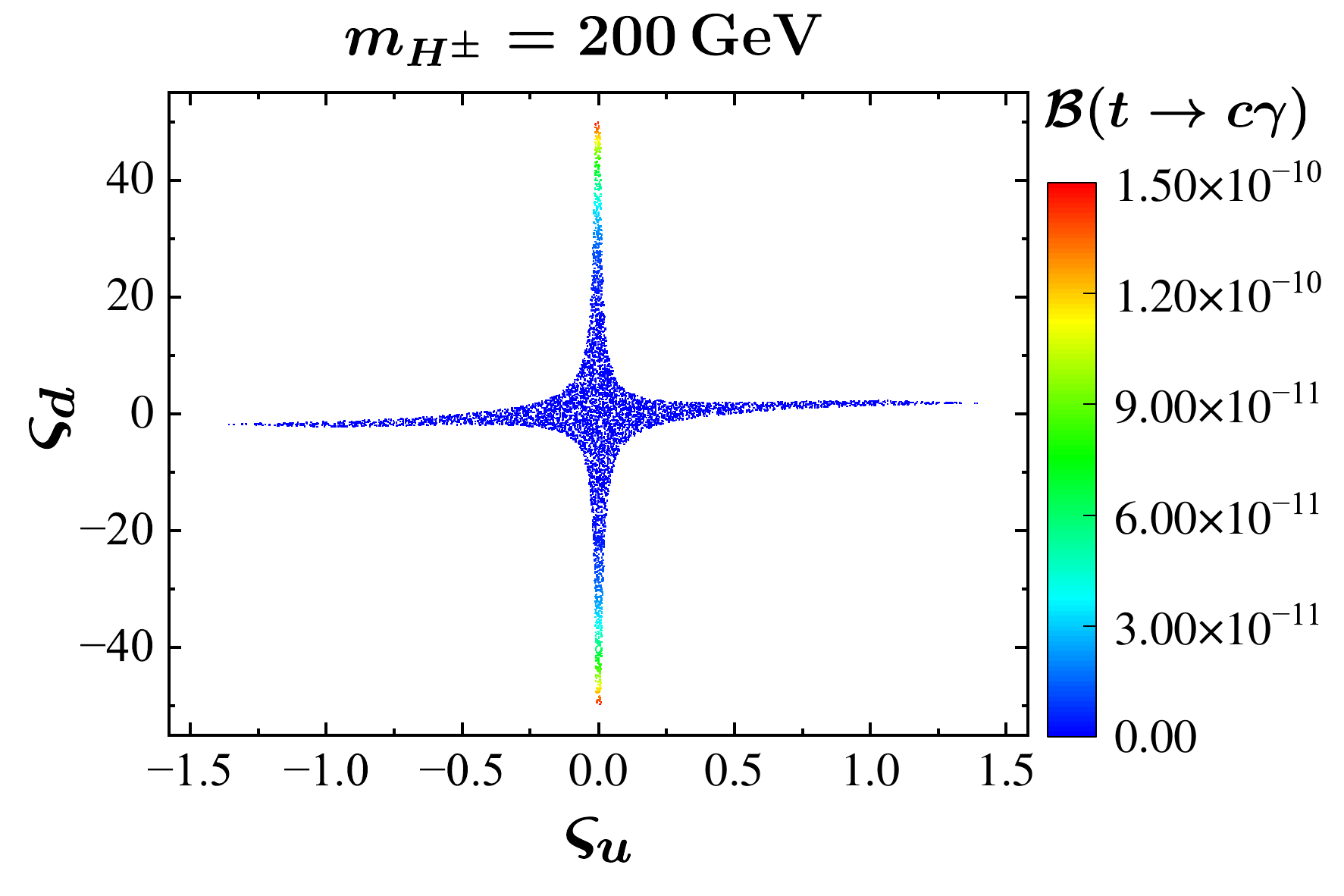}
    \includegraphics[width=0.49\textwidth]{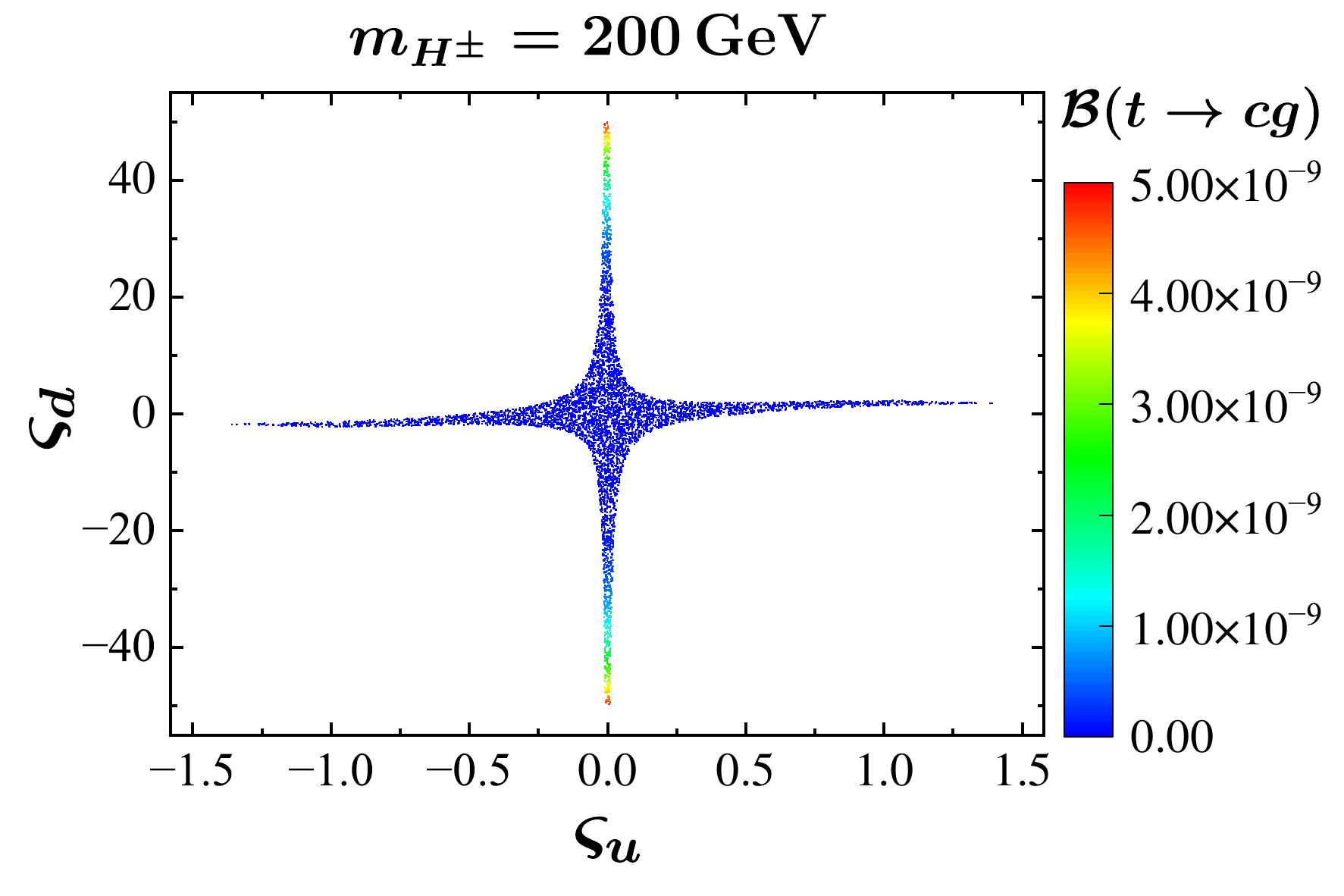}
    \includegraphics[width=0.49\textwidth]{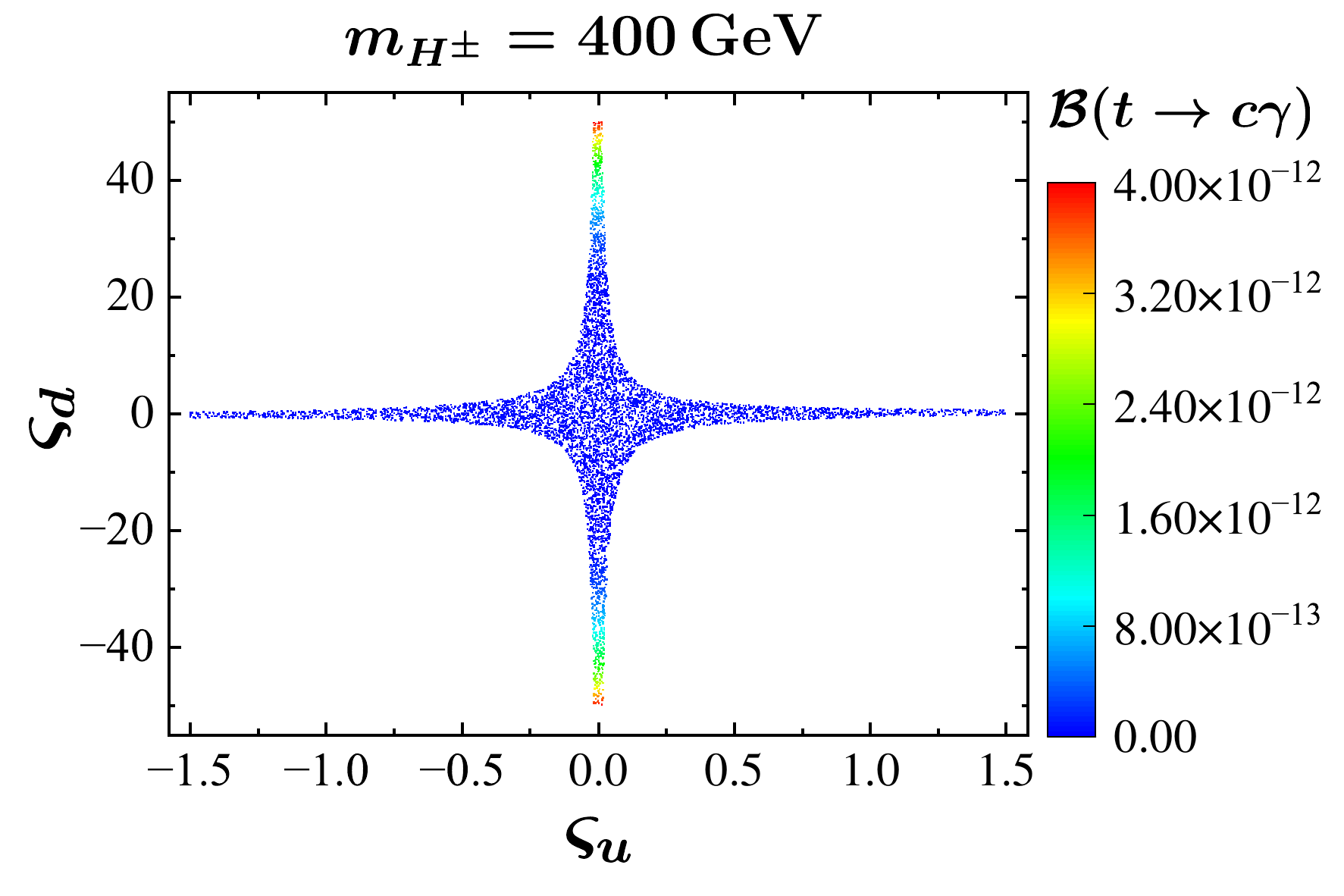}
    \includegraphics[width=0.49\textwidth]{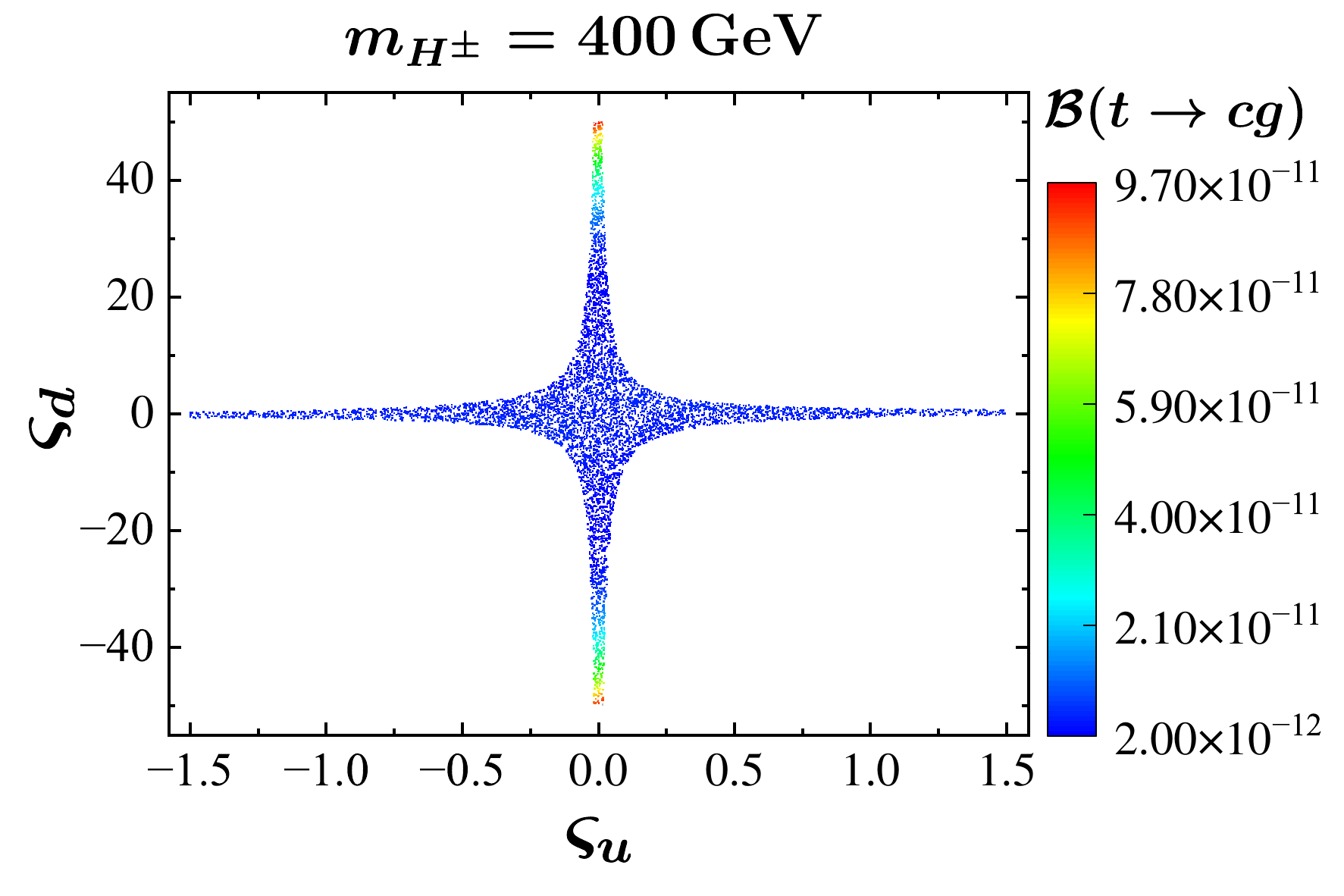}
    \includegraphics[width=0.49\textwidth]{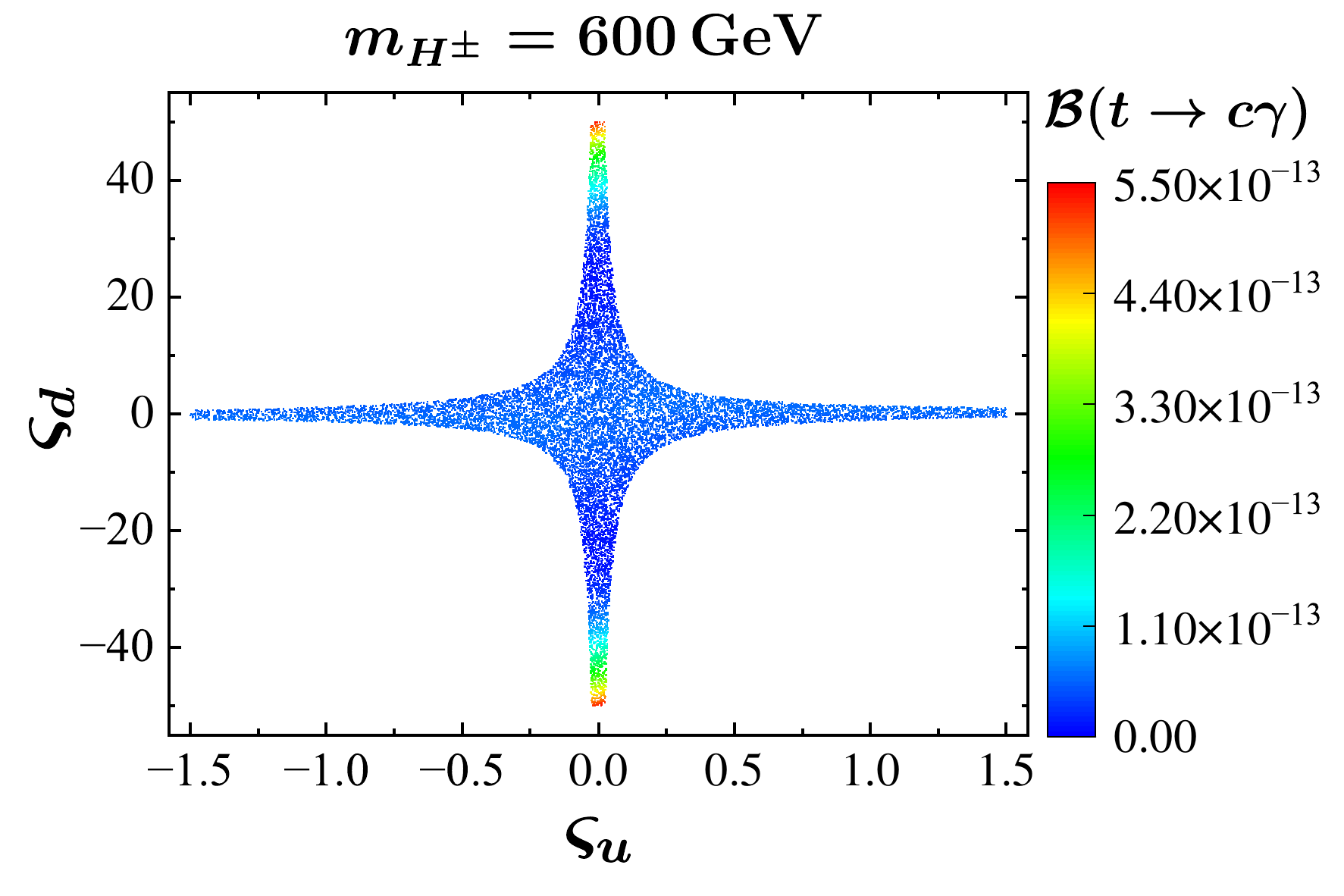}
    \includegraphics[width=0.49\textwidth]{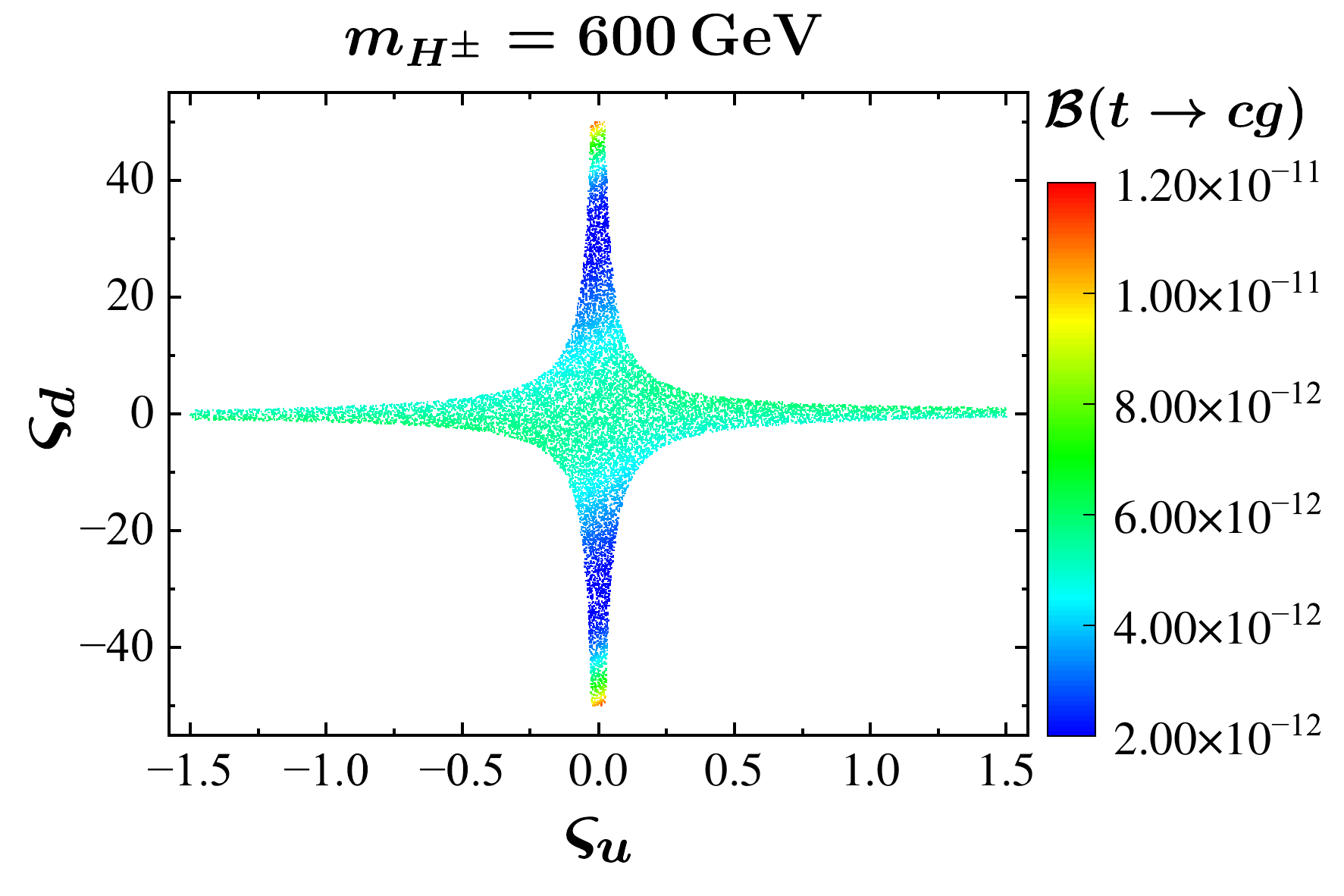}
    \caption{The branching ratios $\mathcal{B}(t\to c \gamma)$ (left column) and $\mathcal{B}(t\to cg)$ (right column) versus the real alignment parameters $\varsigma_u$ and $\varsigma_d$, for three benchmark values of the charged-Higgs mass, $m_{H^{\pm}}=200~\gev$, $400~\gev$, and $600~\gev$, respectively. The color bars indicate the values of the branching ratios. \label{Fig:tc branching ratio}}
\end{figure}

Firstly, with the help of eqs.~\eqref{eq:t2bWNLO} and \eqref{eq:t2bWLO} as well as the input parameters collected in table~\ref{tab:inputs}, the $t \to b W^{+}$ decay width is calculated to be
\begin{align}
	\Gamma(t \to b W^+)=1.35\,(1.48)~\gev,
\end{align}
at the NLO (LO) in QCD, which is well consistent with the measured top-quark total width, $\Gamma_\mathrm{tot}^\mathrm{exp}(t) = 1.42^{+0.19}_{-0.15}~\gev$~\cite{ParticleDataGroup:2024cfk}. This motivates us to choose the condition $m_{H^{\pm}} > m_t -m_b$, so that the tree-level process $t \to b H^+$ is kinematically forbidden and hence provides no contribution to the total top-quark width. Specifically, we require the lower limit of the charged-Higgs mass to be $m_{H^{\pm}} \geq 200~\gev$ for the purpose. 

With the formulae given in subsection~\ref{sec: branching ratio formulas} and the input parameters detailed in subsection~\ref{sec:input}, we are now ready to present the numerical results of the branching ratios of $t \to c\gamma$ and $t \to cg$ decays, as well as their variations with respect to the alignment parameters $\varsigma_{u,d}$, the charged-Higgs mass $m_{H^{\pm}}$, and the relative phase $\varphi$. In Fig.~\ref{Fig:tc branching ratio}, the branching ratios $\mathcal{B}(t\to c \gamma)$ (left column) and $\mathcal{B}(t\to cg)$ (right column) are projected as colors onto the real $\varsigma_{u}-\varsigma_{d}$ plane, for three benchmark values of the charged-Higgs mass, $m_{H^{\pm}}=200~\gev$, $400~\gev$, and $600~\gev$, respectively. It can be seen that the maximum branching ratios of both $t \to c\gamma$ and $t \to cg$ decays are reached in the limits $|\varsigma_u|\to 0$ and $|\varsigma_d|\to 49$, which correspond to the minimum $|\varsigma_u|$ and maximum $|\varsigma_d|$ allowed by the combined constraints of the inclusive and exclusive radiative $b \to s \gamma$ decays, as discussed in subsection~\ref{sec:input}. Since the allowed range of $\varsigma_d$ is much larger than that of $\varsigma_u$, we can find from eq.~(4.3) in ref.~\cite{Cai:2022xha} that the branching ratios can be significantly enhanced when $\varsigma_d$ takes a large value, while the allowed parameter space in the real case requires that a large $\varsigma_d$ must accompany a small $\varsigma_u$ and vice versa, as shown in Figs.~\ref{Fig:paraspace} and \ref{Fig:tc branching ratio}. In the limiting case with $\varsigma_{u,d} = 0$, the SM results would be recovered,
\begin{equation} \label{eq:NbrSM}
	\mathcal{B}^{\text{SM}}(t \to c\gamma) = 5.49\times 10^{-14}\,,  \qquad
	\mathcal{B}^{\text{SM}}(t \to cg) = 5.40\times 10^{-12}\,, 
\end{equation}
which are comparable with those obtained in refs.~\cite{Aguilar-Saavedra:2002lwv,Balaji:2020qjg,Cai:2022xha}, with the marginal differences attributed mainly to the different input parameters, especially to the $\overline{{\mathrm{MS}}}$ running mass of the bottom quark. Notice that neither the branching ratios nor the CP asymmetries are quite sensitive to the uncertainty of the top-quark pole mass, since the leading dependence on $m_t$ gets cancelled when calculating these observables~\cite{Aguilar-Saavedra:2002lwv,Balaji:2020qjg}.

\begin{figure}[t]
    \centering
    \includegraphics[width=0.49\textwidth]{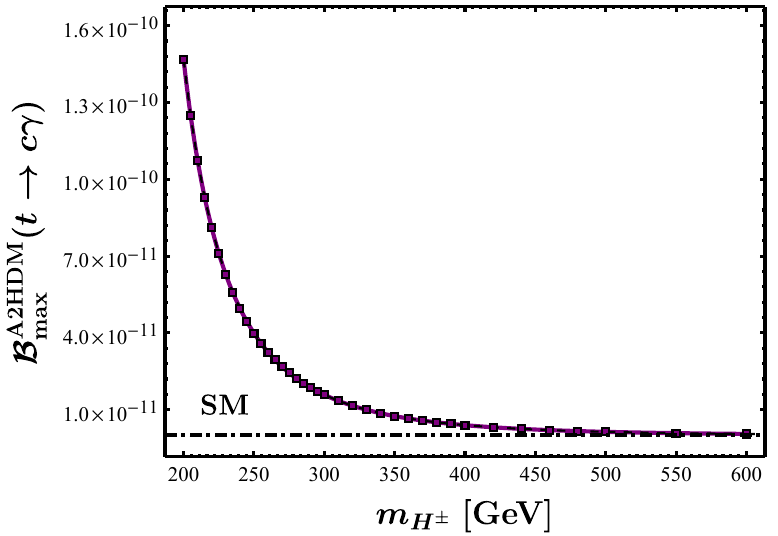}\hspace{0.12cm}
    \includegraphics[width=0.49\textwidth]{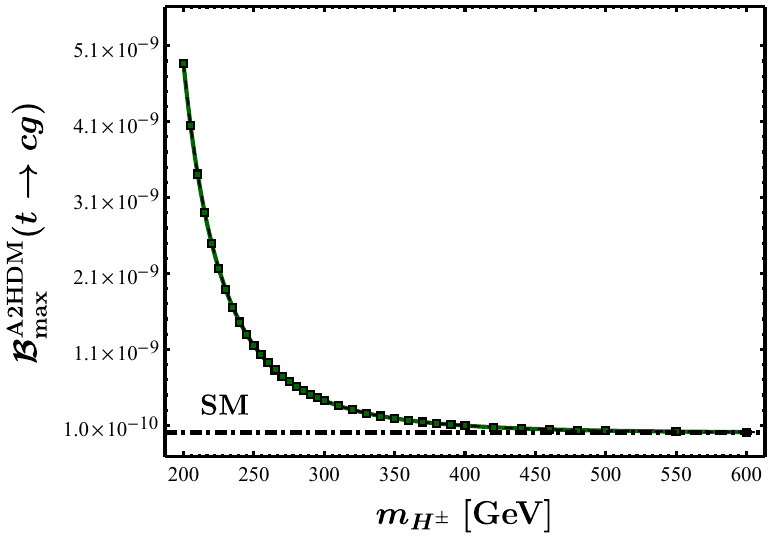}
    \caption{The dependence of the maximum branching ratios of $t\to c \gamma$ (left) and $t\to cg$ (right) decays on the charged-Higgs mass $m_{H^{\pm}}$, with the two real alignment parameters fixed at $(\varsigma_u,\varsigma_d) = (-0.011,49.8)$. The horizontal dash-dotted lines represent the SM predictions given by eq.~\eqref{eq:NbrSM}. \label{Fig:BrmH}} 
\end{figure}

It is also observed that the maximum values of $\mathcal{B}(t\to c \gamma)$ and $\mathcal{B}(t\to cg)$ are always reached around the same set of real alignment parameters, $(\varsigma_u,\varsigma_d) \simeq (-0.011,49.8)$, when the charged-Higgs mass $m_{H^{\pm}}$ varies within the range $[200, 600]~\gev$. To explore the dependence of the maximum branching ratios on $m_{H^{\pm}}$, we fix the alignment parameters at $(\varsigma_u,\varsigma_d) = (-0.011,49.8)$, and show in Fig.~\ref{Fig:BrmH} the variation of $\mathcal{B}_\mathrm{max}^\mathrm{A2HDM}(t\to c \gamma)$ (left) and $\mathcal{B}_\mathrm{max}^\mathrm{A2HDM}(t\to cg)$ (right) with respect to $m_{H^{\pm}}$. For comparison, the SM predictions given by eq.~\eqref{eq:NbrSM} are also shown as horizontal dash-dotted lines in Fig.~\ref{Fig:BrmH}. It is observed that the maximum branching ratios decrease as the charged-Higgs mass $m_{H^{\pm}}$ increases, which is due to the fact that the charged Higgs contributes to these rare FCNC decays in the loop as a virtual propagator and no $m_{H^{\pm}}$ is involved in the fermion-scalar vertices. In particular, the maximum branching ratios would approach the corresponding SM predictions for $m_{H^\pm}\gtrsim 600~\gev$. This is the exact reason why we have chosen the upper limit of the charged-Higgs mass to be $600~\gev$ throughout this work. In table~\ref{tab:num maximumBR}, we have listed the resulting maximum branching ratios for several typical values of $m_{H^{\pm}}$. It is found that the maximum branching ratios that can be reached in the A2HDM are $\mathcal{B}^{\text{A2HDM}}_\text{max}(t \to c\gamma)=1.47\times 10^{-10}$ and $\mathcal{B}^{\text{A2HDM}}_\text{max}(t \to cg)=4.86\times 10^{-9}$, with a charged-Higgs mass of $m_{H^{\pm}}=200~\gev$. These results are about four and three orders of magnitude higher than the corresponding SM predictions given by eq.~\eqref{eq:NbrSM}, but are still below the current experimental upper limits of $\mathcal{O}(10^{-5})$ and $\mathcal{O}(10^{-4})$~\cite{ParticleDataGroup:2024cfk,LHCTop}. They are also out of the expected sensitivities of the HL-LHC~\cite{Cerri:2018ypt,Azzi:2019yne} and the future colliders~\cite{LHeCStudyGroup:2012zhm,ILC:2013jhg,CLICdp:2018esa,FCC:2018byv,CEPCStudyGroup:2018ghi}.

\begin{table}[t]
	\begin{center}	
		\let\oldarraystretch=\arraystretch
		\renewcommand*{\arraystretch}{1.3}
		\tabcolsep=0.15cm
        \begin{adjustbox}{width=0.98\textwidth,center}
		\begin{tabular}{|cccccc|}
			\hline\hline
			$m_{H^\pm}~[\gev$]
			& $200$
			& $300$
			& $400$
			& $500$
			& $600$
			\\\hline
			$\mathcal{B}^{\text{A2HDM}}_{\text{max}}(t \to c\gamma)$
		    & $1.47\times 10^{-10}$
		    & $1.58\times 10^{-11}$
		    & $3.95\times 10^{-12}$
		    & $1.34\times 10^{-12}$ 
		    & $5.33\times 10^{-13}$
			\\
			$\mathcal{B}^{\text{A2HDM}}_{\text{max}}(t \to cg)$
			& $4.86\times 10^{-9}$
			& $4.26\times 10^{-10}$
			& $9.62\times 10^{-11}$
			& $2.98\times 10^{-11}$ 
			& $1.10\times 10^{-11}$
			\\
			\hline\hline
		\end{tabular}
        \end{adjustbox}
		\caption{The maximum branching ratios of $t \to c \gamma$ and $t \to c g$ decays for several typical values of the charged-Higgs mass, with the two real alignment parameters fixed at $(\varsigma_u,\varsigma_d) = (-0.011,49.8)$. \label{tab:num maximumBR}} 
	\end{center}
\end{table}

\begin{figure}[t] 
    \includegraphics[width=0.98\textwidth]{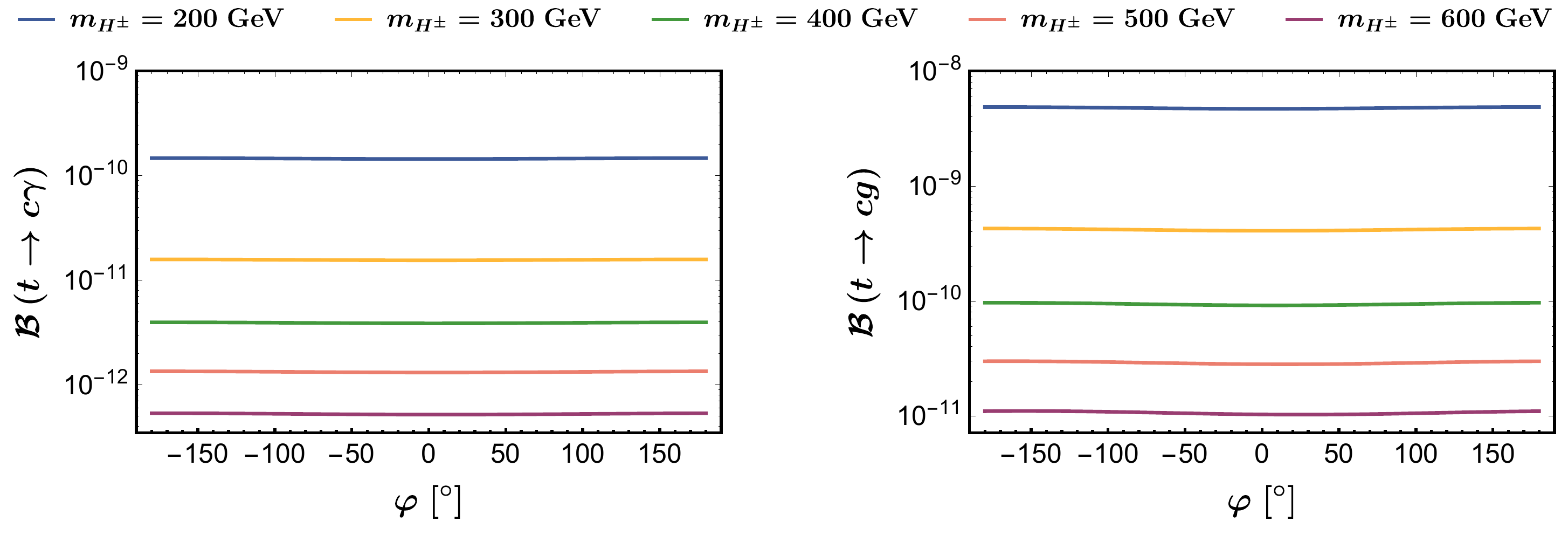}
    \caption{The branching ratios of $t\to c \gamma$ (left) and $t\to cg$ (right) decays as a function of the relative phase $\varphi$ for five different values of the charged-Higgs mass, with the absolute values of the two alignment parameters fixed at $(|\varsigma_u|,|\varsigma_d|) = (0.011,49.8)$. \label{Fig:Brphi}} 
\end{figure}

In the above discussions, the two alignment parameters $\varsigma_{u,d}$ are taken to be real. To explore the dependence of the branching ratios on the relative phase $\varphi$ between $\varsigma_u$ and $\varsigma_d$, we show in Fig.~\ref{Fig:Brphi} the variation of $\mathcal{B}(t\to c \gamma)$ (left) and $\mathcal{B}(t\to cg)$ (right) with respect to $\varphi$, by fixing the absolute values of the two alignment parameters at $(|\varsigma_u|,|\varsigma_d|) = (0.011,49.8)$ for five different charged-Higgs masses. It can be seen that the relative phase $\varphi$ has no significant effect on these two branching ratios. Therefore, the branching ratios in the case of complex alignment parameters will not be discussed in detail here. 

\subsection{CP asymmetries of \texorpdfstring{$t \to c\gamma_{\pm}$}{ttocgammapm} and \texorpdfstring{$t \to c g_{\pm}$}{ttocgpm} decays}

In this subsection, we evaluate the polarized CP asymmetries of the $t \to c\gamma_{\pm}$ and $t \to c g_{\pm}$ decays both within the SM and in the A2HDM, and investigate in detail the variations of these observables with respect to the different model parameters. 

Firstly, let us discuss the polarized CP asymmetries within the SM. In this case, the absorptive parts of the loop kinetic terms $\mathcal{F}_{\alpha}^{L,R}$ generated above the threshold $m_{t}>m_{W}+m_{\alpha}$, and the coherent sum over the amplitudes corresponding to the three different down-type quark flavours $\alpha=d,s,b$ in the loop with complex CKM matrix elements $V_{t\alpha}^\ast V_{c\alpha}$, lead to interference terms, which produce non-vanishing polarized CP asymmetries (cf. eqs.~\eqref{eq: polarized CPp} and \eqref{eq: polarized CPm}) in the decays we are considering. Numerically, we obtain
\begin{equation} \label{eq:NCPSM}
\begin{aligned} 
	\Delta_{\text{CP},+}^{\text{SM}}(t \to c\gamma) = -7.95\times 10^{-11}\,,  \qquad  
	\Delta_{\text{CP},-}^{\text{SM}}(t \to c\gamma) = -4.89\times 10^{-6}\,, \\[0.2cm]
	\Delta_{\text{CP},+}^{\text{SM}}(t \to cg) = -1.49\times 10^{-10}\,,  \qquad  
	\Delta_{\text{CP},-}^{\text{SM}}(t \to cg) = -4.04\times 10^{-6}\,.
\end{aligned}
\end{equation}
These results are generally consistent with the predictions made in refs.~\cite{Aguilar-Saavedra:2002lwv,Balaji:2020qjg}, with the marginal differences being due to the different input parameters, among which the internal down-type quark masses have a greater impact on the final results. It can be seen that the polarized CP asymmetries of $t \to c\gamma_{\pm}$ and $t \to cg_{\pm}$ decays are quite small within the SM. This is due to the smallness of the internal down-type quark masses compared to the top-quark mass, which leads to a more efficient GIM cancellation, as well as the smallness of the Jarlskog invariants $|\mathcal{J}_{\alpha \beta}|=|\operatorname{Im}(V_{t\alpha}^{\ast} V_{c\alpha} V_{t\beta} V_{c\beta}^{\ast})|=\mathcal{O}(10^{-5})$, which constitutes a measure of the strength of CP violation within the SM. The hierarchy $\Delta_{\text{CP},+}^{\text{SM}} \ll \Delta_{\text{CP},-}^{\text{SM}}$ results directly from the angular momentum conservation and the $V-A$ nature of weak interaction, which indicate that the emitted photons (gluons) are predominantly left-handed in the $t \to c\gamma (g)$ decays. The un-polarized CP asymmetries $\Delta_{\text{CP}} =\Delta_{\text{CP},+} + \Delta_{\text{CP},-} \simeq \Delta_{\text{CP},-}$, as defined by eq.~\eqref{eq:Delta_CP_unpolarized}, are also small within the SM~\cite{Deshpande:1990ua,Aguilar-Saavedra:2002lwv}.

\begin{figure}[t]
    \centering 
    \includegraphics[width=0.49\textwidth]{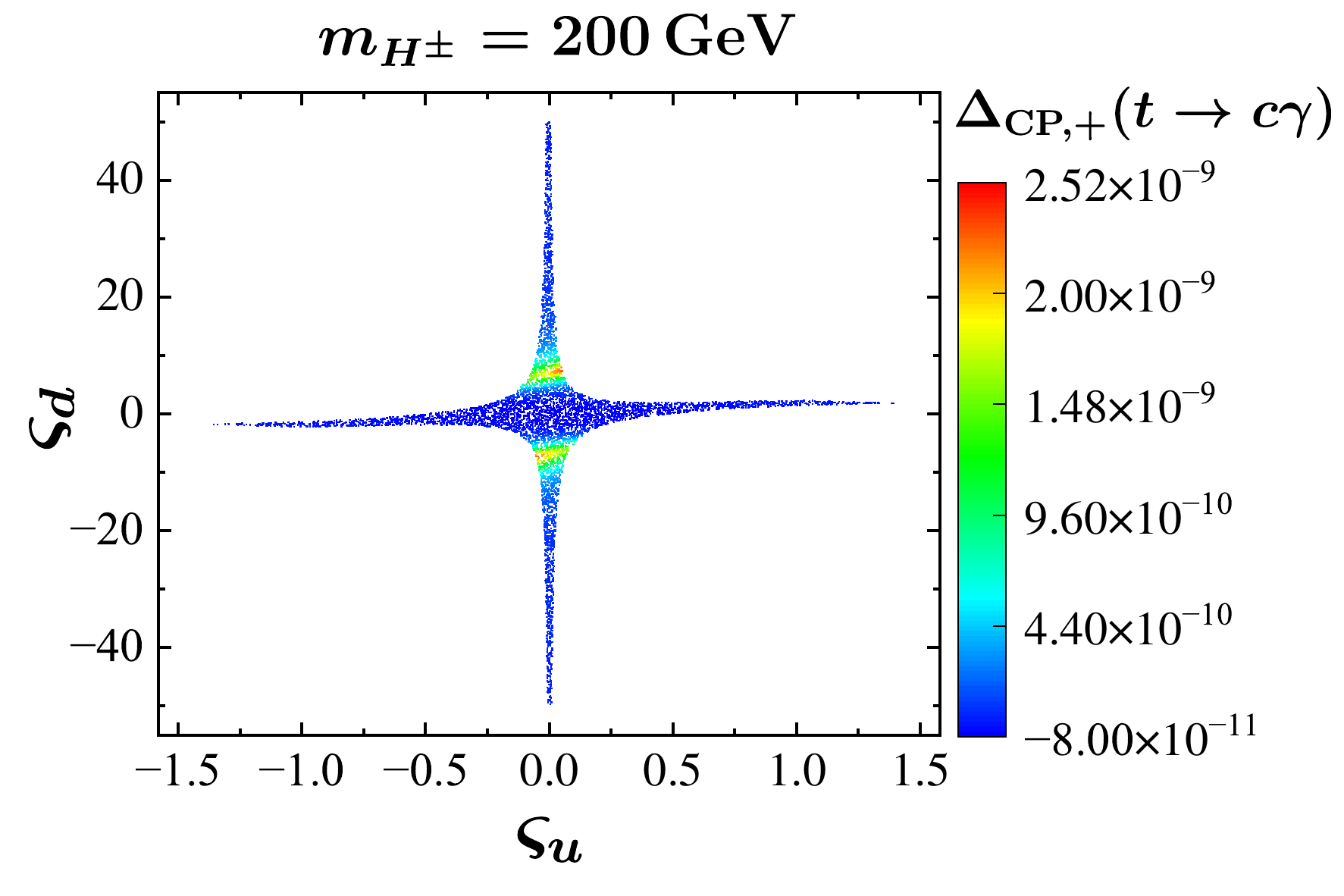}
    \includegraphics[width=0.49\textwidth]{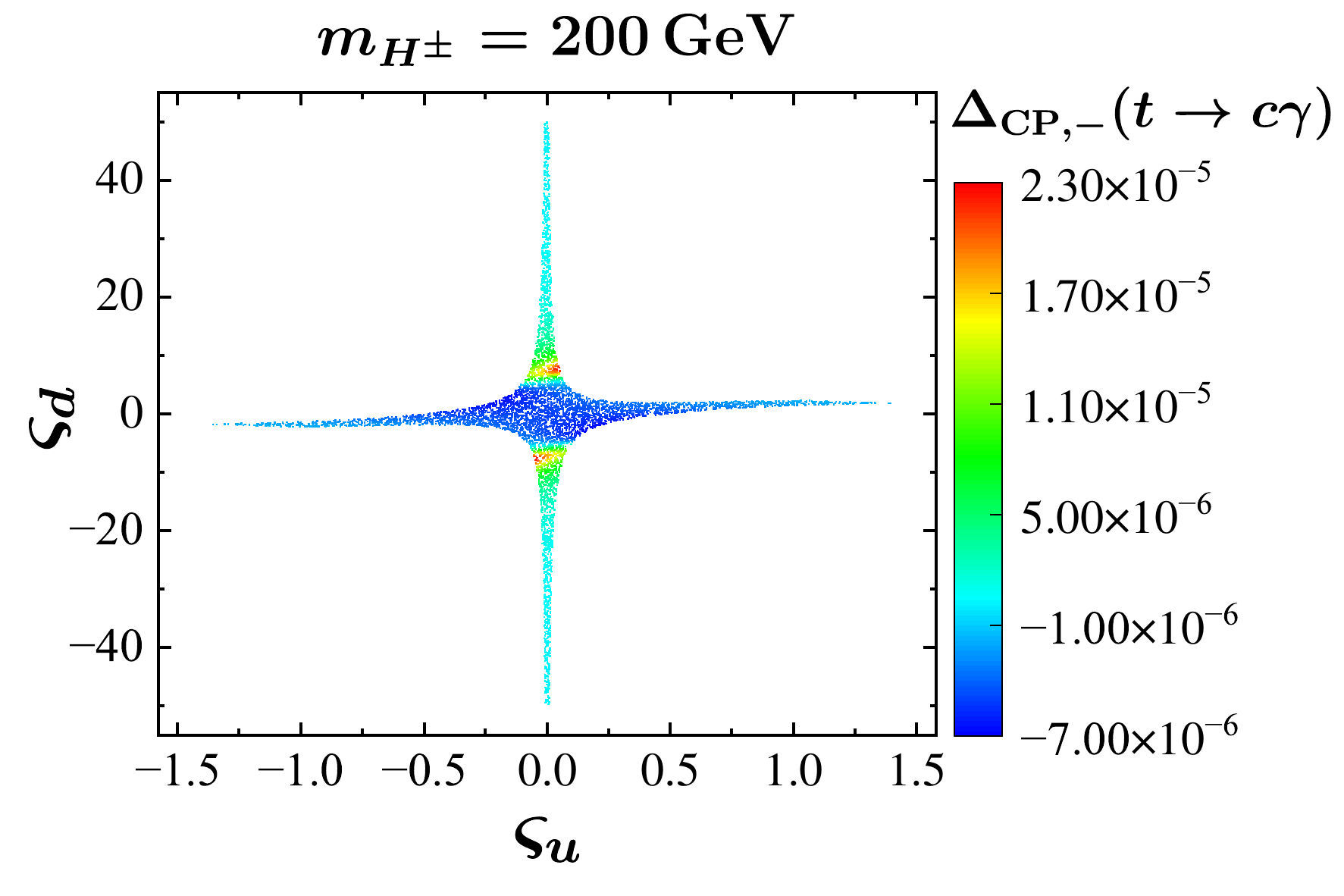}
    \includegraphics[width=0.49\textwidth]{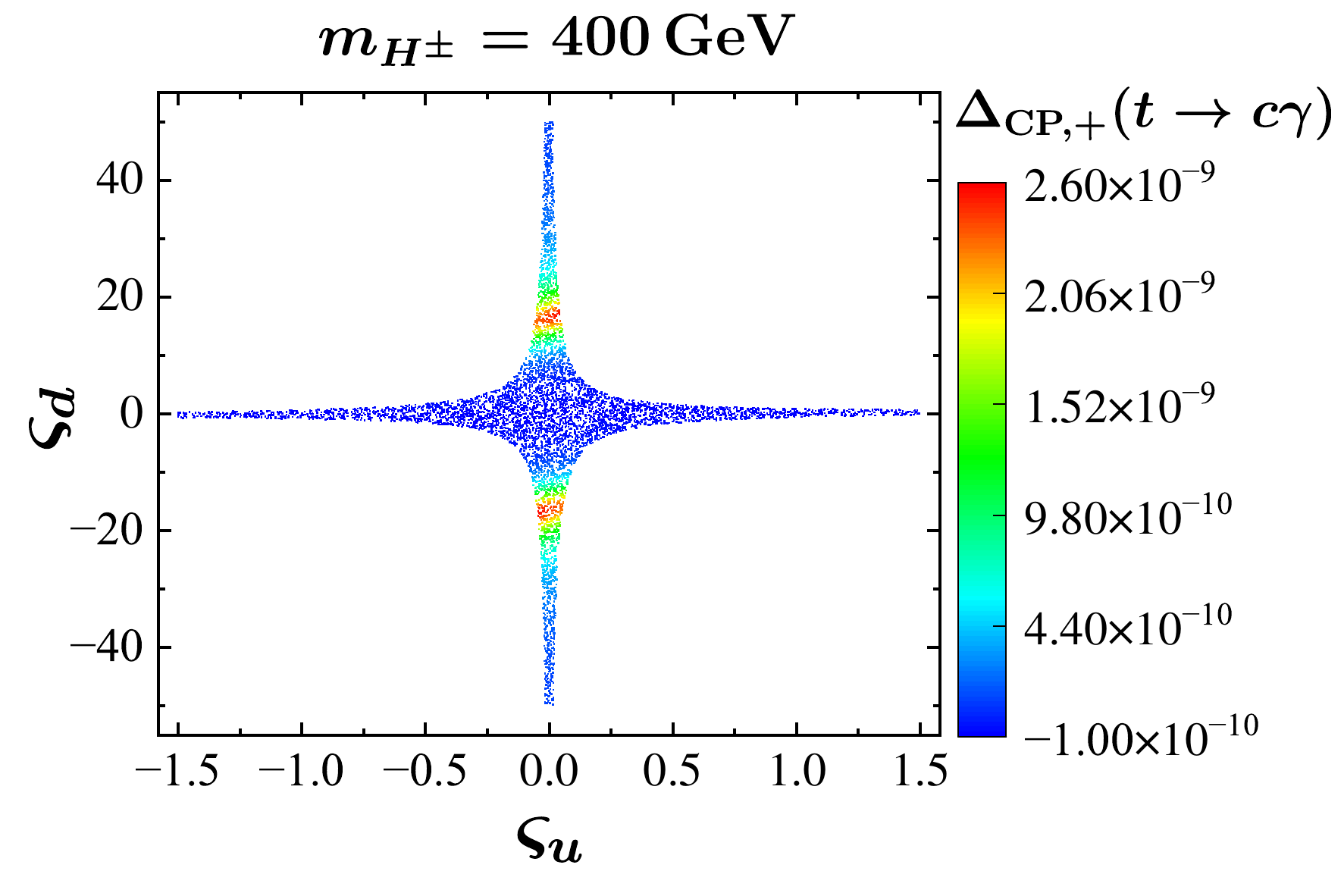}
    \includegraphics[width=0.49\textwidth]{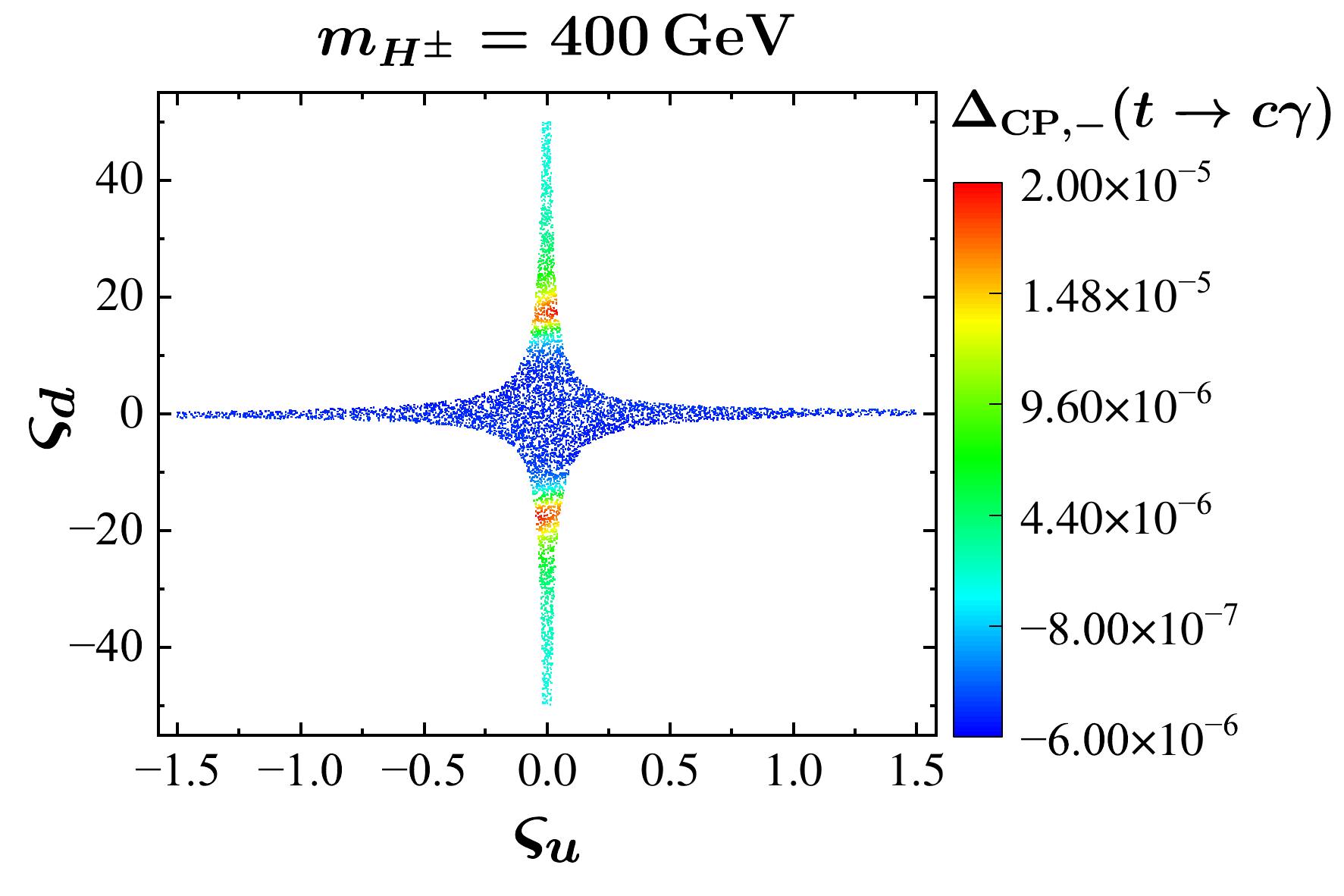}
    \includegraphics[width=0.49\textwidth]{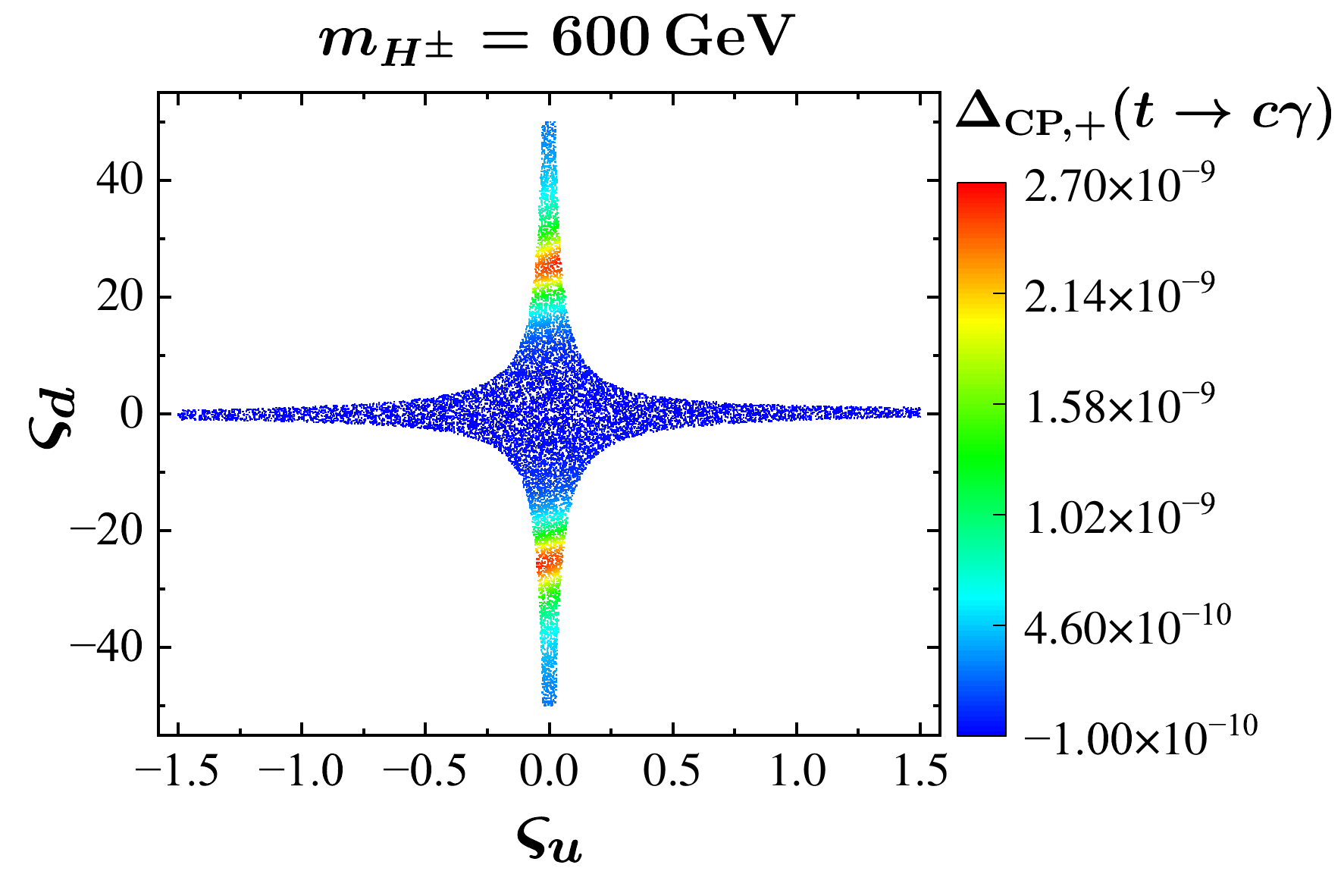}
    \includegraphics[width=0.49\textwidth]{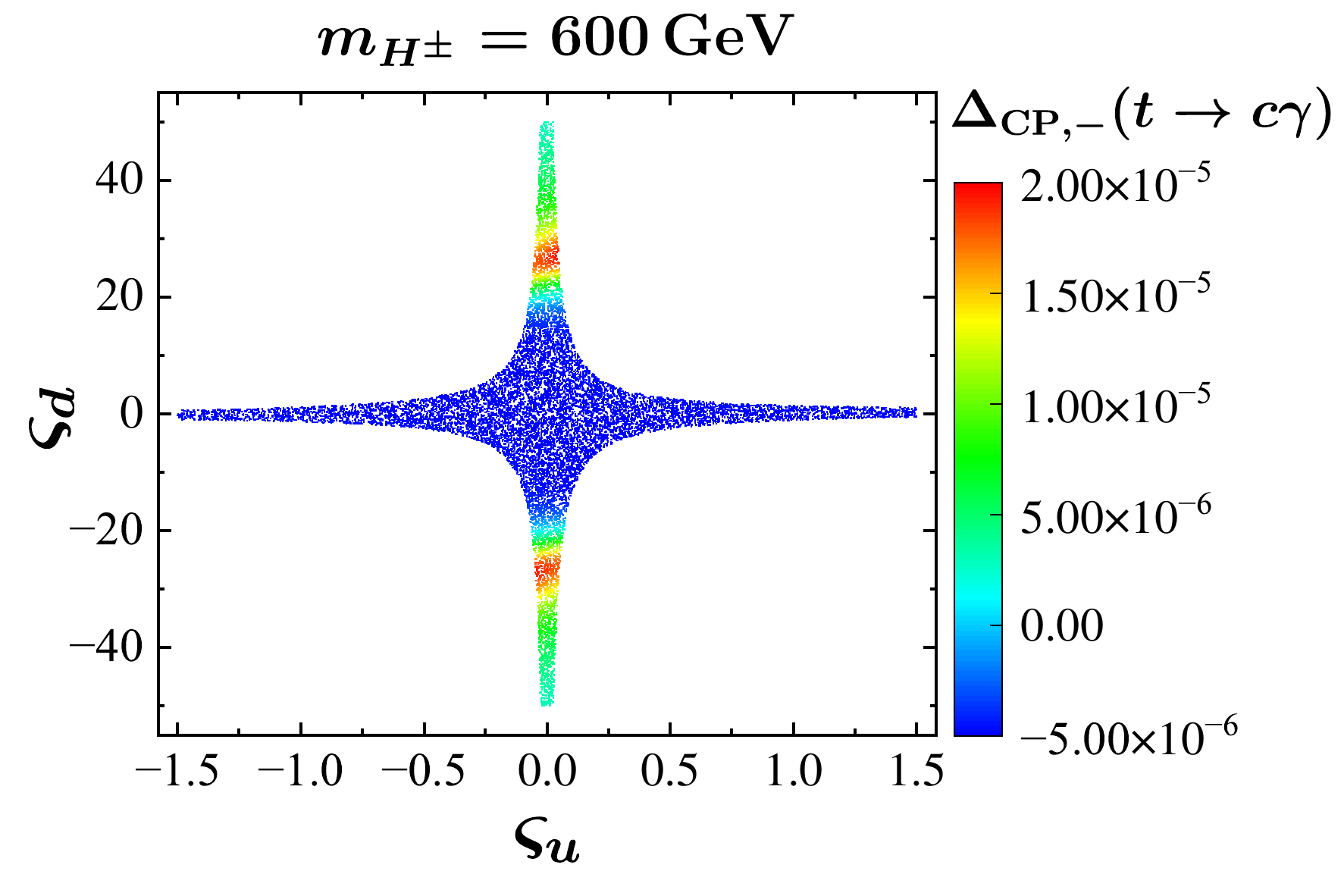}
    \caption{The polarized CP asymmetries $\Delta_{\text{CP},+}(t \to c\gamma)$ (left column) and $\Delta_{\text{CP},-}(t \to c\gamma)$ (right column) versus the real alignment parameters $\varsigma_u$ and $\varsigma_d$, for three benchmark values of the charged-Higgs mass, $m_{H^{\pm}}=200~\gev$, $400~\gev$, and $600~\gev$, respectively. The color bars indicate the values of the CP asymmetries.\label{Fig:tcy cpre}}
\end{figure}

Let us now analyze the polarized CP asymmetries in the A2HDM with real alignment parameters. To this end, we display in Fig.~\ref{Fig:tcy cpre} the two polarized CP asymmetries $\Delta_{\text{CP},+}(t \to c\gamma)$ (left column) and  $\Delta_{\text{CP},-}(t \to c\gamma)$ (right column), which are projected as colors onto the real $\varsigma_{u}-\varsigma_{d}$ plane, for three benchmark values of the charged-Higgs mass, $m_{H^{\pm}}=200~\gev$, $400~\gev$, and $600~\gev$, respectively. It should be noted that the points where $\varsigma_{u}$ and $\varsigma_{d}$ are both zero in the figure correspond to the CP asymmetries in the SM. One can see from the values of the color bars that the resulting ranges of the CP asymmetries do not vary too much with the charged-Higgs mass $m_{H^{\pm}}$, while the red points move in the direction of larger absolute value of $\varsigma_{d}$ as the charged-Higgs mass increases, which means that the absolute value of $\varsigma_{d}$ corresponding to the maximum of the CP asymmetries increases along with $m_{H^{\pm}}$. A similar behaviour is also observed in the $t \to cg_{\pm}$ decays, as shown explicitly by Fig.~\ref{Fig:tcg cppmre} in appendix~\ref{sec:Result diagrams}. The maximum absolute values of the polarized CP asymmetries in the A2HDM with real alignment parameters are given, respectively, as
\begin{equation} \label{eq:NCP reA2HDM}
\begin{aligned}
	|\Delta_{\text{CP},+, \text{real}}^{\text{A2HDM, max}}(t \to c\gamma)| = 2.61 \times 10^{-9}\,,  \qquad  
	|\Delta_{\text{CP},-, \text{real}}^{\text{A2HDM, max}}(t \to c\gamma)| = 2.29 \times 10^{-5}\,, \\[0.2cm]
	|\Delta_{\text{CP},+, \text{real}}^{\text{A2HDM, max}}(t \to cg)| = 1.61 \times 10^{-9}\,,  \qquad  
	|\Delta_{\text{CP},-, \text{real}}^{\text{A2HDM, max}}(t \to cg)| = 2.13 \times 10^{-5}\,. 
\end{aligned}
\end{equation}
We can see that they are not significantly enhanced compared to the corresponding SM predictions given by eq.~\eqref{eq:NCPSM}. This is due to the fact that there are no new sources of CP violation beyond the SM in the case of real alignment parameters and, for moderate values of $\varsigma_u$ and $\varsigma_d$, the NP contributions mediated by the charged-Higgs boson and their interference with the SM contributions have no unexpectedly large impact on the CP asymmetries.

\begin{figure}[t]
    \centering
    \includegraphics[width=0.49\textwidth]{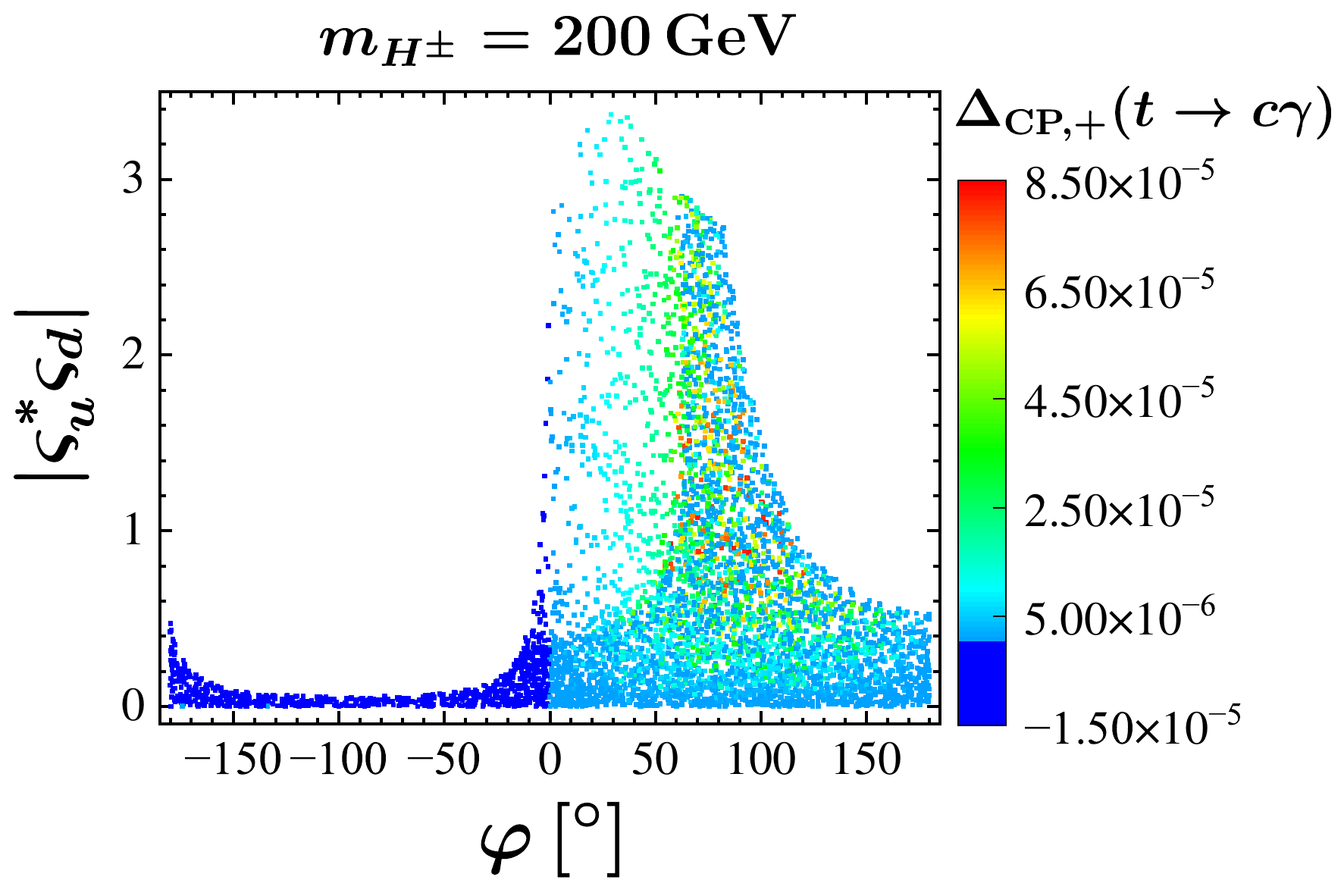}
    \includegraphics[width=0.49\textwidth]{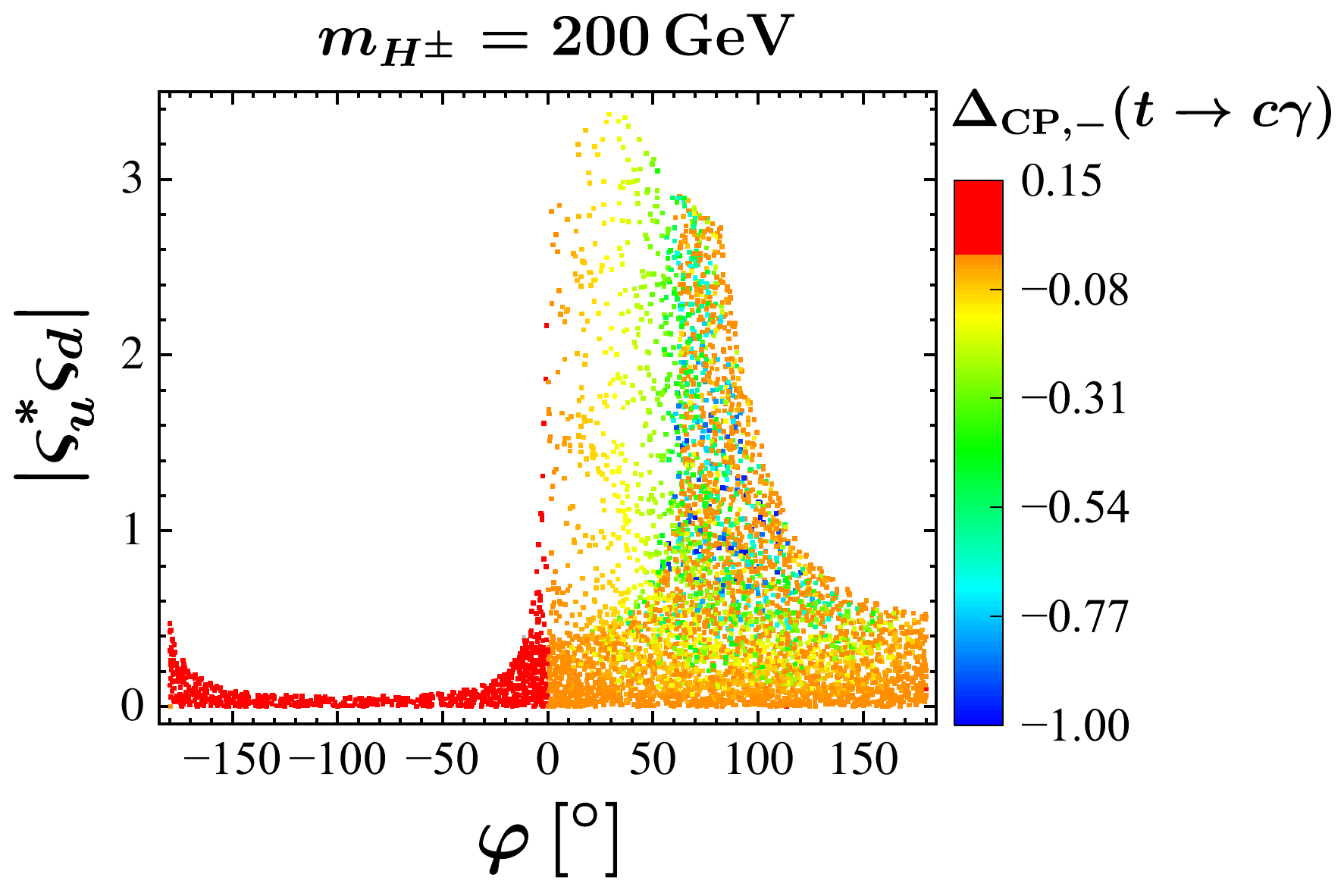}
    \includegraphics[width=0.49\textwidth]{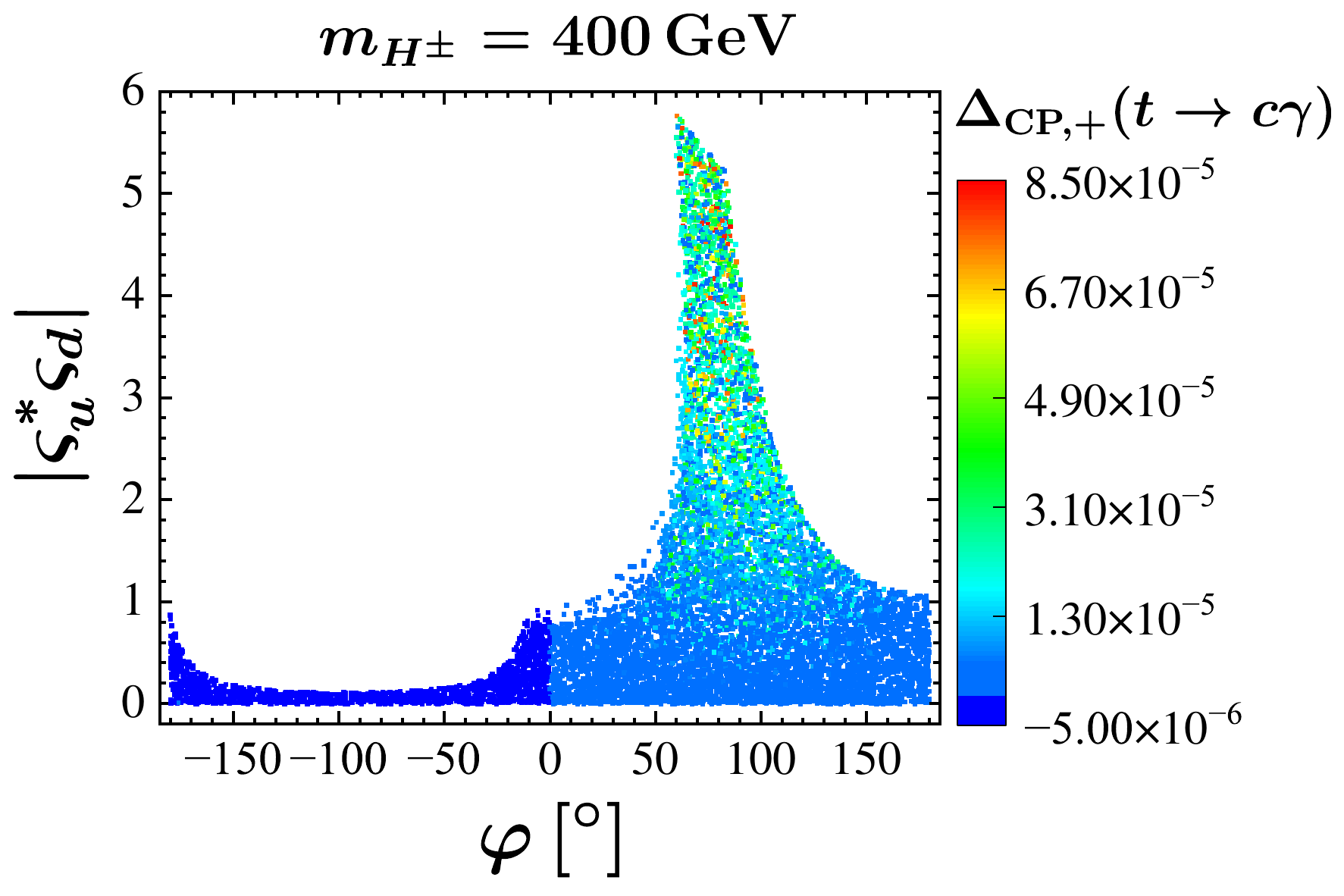}
    \includegraphics[width=0.49\textwidth]{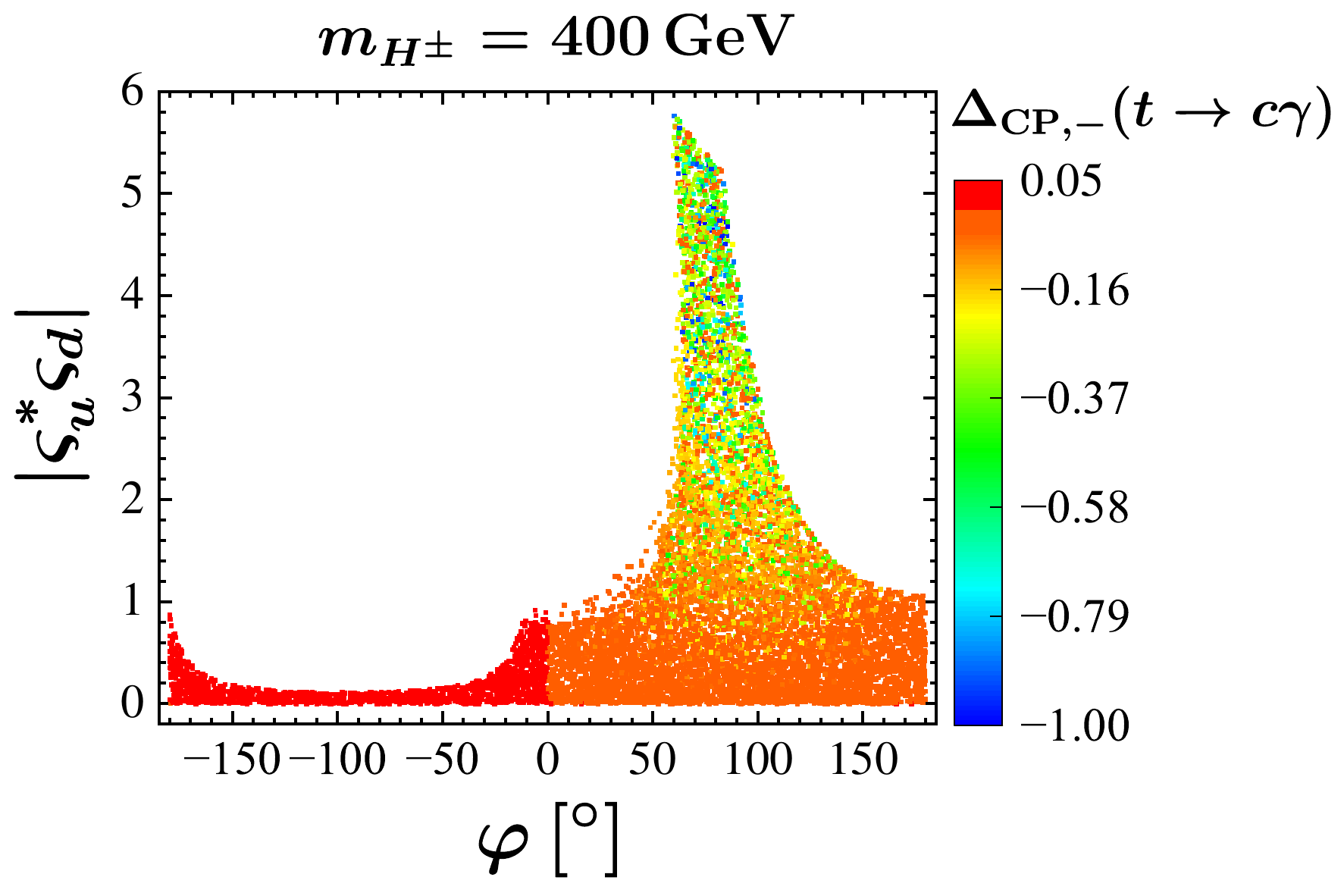}
    \includegraphics[width=0.49\textwidth]{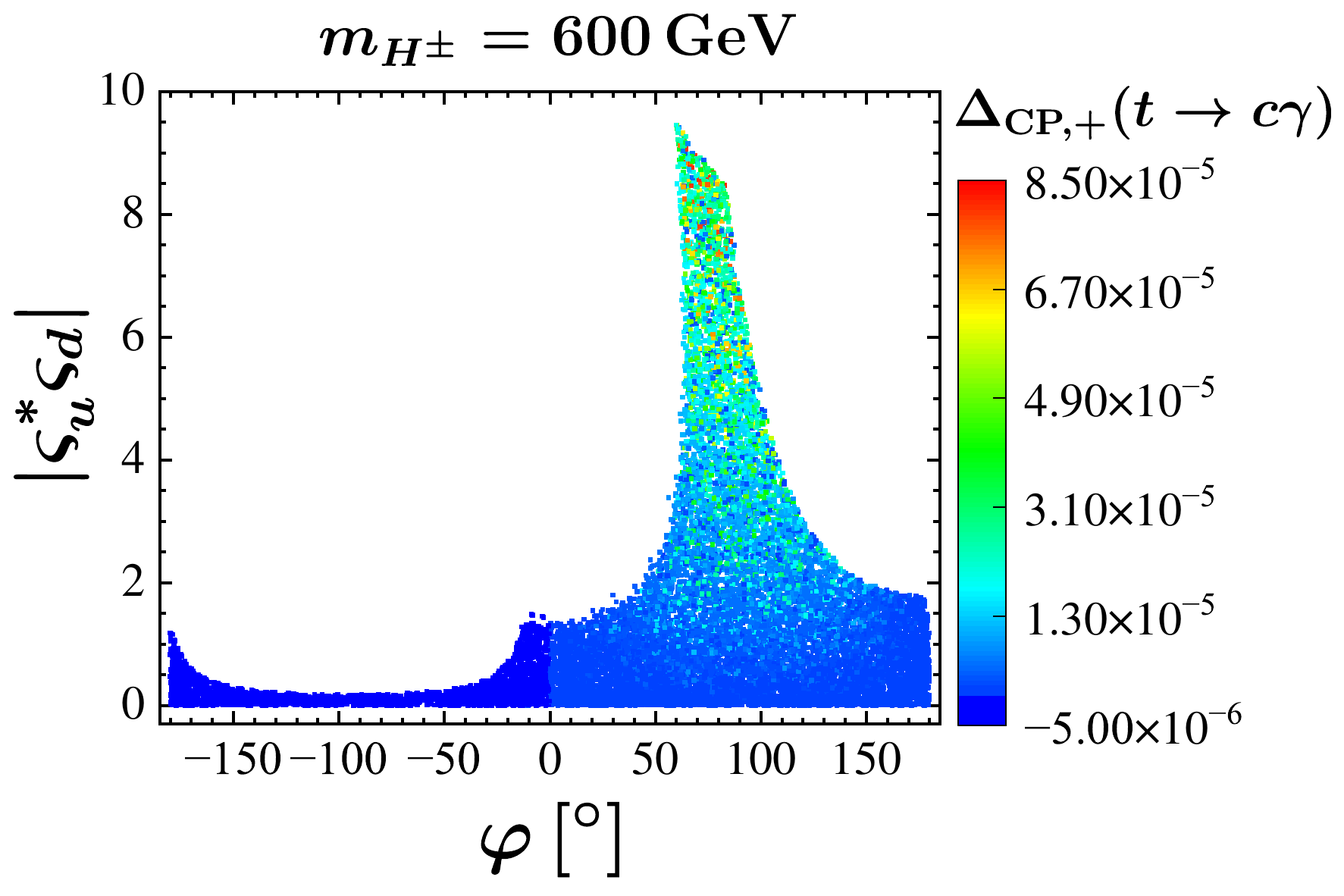}
    \includegraphics[width=0.49\textwidth]{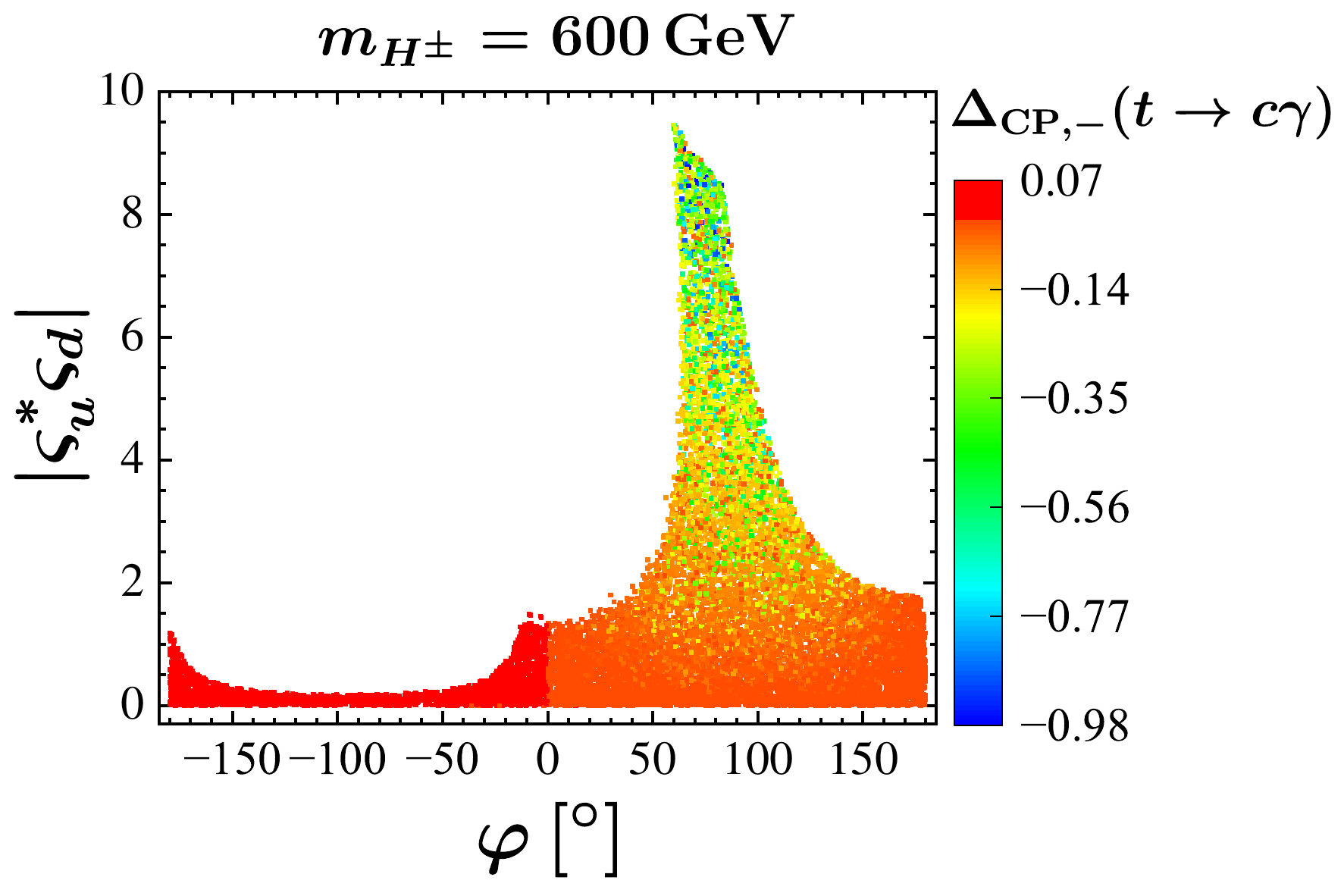}
    \caption{The polarized CP asymmetries $\Delta_{\text{CP},+}(t \to c\gamma)$ (left column) and $\Delta_{\text{CP},-}(t \to c\gamma)$ (right column) versus the product of the alignment parameters $|\varsigma_{u}^{\ast}\varsigma_{d}|$ and their relative phase $\varphi$ for three benchmark values of the charged-Higgs mass, $m_{H^{\pm}}=200~\gev$, $400~\gev$, and $600~\gev$, respectively. Note that, for $\Delta_{\text{CP},+}(t \to c\gamma)$, the dark blue regions represent the negative values, while for $\Delta_{\text{CP},-}(t \to c\gamma)$, the red regions represent the positive values. \label{Fig:tcy cpim}}
\end{figure}

Finally, we focus on the polarized CP asymmetries of $t \to c\gamma_{\pm}$ and $t \to cg_{\pm}$ decays in the A2HDM when the two alignment parameters $\varsigma_{u}$ and $\varsigma_{d}$ are complex. In this case, the relative phase $\varphi$ between $\varsigma_{u}$ and $\varsigma_{d}$ provides another source of CP violation beyond the CKM matrix of the SM. As can be seen from eqs.~\eqref{eq: polarized CPp}, \eqref{eq: polarized CPm} and \eqref{eq:Jarlskog_quantities}, the interference terms associated with the Jarlskog-like quantities $\mathcal{J}_{\alpha \beta}^{\mathbf{SN,L}}$ and $\mathcal{J}_{\alpha \beta}^{\mathbf{SN,R}}$ and hence the two polarized CP asymmetries can be significantly changed depending on the value of $\varphi$. To see this point clearly, we show in Fig.~\ref{Fig:tcy cpim} the polarized CP asymmetries $\Delta_{\text{CP},+}(t \to c\gamma)$ (left column) and $\Delta_{\text{CP},-}(t \to c\gamma)$ (right column), which are projected as colors onto the $|\varsigma_{u}^{\ast}\varsigma_{d}|-\varphi$ plane, for three benchmark values of the charged-Higgs mass, $m_{H^{\pm}}=200~\gev$, $400~\gev$, and $600~\gev$, respectively. It should be noted that for $\Delta_{\text{CP},+}(t \to c\gamma)$ the dark blue points represent the negative values, while for $\Delta_{\text{CP},-}(t \to c\gamma)$ the red points represent the positive values, as indicated by the color bars on the right side of each plot. It is observed that, as in the real case, the ranges of the CP asymmetries do not vary too much with the charged-Higgs mass $m_{H^{\pm}}$. In addition, the variation of $|\Delta_{\text{CP},\pm}|$ with respect to $\varphi$ depends on the allowed parameter space: in the range $\varphi \in [-180^\circ,0^\circ]$, small $|\varsigma_{u}^{\ast}\varsigma_{d}|$ result in small absolute values of $|\Delta_{\text{CP},\pm}|$, while large $|\Delta_{\text{CP},\pm}|$ can be obtained around the range $\varphi \in [50^\circ,150^\circ]$, which correspond to large $|\varsigma_{u}^{\ast}\varsigma_{d}|$. As a consequence, the two polarized CP asymmetries in the complex case can be significantly enhanced relative to both the SM and the real case. In particular, the maximum absolute value of the CP asymmetry $\Delta_{\text{CP},-}$ can even reach up to $\mathcal{O}(1)$ around the range $\varphi \in [70^\circ,100^\circ]$, as shown by the blue points in the right column of Fig.~\ref{Fig:tcy cpim}. Similar observations also apply to the CP asymmetries of $t \to cg_{\pm}$ decays, as demonstrated in Fig.~\ref{Fig:tcg cppmim}. Explicitly, the maximum absolute values of the polarized CP asymmetries that can be reached in the A2HDM with complex alignment parameters are given, respectively, as
\begin{equation} \label{eq:NCP imA2HDM}
\begin{aligned}
	|\Delta_{\text{CP},+,\text{complex}}^{\text{A2HDM, max}}(t \to c\gamma)| = 8.36 \times 10^{-5}\,,  \qquad  
	|\Delta_{\text{CP},-,\text{complex}}^{\text{A2HDM, max}}(t \to c\gamma)| = 0.99\,, \\[0.2cm]
	|\Delta_{\text{CP},+,\text{complex}}^{\text{A2HDM, max}}(t \to cg)| = 8.10 \times 10^{-5}\,,  \qquad  
	|\Delta_{\text{CP},-,\text{complex}}^{\text{A2HDM, max}}(t \to cg)| = 0.99 \,, 
\end{aligned} 
\end{equation}
which are about $4\sim 6$ orders of magnitude larger than the corresponding predictions within the SM (cf. eq.~\eqref{eq:NCPSM}) and in the real case (cf. eq.~\eqref{eq:NCP reA2HDM}).

To better understand the significant enhancements of the polarized CP asymmetries in the complex case, we should recall again that the dominant charged-Higgs effects induced by the relative phase $\varphi$ are encoded in the three Jarlskog-like quantities $\textstyle \mathcal{J}_{\alpha \beta}^{\mathbf{SN,L}}$, $\textstyle \mathcal{J}_{\alpha \beta}^{\mathbf{SN,R}}$, and $\textstyle \mathcal{R}_{\alpha \beta}^{\mathbf{SN}}$, defined in eq.~\eqref{eq:Jarlskog_quantities}. They affect the third terms in the numerators and denominators of eqs.~\eqref{eq: polarized CPp} and \eqref{eq: polarized CPm}, respectively. The third terms in the numerators of eqs.~\eqref{eq: polarized CPp} and \eqref{eq: polarized CPm} can be expressed, explicitly, as
\begin{eqnarray}
	\sum_{\alpha, \beta} \mathcal{J}_{\alpha \beta}^{\mathbf{SN,L}} \operatorname{Im}\left(\mathcal{F}_{\alpha}^{L} \mathcal{N}_{\beta}^{L\ast}\right) 
	& = & \sum_{\alpha, \beta} \left[ |\varsigma _{u}||\varsigma _{d}| \mathcal{N}_{\beta}^{L\ast} \operatorname{Im}\left( \mathcal{F}_{\alpha}^{L} \right) \operatorname{Im}\left( V_{t\alpha }^{\ast } V_{c\alpha } V_{t\beta  } V_{c\beta  }^{\ast } \right) \cos{\varphi} \right. \nonumber \\
	&& \hspace{0.48cm} \left. - |\varsigma _{u}||\varsigma _{d}| \mathcal{N}_{\beta}^{L\ast} \operatorname{Im}\left( \mathcal{F}_{\alpha}^{L} \right) \operatorname{Re} \left( V_{t\alpha }^{\ast } V_{c\alpha } V_{t\beta  } V_{c\beta  }^{\ast } \right) \sin{\varphi} \right] \,, \label{eq: third CPp} \\[0.3cm]
	\sum_{\alpha, \beta} \mathcal{J}_{\alpha \beta}^{\mathbf{SN,R}} \operatorname{Im}\left(\mathcal{F}_{\alpha}^{R} \mathcal{N}_{\beta}^{R\ast}\right) 
	& = & \sum_{\alpha, \beta} \left[ |\varsigma _{u}||\varsigma _{d}| \mathcal{N}_{\beta}^{R\ast} \operatorname{Im}\left( \mathcal{F}_{\alpha}^{R} \right) \operatorname{Im}\left( V_{t\alpha }^{\ast } V_{c\alpha } V_{t\beta  } V_{c\beta  }^{\ast } \right) \cos{\varphi} \right. \nonumber\\
	&& \hspace{0.45cm} \left. + |\varsigma _{u}||\varsigma _{d}| \mathcal{N}_{\beta}^{R\ast} \operatorname{Im}\left( \mathcal{F}_{\alpha}^{R} \right) \operatorname{Re} \left( V_{t\alpha }^{\ast } V_{c\alpha } V_{t\beta  } V_{c\beta  }^{\ast } \right) \sin{\varphi} \right] \,, \label{eq: third CPm}
\end{eqnarray} 
where the loop kinetic terms $\mathcal{N}_{\beta}^{L,R}$ mediated by the charged Higgs are real since we have assumed $m_{H^{\pm}}>m_{t}$, and $\operatorname{Im}(\mathcal{F}_{\alpha}^{L,R})$ are the imaginary parts of the loop integrals associated with the SM contributions. For the three different down-type quarks ($\alpha,\beta=d,s,b$) running in the loops, the resulting values of $\operatorname{Im}(\mathcal{F}_{\alpha}^{L,R})$ are approximately equal to each other. When summed over $\alpha$ and $\beta$, the terms associated with $\cos{\varphi}$ are largely cancelled out, due to the properties of the Jarlskog invariants, $|\mathcal{J}_{\alpha \beta}| = |\operatorname{Im}(V_{t\alpha }^{\ast } V_{c\alpha } V_{t\beta  } V_{c\beta  }^{\ast })| \approx 3.09 \times10^{-5}$ and $\mathcal{J}_{\alpha \beta}=-\mathcal{J}_{\beta \alpha}$, as well as the approximate equalities of $\operatorname{Im}(\mathcal{F}_{\alpha}^{L,R})$.\footnote{Making use of the relations among the Jarlskog invariants, $\mathcal{J}_{ds}= \mathcal{J}_{bd}= \mathcal{J}_{sb}= -\mathcal{J}_{sd}= -\mathcal{J}_{db}= -\mathcal{J}_{bs}$, as well as the numerical approximations of the SM form factors, $\operatorname{Im}(\mathcal{F}_{d}^{L,R})\approx \operatorname{Im}(\mathcal{F}_{s}^{L,R})\approx \operatorname{Im}(\mathcal{F}_{b}^{L,R})$, we can obtain $\mathcal{J}_{ds}(\mathcal{N}_{s}^{L\ast,R\ast} \operatorname{Im}\mathcal{F}_{d}^{L,R} ) \approx -\mathcal{J}_{bs}(\mathcal{N}_{s}^{L\ast,R\ast} \operatorname{Im}\mathcal{F}_{b}^{L,R} )$, $\mathcal{J}_{db}(\mathcal{N}_{b}^{L\ast,R\ast} \operatorname{Im}\mathcal{F}_{d}^{L,R}) \approx -\mathcal{J}_{sb}(\mathcal{N}_{b}^{L\ast,R\ast} \operatorname{Im}\mathcal{F}_{s}^{L,R} )$, $\mathcal{J}_{bd}(\mathcal{N}_{d}^{L\ast,R\ast} \operatorname{Im}\mathcal{F}_{b}^{L,R} ) \approx -\mathcal{J}_{sd}(\mathcal{N}_{d}^{L\ast,R\ast} \operatorname{Im}\mathcal{F}_{s}^{L,R})$. Therefore, the terms of the numerators of $\Delta_{\text{CP},\pm}$ associated with $\cos{\varphi}$ are largely cancelled out when summed over the three down-type quark flavours $\alpha,\beta=d,s,b$.} There are also strong cancellations in the first terms of the numerators of $\Delta_{\text{CP},\pm}$ (cf. eqs.~\eqref{eq: polarized CPp} and \eqref{eq: polarized CPm}), when contributions from the three down-type quark flavours are summed over. On the other hand, the cancellations among the different terms associated with $\sin{\varphi}$ are not so efficient, due to the variances of $\mathcal{R}_{\alpha \beta}=\operatorname{Re}(V_{t\alpha }^{\ast } V_{c\alpha } V_{t\beta } V_{c\beta  }^{\ast })$ with respect to the down-type quark flavours $\alpha=d,s,b$ and $\beta=d,s,b$. Therefore, the numerators of $\Delta_{\text{CP},\pm}$ are dominated by the terms associated with $\sin{\varphi}$. As the products $\mathcal{N}_{\beta}^{L\ast} \operatorname{Im}\left( \mathcal{F}_{\alpha}^{L} \right)$ and $\mathcal{N}_{\beta}^{R\ast} \operatorname{Im}\left( \mathcal{F}_{\alpha}^{R} \right)$ have numerically the same signs, the terms associated with $\sin{\varphi}$ in eqs.~\eqref{eq: third CPp} and \eqref{eq: third CPm} must have opposite signs. This in turn implies that the two polarized CP asymmetries $\Delta_{\text{CP},+}$ and $\Delta_{\text{CP},-}$ in the complex case have generally opposite signs, except when the terms associated with $\sin{\varphi}$ become smaller than the other terms in the numerators of $\Delta_{\text{CP},\pm}$, as can be clearly seen from Figs.~\ref{Fig:tcy cpim} and \ref{Fig:tcg cppmim}. Among the different terms in the denominators of $\Delta_{\text{CP},+}$ and $\Delta_{\text{CP},-}$, as given by eq.~\eqref{eq: Denominator}, the third one can be rewritten, explicitly, as
\begin{eqnarray} \label{eq: third Denominator}
        \sum_{\alpha, \beta} \mathcal{R}_{\alpha \beta}^{\mathbf{SN}} \left[ \operatorname{Re}\Big(\mathcal{F}_{\beta }^{R} \mathcal{N}_{\alpha}^{R\ast}\Big) m_{t}^{2} + \operatorname{Re}\Big(\mathcal{F}_{\alpha }^{L} \mathcal{N}_{\beta }^{L\ast}\Big) m_{c}^{2} \right] 
	& = & \sum_{\alpha, \beta} |\varsigma _{u}||\varsigma _{d}| \left[ \operatorname{Re}\left( V_{t\alpha }^{\ast } V_{c\alpha } V_{t\beta  } 
        V_{c\beta}^{\ast } \right)\cos{\varphi} \nonumber \right. \\[0.15cm]
        &   & \hspace{-6cm} \left. +\operatorname{Im} \left( V_{t\alpha }^{\ast } V_{c\alpha } V_{t\beta  } V_{c\beta  }^{\ast } \right) \sin{\varphi} \right] \left[\mathcal{N}_{\alpha}^{R\ast} \operatorname{Re}\left( \mathcal{F}_{\beta}^{R} \right) m_{t}^{2}+ \mathcal{N}_{\beta}^{L\ast} \operatorname{Re}\left( \mathcal{F}_{\alpha}^{L} \right) m_{c}^{2} \right] \,,
\end{eqnarray}
where the terms associated with $\sin{\varphi}$ are largely cancelled out, once the summation over the down-type quark flavours is performed. 

To clearly display the dependence of the polarized CP asymmetries on the model parameters $|\varsigma_{u}|$, $|\varsigma_{d}|$ and $\varphi$, let us fix the charged-Higgs mass at $m_{H^{\pm}}=200~\gev$, and define $\Delta_{\text{CP},+}^{200}(t \to c\gamma)=N_{200,+}/D_{200}$ and $\Delta_{\text{CP},-}^{200}(t \to c\gamma)=N_{200,-}/D_{200}$, with the numerical results of $N_{200,+}$, $N_{200,-}$, and $D_{200}$ given, respectively, as
\begin{eqnarray}
     N_{200,+} &=& |\varsigma_{u}||\varsigma_{d}|\sin{\varphi} \times2.37 \times 10^{-23}\left(1-2.39 \times 10^{-50} |\varsigma_{d}|^{2} + 6.48 \times 10^{-50} |\varsigma_{u}|^{2} \right) \notag\\[0.15cm]
     &-&|\varsigma_{u}||\varsigma_{d}|\cos{\varphi} \times2.42 \times 10^{-29}\left(1 + 3.35 \times 10^{-49} |\varsigma_{d}|^{2} - 3.84 \times 10^{-50} |\varsigma_{u}|^{2} \right)\notag\\[0.15cm]
     &-&5.77\times 10^{-29} \left(1-1.26 \times 10^{-1}|\varsigma_{d}|^{2} -2.95\times 10^{-1}|\varsigma_{u}|^{2} -2.66 \times 10^{-51} |\varsigma_{d}|^{4} \right. \notag\\[0.15cm]
	&& \left. \hspace{2.5cm} - 6.07 \times 10^{-50}|\varsigma_{u}|^{4}+1.40 \times 10^{-49}|\varsigma_{d}|^{2} |\varsigma_{u}|^{2}\right) \,, \\[0.2cm]
     N_{200,-}&=&-|\varsigma_{u}||\varsigma_{d}|\sin{\varphi}\times2.83 \times 10^{-19}\left(1-1.32 \times 10^{-51} |\varsigma_{d}|^{2}-8.82 \times 10^{-51} |\varsigma_{u}|^{2}\right) \notag\\[0.15cm]
	&& -|\varsigma_{u}||\varsigma_{d}|\cos{\varphi}\times6.76 \times 10^{-25} \left(1+3.77 \times 10^{-51} |\varsigma_{d}|^{2}+3.93 \times 10^{-51} 
     |\varsigma_{u}|^{2} \right)\notag\\[0.15cm]
	&& -3.55\times 10^{-24} \left(1-3.58 \times 10^{-2}|\varsigma_{d}|^{2}+5.66\times 10^{-6}|\varsigma_{u}|^{2}+2.01 \times 10^{-55} |\varsigma_{d}|^{4} \right. \notag\\[0.15cm]
	&& \left. \hspace{2.8cm} +4.56 \times 10^{-56}|\varsigma_{u}|^{4}+7.39 \times 10^{-52}|\varsigma_{d}|^{2} |\varsigma_{u}|^{2} \right) \,, \\[0.2cm]
     D_{200}&=&|\varsigma_{u}||\varsigma_{d}|\sin{\varphi}\times2.23\times 10^{-24} \left(1-4.35\times 10^{-2} |\varsigma_{d}|^{2}+1.69\times 10^{-5} |\varsigma_{u}|^{2} \right) \notag\\[0.15cm] 
	&+&|\varsigma_{u}||\varsigma_{d}|\cos{\varphi}\times5.83 \times 10^{-19} \left(1-2.35\times 10^{-2} |\varsigma_{d}|^{2}+6.15\times 10^{-5}|\varsigma_{u}|^{2} \right) \notag\\[0.15cm]
	&+&7.25\times 10^{-19} \left(1-3.80\times 10^{-2}|\varsigma_{d}|^{2}+1.40\times10^{-5}|\varsigma_{u}|^{2}+4.46\times 10^{-4}|\varsigma_{d}|^{4} \right. \notag\\[0.15cm]
	&& \left. \hspace{2.65cm} +1.52\times 10^{-5}|\varsigma_{u}|^{4}+1.99\times 10^{-1}|\varsigma_{d}|^{2} |\varsigma_{u}|^{2} \right) \,.
\end{eqnarray}
From the numerical coefficients of the different alignment parameters, we can see that the numerators of the CP asymmetries are dominated by the terms associated with $\sin{\varphi}$, while the denominators by the terms that are independent of $\sin{\varphi}$. As the terms associated with $\sin{\varphi}$ in eqs.~\eqref{eq: third CPp}--\eqref{eq: third Denominator} are non-vanishing only when the two alignment parameters $\varsigma_{u}$ and $\varsigma_{d}$ are complex with a non-zero $\varphi$, it is expected that large enhancements of the CP asymmetries can be generated only in the complex case, due to the large contributions of the terms associated with $\sin{\varphi}$ in the numerators, while the similar term in the denominator has only a marginal impact.

As can be seen from eqs.~\eqref{eq:NCP reA2HDM} and \eqref{eq:NCP imA2HDM}, the same hierarchy $\Delta_{\text{CP},+} \ll \Delta_{\text{CP},-}$ as observed within the SM also holds in the A2HDM with both real and complex alignment parameters. The polarization-independent CP asymmetries of $t \to c \gamma$ and $t \to cg$ decays can then be approximated as $\Delta_{\text{CP}} \approx \Delta_{\text{CP},-}$. Comparing Figs.~\ref{Fig:tc branching ratio}, \ref{Fig:tcy cpre}, and \ref{Fig:tcg cppmre}, we can see that the branching ratios and CP asymmetries of these two decay modes follow different trends with the variations of $\varsigma_{u}$ and $\varsigma_{d}$ in the real parameter space, where large CP asymmetries correspond to small branching ratios. In the complex parameter space, the situation is similar; for a given complex parameter space where $|\Delta_{\text{CP}}|$ approach to $\mathcal{O}(1)$, the predicted branching ratios would be very small. For example, for the decay $t \to c \gamma$, $|\Delta_{\text{CP}}|$ can maximally attain $\mathcal{O}(10^{-4})$ for $\mathcal{B}\sim \mathcal{O}(10^{-10})$, $\mathcal{O}(10^{-3})$ for $\mathcal{B}\sim \mathcal{O}(10^{-11})$, $\mathcal{O}(10^{-2})$ for $\mathcal{B}\sim \mathcal{O}(10^{-12})$, and $\mathcal{O}(10^{-1})$ for $\mathcal{B}\sim \mathcal{O}(10^{-14}-10^{-13})$; for the decay $t \to cg$, on the other hand, $|\Delta_{\text{CP}}|$ can maximally attain $\mathcal{O}(10^{-3})$ for $\mathcal{B}\sim \mathcal{O}(10^{-9})$, $\mathcal{O}(10^{-2})$ for $\mathcal{B}\sim \mathcal{O}(10^{-10})$ and $\mathcal{O}(10^{-1})$ for $\mathcal{B}\sim \mathcal{O}(10^{-12}-10^{-11})$. Although the maximum branching ratios of $t \to c \gamma$ and $t \to cg$ decays predicted in the A2HDM are still below the current experimental upper limits of $\mathcal{O}(10^{-5})$ and $\mathcal{O}(10^{-4})$~\cite{ParticleDataGroup:2024cfk,LHCTop}, the CP asymmetries of these two decay modes can receive significant enhancements in the same model with complex alignment parameters.

\section{Conclusion}
\label{sec:conclusion}

In this paper, we have performed a complete one-loop calculation of the $t \to c \gamma$ and $t \to c g$ decays in the A2HDM, focusing primarily on how large the CP asymmetries of these rare FCNC decays are possible in the model, after taking into account the most relevant constraints on the parameter space. Our main conclusions can be summarized as follows:
\begin{itemize}
	 \item We have updated the SM predictions for the branching ratios and CP asymmetries of $t \to c \gamma(g)$ decays, which are given by eqs.~\eqref{eq:NbrSM} and \eqref{eq:NCPSM}, respectively. Our SM results are generally consistent with those obtained in refs.~\cite{Aguilar-Saavedra:2002lwv,Balaji:2020qjg}. It should be mentioned that these observables are mostly sensitive to the internal bottom-quark mass, for which the $\overline{\rm MS}$ running mass has been chosen. 
	
    \item In the A2HDM, due to the absence of tree-level FCNC interactions, the additional charged Higgs contributes to the $t \to c \gamma(g)$ decays firstly at the one-loop level. When the two alignment parameters $\varsigma_{u,d}$ are real, the branching ratios of $t \to c \gamma$ and $t \to cg$ decays can maximally reach up to $1.47\times10^{-10}$ and $4.86\times10^{-9}$, respectively. These results are about four and three orders of magnitude higher than the corresponding SM predictions, but are still below the current experimental upper limits. They are also out of the expected sensitivities of the HL-LHC and the future colliders. When $\varsigma_{u}$ and $\varsigma_{d}$ are complex, on the other hand, the branching ratios are found to be almost independent of the relative phase $\varphi$ defined by $\varsigma_{u}^{\ast} \varsigma_{d}=|\varsigma_{u}||\varsigma_{d}|e^{-i\varphi}$, within the parameter space allowed by the current experimental data. 
	
    \item The polarized CP asymmetries of $t \to c \gamma(g)$ decays have been investigated in the A2HDM with both real and complex alignment parameters, under the assumption of $m_{H^{\pm}} > m_t$ to avoid the presence of the decay $t\to b H^{+}$. In the real case, the sources of CP violation are the same as within the SM, and the predicted CP asymmetries do not show any significant enhancements compared to the SM results. When $\varsigma_{u}$ and $\varsigma_{d}$ are complex, however, the relative phase $\varphi$ provides another source of CP violation beyond the CKM matrix of the SM. The resulting CP asymmetries are found to be very sensitive to the relative phase $\varphi$: when $\varphi$ varies within the range $[50^\circ,150^\circ]$, the magnitudes of the CP asymmetries can be significantly enhanced relative to both the SM and the real case; in particular, the maximum absolute values of the CP asymmetries can even reach up to $\mathcal{O}(1)$ for these two decay modes, in the range $\varphi \in [70^\circ,100^\circ]$. The variations of these observables with respect to the model parameters and the mechanism underlying these enhancements have been discussed in detail.

    \item Due to the angular momentum conservation and the $V-A$ nature of weak interaction, the emitted photons (gluons) are predominantly left-handed in the $t \to c\gamma (g)$ decays. This indicates the hierarchy $\Delta_{\text{CP},+} \ll \Delta_{\text{CP},-}$ between the two polarized CP asymmetries, and the polarization-independent CP asymmetries can be approximated as $\Delta_{\text{CP}} \approx \Delta_{\text{CP},-}$. The branching ratios and CP asymmetries show different trends with the variations of $\varsigma_{u}$ and $\varsigma_{d}$: when the model parameters are set to obtain large CP asymmetries, the corresponding branching ratios would be small. 
\end{itemize}

These interesting observations may motivate us to perform further detailed studies of these rare FCNC top-quark decays from both the theoretical and experimental aspects. In particular, as new sources of CP violation beyond the SM are usually required to explain the baryon asymmetry of the Universe, it is of paramount importance to develop new search strategies of CP violation in the top-quark sector. It is also conceivable that much more advanced data analysis techniques will be available by the time when new energy-frontier colliders start to operate, which can even render some producible decays like the ones discussed here observable.

\acknowledgments
This work is supported by the National Natural Science Foundation of China under Grant Nos.~12475094, 12135006, 12075097 and 12347175, the China Postdoctoral Foundation under Grant No.~GZB20230195, as well as by the Fundamental Research Funds for the Central Universities under Grant Nos.~CCNU22LJ004 and CCNU24AI003.

\appendix

\section{\boldmath Feynman rules for \texorpdfstring{$t \to c \gamma (g)$}{ttocgammag} decays in the A2HDM}
\label{sec:Feynman rules}

The Feynman rules needed to calculate the charged-Higgs contributions to the $t \to c \gamma (g)$ decays in the A2HDM are listed below.

\begin{figure}[h]
	\centering 
	\hspace{2cm}
	\begin{minipage}{0.4\textwidth}
		\tikzset{every picture/.style={line width=0.75pt}}
		\begin{tikzpicture}[x=0.75pt,y=0.75pt,yscale=-1,xscale=1]
			\draw [line width=1.0]  [dash pattern={on 4.5pt off 4.5pt}]  (280,110) -- (371,110) ;
			\draw [shift={(329.9,110)}, rotate = 180] [fill={rgb, 255:red, 0; green, 0; blue, 0 }  ][line width=0.08]  [draw opacity=0] (8.75,-4.2) -- (0,0) -- (8.75,4.2) -- (5.81,0) -- cycle    ;
			\draw (311,114.4) node [anchor=north west][inner sep=0.75pt]  [font=\normalsize]  {$H^{\pm }$};
			\draw (316,88.4) node [anchor=north west][inner sep=0.75pt]  [font=\normalsize]  {$p$};
		\end{tikzpicture}
	\end{minipage} 
	\hfill
	\hspace{-9cm}	
	\begin{minipage}{0.6\textwidth}
		\begin{equation}
			\frac{i}{p^{2}-m_{H^{\pm}}^{2}+i\epsilon} \nonumber
		\end{equation}
	\end{minipage}
\end{figure}

\begin{figure}[h]
	\hspace{2cm}
	\begin{minipage}{0.4\textwidth}
		\tikzset{every picture/.style={line width=0.75pt}} 
		\begin{tikzpicture}[x=0.75pt,y=0.75pt,yscale=-1,xscale=1]
			\draw [line width=1.0]  [dash pattern={on 4.5pt off 4.5pt}]  (149.57,110) -- (190,110.5) ;
			\draw    (190,110.5) -- (235.57,156) ;
			\draw [shift={(209.88,130.35)}, rotate = 44.96] [fill={rgb, 255:red, 0; green, 0; blue, 0 }  ][line width=0.08]  [draw opacity=0] (7.14,-3.43) -- (0,0) -- (7.14,3.43) -- (4.74,0) -- cycle    ; 
			\draw    (190,110.5) -- (234.57,66) ;
			\draw [shift={(214.13,86.41)}, rotate = 135.05] [fill={rgb, 255:red, 0; green, 0; blue, 0 }  ][line width=0.08]  [draw opacity=0] (7.14,-3.43) -- (0,0) -- (7.14,3.43) -- (4.74,0) -- cycle    ;
			\draw (146,115.4) node [anchor=north west][inner sep=0.75pt]  [font=\normalsize]  {$H^{+}$};
			\draw (201,65.4) node [anchor=north west][inner sep=0.75pt]    {$u_{\alpha }$};
			\draw (201,137.4) node [anchor=north west][inner sep=0.75pt]    {$d_{\beta }$};
		\end{tikzpicture}
	\end{minipage} 
	\hfill
	\hspace{-3cm}	
	\begin{minipage}{0.6\textwidth}
		\begin{equation}
			\frac{ig}{\sqrt{2}m_{W}}\left( \varsigma_{u} m_{u_{\alpha}} V_{\alpha \beta} P_{L} - \varsigma_{d} m_{d_{\beta}} V_{\alpha \beta} P_{R} \right) \nonumber
		\end{equation}
	\end{minipage}
\end{figure}

\begin{figure}[h]
	\hspace{2cm}
	\begin{minipage}{0.4\textwidth}
		\tikzset{every picture/.style={line width=0.75pt}} 
		\begin{tikzpicture}[x=0.75pt,y=0.75pt,yscale=-1,xscale=1]
			\draw [line width=1.0]  [dash pattern={on 4.5pt off 4.5pt}]  (149.57,110) -- (190,110.5) ;
			\draw    (190,110.5) -- (235.57,156) ;
			\draw [shift={(214.63,135.09)}, rotate = 224.96] [fill={rgb, 255:red, 0; green, 0; blue, 0 }  ][line width=0.08]  [draw opacity=0] (7.14,-3.43) -- (0,0) -- (7.14,3.43) -- (4.74,0) -- cycle    ; 
			\draw    (190,110.5) -- (234.57,66) ;
			\draw [shift={(209.38,91.15)}, rotate = 315.05] [fill={rgb, 255:red, 0; green, 0; blue, 0 }  ][line width=0.08]  [draw opacity=0] (7.14,-3.43) -- (0,0) -- (7.14,3.43) -- (4.74,0) -- cycle    ;
			\draw (146,115.4) node [anchor=north west][inner sep=0.75pt]  [font=\normalsize]  {$H^{-}$};
			\draw (201,65.4) node [anchor=north west][inner sep=0.75pt]    {$u_{\alpha }$};
			\draw (201,137.4) node [anchor=north west][inner sep=0.75pt]    {$d_{\beta }$};
		\end{tikzpicture}
	\end{minipage} 
	\hfill
	\hspace{-3cm}	
	\begin{minipage}{0.6\textwidth}
		\begin{equation}
			\frac{ig}{\sqrt{2}m_{W}} \left( \varsigma_{u}^{*} m_{u_{\alpha}} V_{\alpha \beta}^{*} P_{R} - \varsigma_{d}^{*} m_{d_{\beta}} V_{\alpha \beta}^{*} P_{L} \right) \nonumber
		\end{equation}
	\end{minipage}
\end{figure}
\noindent Here $g$ is the $SU(2)_L$ gauge coupling, and $P_{L,R}=(1\mp\gamma_5)/2$ are the left- and right-handed chiral projectors.

\section{\boldmath Polarized decay amplitudes of \texorpdfstring{$t \to c \gamma_{\pm}$}{ttocgammapm} decays}
\label{sec:Polarized amplitudes}

In this appendix, we detail the derivation of the polarized $t(p_{\text{i}}) \to c(p_{\text{f}}) \gamma_{\pm}(q)$ and $\bar{t}(p_{\text{i}}) \to \bar{c}(p_{\text{f}}) \gamma_{\pm}(q)$ amplitudes, using the explicit representation of the Dirac spinors and $\gamma$ matrices. The same derivation presented here can also be applied to the $t \to cg$ process. 

To simplify the derivation, let us work in the rest frame of the initial top or the anti-top quark, and assume that the photon is released along the $+z$ direction, although the final results should be frame-independent and can be generalized to any inertial reference frame via spatial rotations and Lorentz boosts. In such a kinematic configuration, the four-momenta of the initial- and final-state particles can then be written, respectively, as
\begin{eqnarray}
	p_{\text{i}}^{\mu}=(m_t,0,0,0)\,, \qquad
	q^\mu=(q,0,0,q)\,,\qquad
	p_{\text{f}}^\mu = (E_{\text{f}}, 0,0,-q)\,,
\end{eqnarray}
where, due to energy-momentum conservation, $q$ and $E_{\text{f}}$ are given by
\begin{eqnarray}
	q=\frac{m_t^2-m_c^2}{2m_t}\,, \qquad
	E_{\text{f}}=\frac{m_t^2+m_c^2}{2m_t}\,.
\end{eqnarray}
As the angular momentum along the $z$ direction must be conserved, $S_z(t) = S_z(c) + S_z(\gamma)$, with $S_z = \pm 1/2$ for a fermion and $S_z = \pm 1$ for an on-shell photon, we have only two types of polarized amplitudes, $i\mathcal{M}(t_{+\frac{1}{2}} \to c_{-\frac{1}{2}} + \gamma _{+})$ and $i\mathcal{M}(t_{-\frac{1}{2}} \to c_{+\frac{1}{2}} + \gamma _{-})$. For an on-shell photon moving in the $+z$ direction, the two physical polarization four-vectors can be written as~\cite{Peskin:1995ev}
\begin{eqnarray} \label{eq:polarization_vectors}
	\varepsilon_+^\mu = \frac{1}{\sqrt{2}} (0,1,i,0)\,,\qquad
	\varepsilon_-^\mu = \frac{1}{\sqrt{2}} (0,1,-i,0)\,, 
\end{eqnarray}
corresponding to $S_z(\gamma)=+1$ (with a righ-handed helicity) and $S_z(\gamma)=-1$ (with a left-handed helicity), respectively.

For the Dirac $\gamma$ matrices, we choose the chiral (spinorial) representation with
\begin{eqnarray}
	\gamma^\mu =  \begin{pmatrix}
		0 & \sigma^\mu  \\
		\bar{\sigma}^\mu & 0
	\end{pmatrix}\,,  \quad 
	\sigma^{\mu\nu} = \frac{i}{2} [\gamma^\mu,\gamma^\nu]\,, \quad
	\gamma_5 \equiv i\gamma^0\gamma^1\gamma^2\gamma^3 =  \begin{pmatrix}
		-\mathbf{1} & 0  \\
		0 & \mathbf{1}
	\end{pmatrix}\,, 
\end{eqnarray}
where $\sigma^\mu=(\mathbf{1},\sigma^1, \sigma^2, \sigma^3)$ and $\bar{\sigma}^\mu=(\mathbf{1},-\sigma^1, -\sigma^2, -\sigma^3)$, with $\sigma^i$ ($i=1,2,3$) being the Pauli matrices. The normalized Dirac spinors for a particle and an anti-particle with momentum $p=(p_0, \vec{p})$ can then be represented, respectively, by
\begin{eqnarray}
	u_S(p) = \begin{pmatrix}
		\sqrt{p \cdot \sigma}\,\xi_S  \\[0.1cm]
		\sqrt{p \cdot \bar{\sigma}}\,\xi_S
	\end{pmatrix} \,, \qquad
	v_S(p) = \begin{pmatrix}
		\sqrt{p \cdot \sigma}\,\eta_S  \\[0.1cm]
		-\sqrt{p \cdot \bar{\sigma}}\,\eta_S
	\end{pmatrix} \,,
\end{eqnarray}
where $\xi_S$ and $\eta_S$ are the two-component spinors normalized to unity, and the subscript $S=\pm 1/2$ characters the polarization of the fermion. Specific to the polarized $t_{\pm\frac{1}{2}} \to c_{\mp\frac{1}{2}} + \gamma _{\pm}$ and $\bar{t}_{\pm\frac{1}{2}} \to \bar{c}_{\mp\frac{1}{2}} + \gamma _{\pm}$ decays as viewed in the top and the anti-top rest frame respectively, the Dirac spinors for the initial fermions are given explicitly by
\begin{equation}
\begin{aligned}
	u_{+\frac{1}{2}}(p_{\text{i}}) &= \sqrt{m_t} \begin{pmatrix}
		\xi_{+\frac{1}{2}}  \\[0.15cm]
		\xi_{+\frac{1}{2}}
	\end{pmatrix} \,, \qquad &&
	u_{-\frac{1}{2}}(p_{\text{i}}) = \sqrt{m_t} \begin{pmatrix}
		\xi_{-\frac{1}{2}}  \\[0.15cm]
		\xi_{-\frac{1}{2}}
	\end{pmatrix} \,, \\[0.2cm]
	v_{+\frac{1}{2}}(p_{\text{i}}) &= \sqrt{m_t} \begin{pmatrix}
		\eta_{+\frac{1}{2}}  \\[0.15cm]
		-\eta_{+\frac{1}{2}}
	\end{pmatrix} \,, \qquad &&
	v_{-\frac{1}{2}}(p_{\text{i}}) = \sqrt{m_t} \begin{pmatrix}
		\eta_{-\frac{1}{2}}  \\[0.15cm]
		-\eta_{-\frac{1}{2}}
	\end{pmatrix} \,,
\end{aligned} 
\end{equation}
while that for the final charm and anti-charm quarks moving in the $-z$ direction by
\begin{equation} \label{eq:spinor}
\begin{aligned}
	u_{+\frac{1}{2}}(p_{\text{f}}) &= \begin{pmatrix}
		\sqrt{E_{\text{f}}+q}\,\xi_{+\frac{1}{2}}  \\[0.15cm]
		\sqrt{E_{\text{f}}-q}\,\xi_{+\frac{1}{2}}
	\end{pmatrix} \,, \qquad &&
	u_{-\frac{1}{2}}(p_{\text{f}}) = \begin{pmatrix}
		\sqrt{E_{\text{f}}-q}\,\xi_{-\frac{1}{2}}  \\[0.15cm]
		\sqrt{E_{\text{f}}+q}\,\xi_{-\frac{1}{2}}
	\end{pmatrix} \,, \\[0.2cm]
	v_{+\frac{1}{2}}(p_{\text{f}}) &= \begin{pmatrix}
		\sqrt{E_{\text{f}}-q}\,\eta_{+\frac{1}{2}}  \\[0.15cm]
		-\sqrt{E_{\text{f}}+q}\,\eta_{+\frac{1}{2}}
	\end{pmatrix} \,, \qquad && 
	v_{-\frac{1}{2}}(p_{\text{f}}) = \begin{pmatrix}
		\sqrt{E_{\text{f}}+q}\,\eta_{-\frac{1}{2}}  \\[0.15cm]
		-\sqrt{E_{\text{f}}-q}\,\eta_{-\frac{1}{2}}
	\end{pmatrix} \,,
\end{aligned}
\end{equation}
with
\begin{equation}
	\xi_{+\frac{1}{2}}  = \eta_{-\frac{1}{2}}  = \begin{pmatrix}
		1  \\ 0
	\end{pmatrix} \,, \qquad
	\xi_{-\frac{1}{2}}  = \eta_{+\frac{1}{2}}  = \begin{pmatrix}
		0  \\ 1
	\end{pmatrix} \,.
\end{equation}
It can be seen that, in the $m_c \to 0$ limit, the spinors $u_{+\frac{1}{2}}(p_{\text{f}})$ and $u_{-\frac{1}{2}}(p_{\text{f}})$ are purely left- and right-handed respectively, when the charm quark moves along the $-z$ direction. 

Equipped with all the above formulae, we can now write down the explicit expressions of the amplitudes in eqs.~\eqref{eq:decay_amplitude} and \eqref{eq:CP amplitude}, with definite spins of the initial and final states. The non-vanishing polarized amplitudes are given, respectively, by 
\begin{equation} \label{eq:amplitudes_spin}
\begin{aligned}
	\mathcal{M}(t_{+\frac{1}{2}} \to c_{-\frac{1}{2}} + \gamma_+) &= +\sqrt{2} f^{L}_{\text{fi},\gamma} 
        \left(m_t^2 - m_c^2\right) \,, \\[0.15cm]
	\mathcal{M}(t_{-\frac{1}{2}} \to c_{+\frac{1}{2}} + \gamma_-) &= - \sqrt{2} f^{R}_{\text{fi},\gamma} 
        \left(m_t^2 - m_c^2\right)\,, \\[0.15cm]
	\mathcal{M}(\bar{t}_{+\frac{1}{2}} \to \bar{c}_{-\frac{1}{2}} + \gamma_+) &= -\sqrt{2} \bar{f}^{L}_{\text{if},\gamma} 
        \left(m_t^2 - m_c^2\right) \,, \\[0.15cm]
	\mathcal{M}(\bar{t}_{-\frac{1}{2}} \to \bar{c}_{+\frac{1}{2}} + \gamma_-) &= +\sqrt{2} \bar{f}^{R}_{\text{if},\gamma} 
        \left(m_t^2 - m_c^2\right) \,, 
\end{aligned}
\end{equation}
while all the other ones are zero because of the angular momentum conservation. It should be noted that the same results for the polarized amplitudes as given by eq.~\eqref{eq:amplitudes_spin} can also be obtained in any inertial reference frame, as required by the Lorentz invariance.

\section{\boldmath Loop kinetic terms for \texorpdfstring{$t \to c \gamma (g)$}{ttocgamma(g)} decays}
\label{sec:loop kinetic terms}

In this appendix, we give the explicit expressions of the loop kinetic terms $\mathcal{F}^{L,R}$ and $\mathcal{N}^{L,R}$ present in eq.~\eqref{eq:effective vertex2}. For the decay $t\to c\gamma$, they are given, respectively, by
\begin{eqnarray}
    \mathcal{F}_{\alpha}^{R}&= &-\frac{e G_{F}}{12 \sqrt{2} \pi^{2}} \Bigl\{3(m_{\alpha }^2+2m_{W}^2)C_{22}(m_c^2,0,m_t^2,m_{\alpha }^2,m_{W}^2,m_{W}^2) \nonumber\\[0.15cm]
    &&+ (m_{\alpha }^2+2m_{W}^2)C_{22}(m_c^2,0,m_t^2,m_{W}^2,m_{\alpha}^2,m_{\alpha}^2) \nonumber\\[0.15cm]
    &&+3(m_{\alpha }^2+2m_{W}^2+m_{c}^{2})C_{12}(m_c^2,0,m_t^2,m_{\alpha }^2,m_{W}^2,m_{W}^2) \nonumber\\[0.15cm]
    &&+(m_{\alpha }^2+2m_{W}^2+m_{c}^{2})C_{12}(m_c^2,0,m_t^2,m_{W}^2,m_{\alpha}^2,m_{\alpha}^2) \nonumber\\[0.15cm]
    &&+m_{c}^{2}\Bigl[3C_{11}(m_c^2,0,m_t^2,m_{\alpha }^2,m_{W}^2,m_{W}^2) +C_{11}(m_c^2,0,m_t^2,m_{W}^2,m_{\alpha}^2,m_{\alpha}^2)\Bigr] \nonumber\\[0.15cm] 
    &&+3m_{\alpha }^2 C_{0}(m_c^2,0,m_t^2,m_{\alpha }^2,m_{W}^2,m_{W}^2) +2m_{W}^2C_{0}(m_c^2,0,m_t^2,m_{W}^2,m_{\alpha}^2,m_{\alpha}^2) \nonumber\\[0.15cm]
    &&+3(m_{\alpha }^2-2m_{W}^2+m_{c}^{2})C_{1}(m_c^2,0,m_t^2,m_{\alpha }^2,m_{W}^2,m_{W}^2) \nonumber\\[0.15cm]
    &&+(-m_{\alpha }^2+2m_{W}^2+m_{c}^{2})C_{1}(m_c^2,0,m_t^2,m_{W}^2,m_{\alpha}^2,m_{\alpha}^2) \nonumber\\[0.15cm]
    &&+6m_{\alpha }^2 C_{2}(m_c^2,0,m_t^2,m_{\alpha }^2,m_{W}^2,m_{W}^2) +4m_{W}^2C_{2}(m_c^2,0,m_t^2,m_{W}^2,m_{\alpha}^2,m_{\alpha}^2) \nonumber\\[0.15cm]
    &&+ \varsigma_{d}\varsigma_{d}^{\ast} m_{\alpha}^{2} \Bigl[ 3C_{22}(m_c^2,0,m_t^2,m_{\alpha }^2,m_{H^{\pm }}^2,m_{H^{\pm }}^2) +C_{22}(m_c^2,0,m_t^2,m_{H^{\pm }}^2,m_{\alpha}^2,m_{\alpha}^2) \nonumber\\[0.15cm]
    &&+ 3C_{12}(m_c^2,0,m_t^2,m_{\alpha }^2,m_{H^{\pm }}^2,m_{H^{\pm }}^2) +C_{12}(m_c^2,0,m_t^2,m_{H^{\pm }}^2,m_{\alpha}^2,m_{\alpha}^2) \nonumber\\[0.15cm]
    &&+ 3C_{2}(m_c^2,0,m_t^2,m_{\alpha }^2,m_{H^{\pm }}^2,m_{H^{\pm }}^2) +C_{2}(m_c^2,0,m_t^2,m_{H^{\pm }}^2,m_{\alpha}^2,m_{\alpha}^2) \Bigr] \nonumber\\[0.15cm] 
    &&+ \varsigma_{u}\varsigma_{u}^{\ast} m_{c}^{2} \Bigl[ 3C_{12}(m_c^2,0,m_t^2,m_{\alpha }^2,m_{H^{\pm }}^2,m_{H^{\pm }}^2) +C_{12}(m_c^2,0,m_t^2,m_{H^{\pm }}^2,m_{\alpha}^2,m_{\alpha}^2) \nonumber\\[0.15cm]
    &&+ 3C_{11}(m_c^2,0,m_t^2,m_{\alpha }^2,m_{H^{\pm }}^2,m_{H^{\pm }}^2) +C_{11}(m_c^2,0,m_t^2,m_{H^{\pm }}^2,m_{\alpha}^2,m_{\alpha}^2) \nonumber\\[0.15cm]
    &&+ 3C_{1}(m_c^2,0,m_t^2,m_{\alpha }^2,m_{H^{\pm }}^2,m_{H^{\pm }}^2) +C_{1}(m_c^2,0,m_t^2,m_{H^{\pm }}^2,m_{\alpha}^2,m_{\alpha}^2) \Bigr]\Bigr\}\,, \\[0.2cm]
	\mathcal{N}^{R}_{\alpha} &=& \mathcal{N}^{L}_{\alpha} \nonumber\\[0.15cm]
    &=& -\frac{e G_{F}}{12 \sqrt{2} \pi^{2}} m_{\alpha}^2 \Bigl[3C_{0}(m_c^2,0,m_t^2,m_{\alpha }^2,m_{H^{\pm }}^2,m_{H^{\pm }}^2)  \nonumber\\[0.15cm]
    &&+3C_{1}(m_c^2,0,m_t^2,m_{\alpha }^2,m_{H^{\pm }}^2,m_{H^{\pm }}^2)-C_{1}(m_c^2,0,m_t^2,m_{H^{\pm }}^2,m_{\alpha}^2,m_{\alpha}^2) \nonumber\\[0.15cm]
    &&+3C_{2}(m_c^2,0,m_t^2,m_{\alpha }^2,m_{H^{\pm }}^2,m_{H^{\pm }}^2)-C_{2}(m_c^2,0,m_t^2,m_{H^{\pm }}^2,m_{\alpha}^2,m_{\alpha}^2)\Bigr]\,.
\end{eqnarray}
For the decay $t\to c g$, on the other hand, we have explicitly
\begin{eqnarray}
	\mathcal{F}_{\alpha}^{R} &=& \frac{g_{s}G_{F}}{4 \sqrt{2} \pi^{2}} \Bigl\{ (m_{\alpha }^2+2m_{W}^2)C_{22}(m_c^2,0,m_t^2,m_{W}^2,m_{\alpha}^2,m_{\alpha}^2) \nonumber\\[0.15cm]
    &&+(m_{\alpha }^2+2m_{W}^2+m_{c}^{2})C_{12}(m_c^2,0,m_t^2,m_{W}^2,m_{\alpha}^2,m_{\alpha}^2) \nonumber\\[0.15cm]
    &&+m_{c}^{2}C_{11}(m_c^2,0,m_t^2,m_{W}^2,m_{\alpha}^2,m_{\alpha}^2)+2m_{W}^{2}C_{0}(m_c^2,0,m_t^2,m_{W}^2,m_{\alpha}^2,m_{\alpha}^2)\nonumber\\[0.15cm]
    &&+(-m_{\alpha }^2+2m_{W}^2+m_{c}^{2})C_{1}(m_c^2,0,m_t^2,m_{W}^2,m_{\alpha}^2,m_{\alpha}^2) \nonumber\\[0.15cm]
    &&+4m_{W}^2 C_{2}(m_c^2,0,m_t^2,m_{W}^2,m_{\alpha}^2,m_{\alpha}^2)\nonumber\\[0.15cm]
    &&+ \varsigma_{d}\varsigma_{d}^{\ast} m_{\alpha}^{2} \Bigl[ C_{22}(m_c^2,0,m_t^2,m_{H^{\pm }}^2,m_{\alpha}^2,m_{\alpha}^2)  +C_{12}(m_c^2,0,m_t^2,m_{H^{\pm }}^2,m_{\alpha}^2,m_{\alpha}^2) \nonumber\\[0.15cm]
    && +C_{2}(m_c^2,0,m_t^2,m_{H^{\pm }}^2,m_{\alpha}^2,m_{\alpha}^2) \Bigr] \nonumber\\[0.15cm] 
    &&+ \varsigma_{u}\varsigma_{u}^{\ast} m_{c}^{2} \Bigl[ C_{12}(m_c^2,0,m_t^2,m_{H^{\pm }}^2,m_{\alpha}^2,m_{\alpha}^2)+C_{11}(m_c^2,0,m_t^2,m_{H^{\pm }}^2,m_{\alpha}^2,m_{\alpha}^2) \nonumber\\[0.15cm]
    &&+C_{1}(m_c^2,0,m_t^2,m_{H^{\pm }}^2,m_{\alpha}^2,m_{\alpha}^2) \Bigr] \Bigr\}\,, \\[0.2cm]
	\mathcal{N}^{R}_{\alpha} &=& \mathcal{N}^{L}_{\alpha} = -\frac{g_{s} G_{F}}{4 \sqrt{2} \pi^{2}} m_{\alpha}^{2} \nonumber\\[0.15cm]
    && \hspace{1.1cm} \times \Bigl[ C_{1}(m_c^2,0,m_t^2,m_{H^{\pm }}^2,m_{\alpha}^2,m_{\alpha}^2)+C_{2}(m_c^2,0,m_t^2,m_{H^{\pm }}^2,m_{\alpha}^2,m_{\alpha}^2)\Bigr]\,.
\end{eqnarray}
Some of the scalar loop functions, such as $C_{11}$, $C_{12}$ and $C_{22}$, can be further reduced to the more basic loop integrals~\cite{Denner:2019vbn,Denner:1991kt,Passarino:1978jh,tHooft:1978jhc}. The loop kinetic terms $\mathcal{F}_{\alpha}^{L}$ can be obtained from  $\mathcal{F}_{\alpha}^{R}$ by interchanging the quark masses $m_{c}$ and $m_{t}$, for both the $t \to c \gamma$ and $t \to c g$ decays. 

\section{\boldmath Figures for the CP asymmetries of \texorpdfstring{$t \to cg_{\pm}$}{ttocgpm} decays} 
\label{sec:Result diagrams}

\begin{figure}[hb]
    \centering 
    \includegraphics[width=0.428\textwidth]{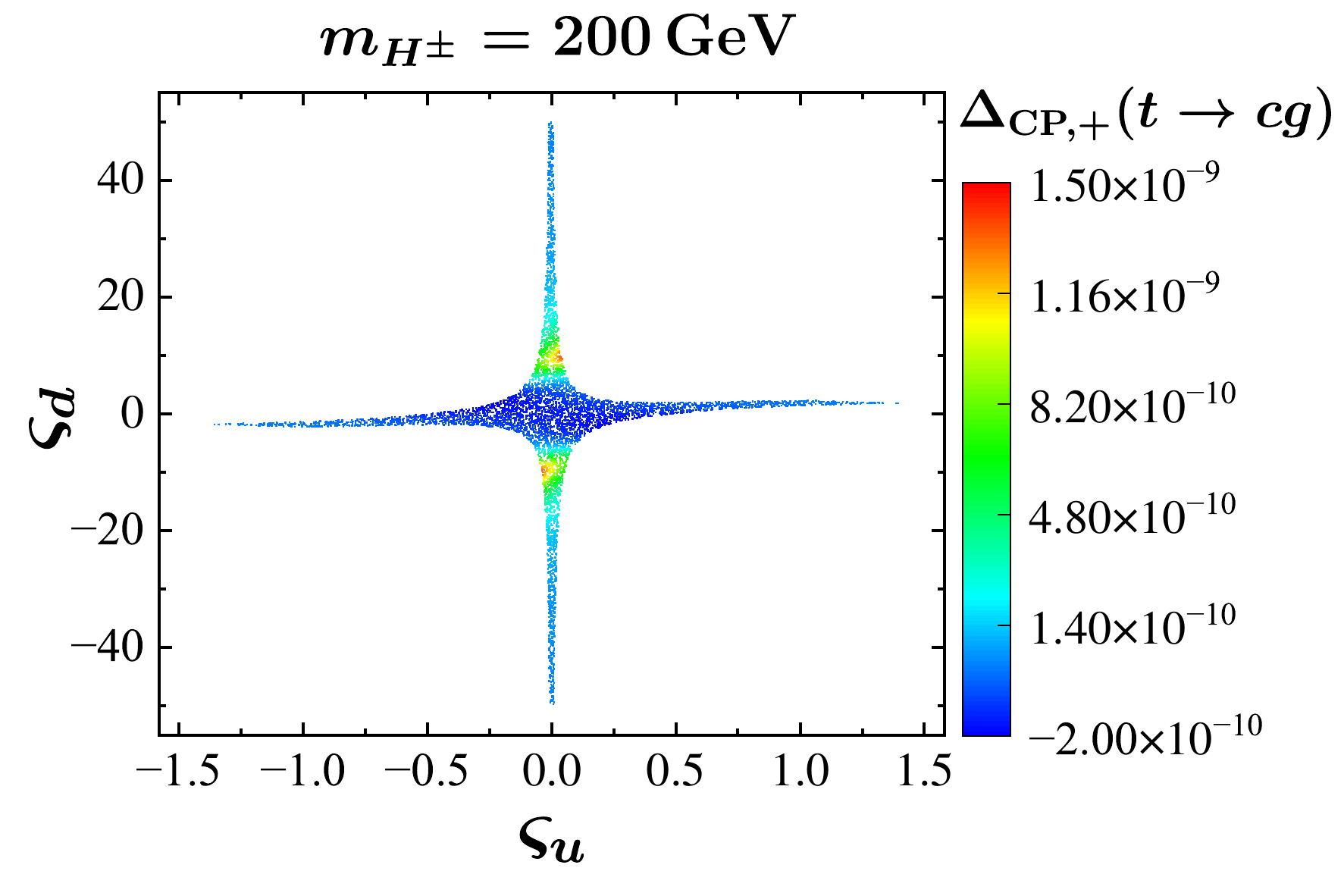}
    \includegraphics[width=0.428\textwidth]{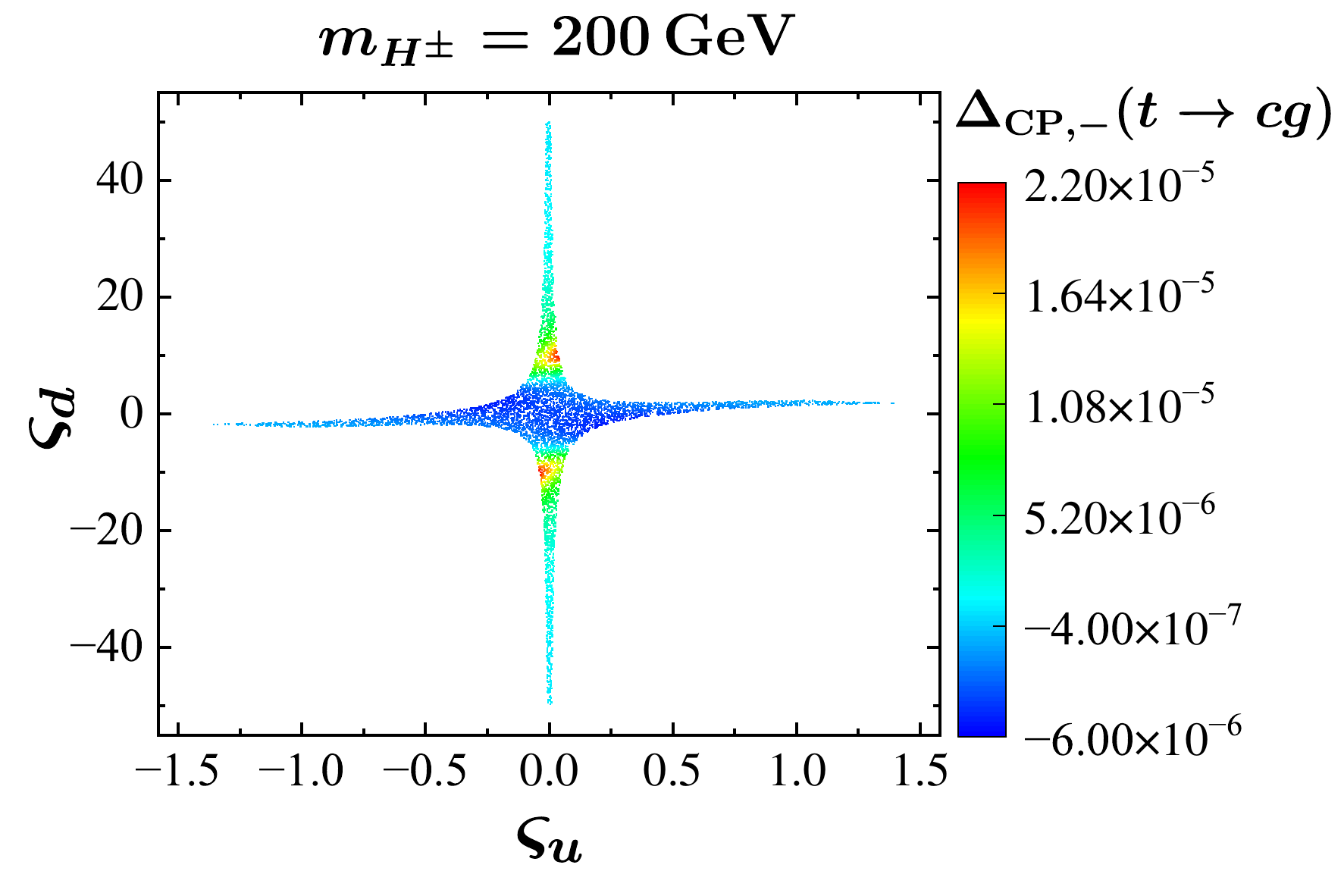}
    \includegraphics[width=0.428\textwidth]{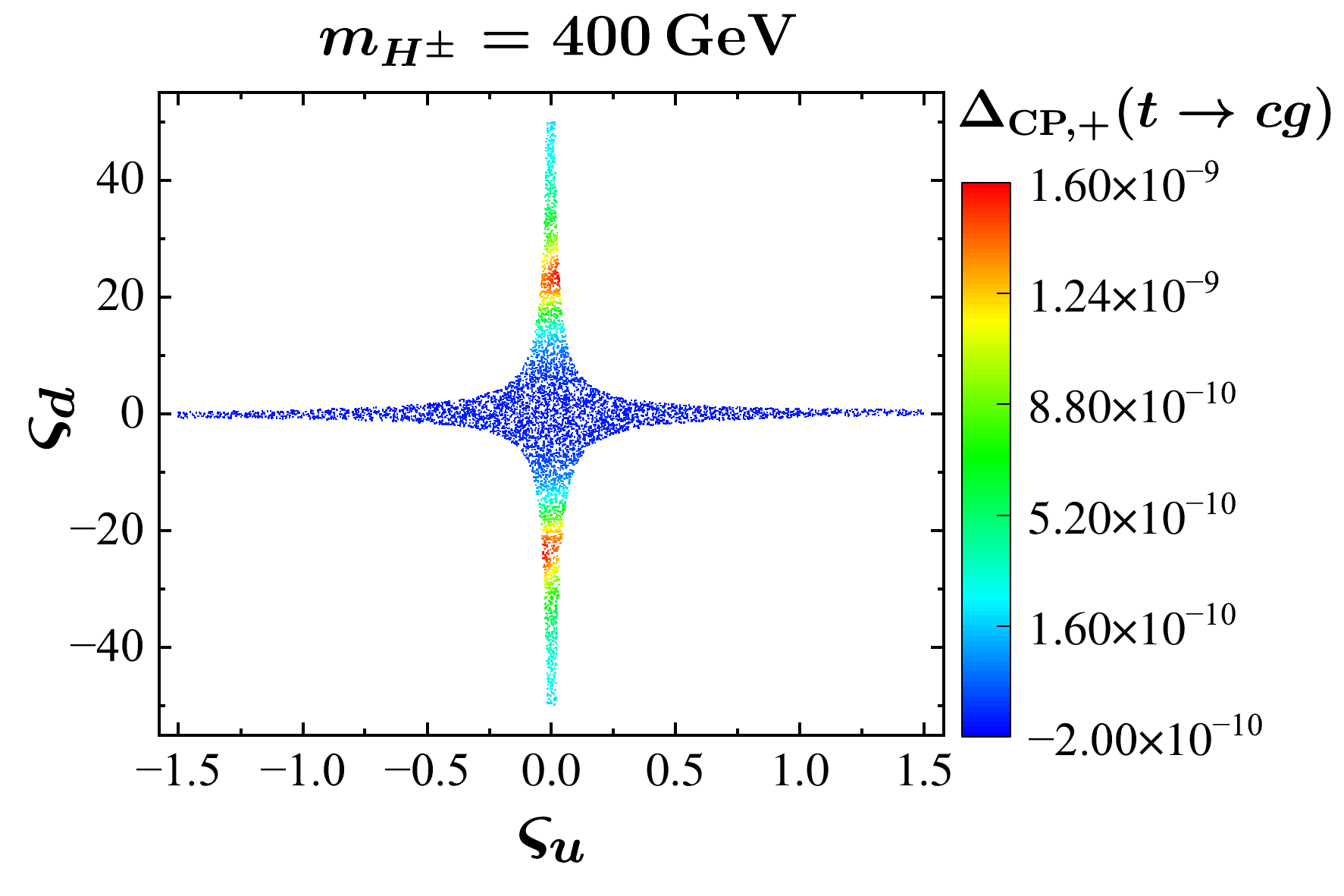}
    \includegraphics[width=0.428\textwidth]{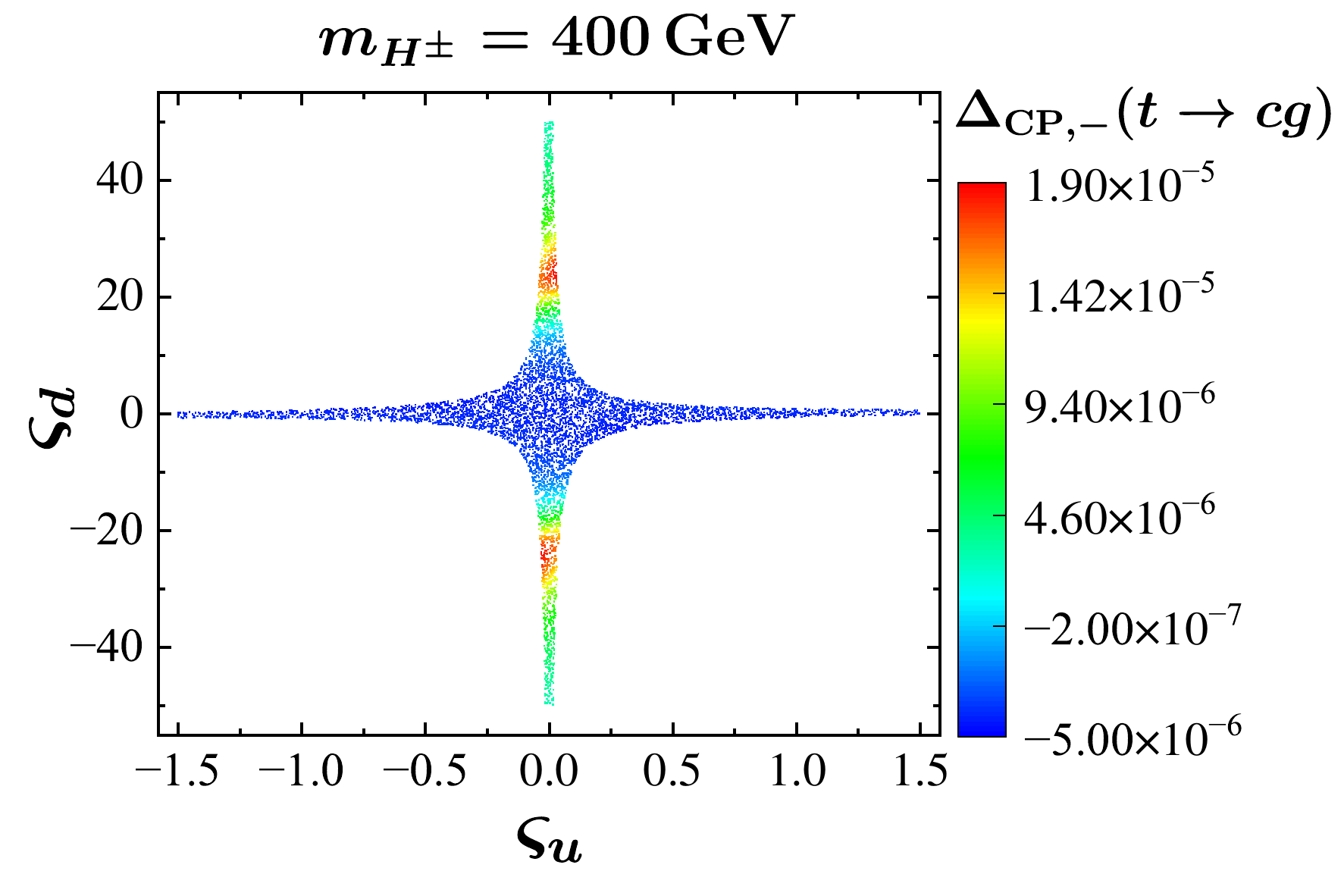}
    \includegraphics[width=0.428\textwidth]{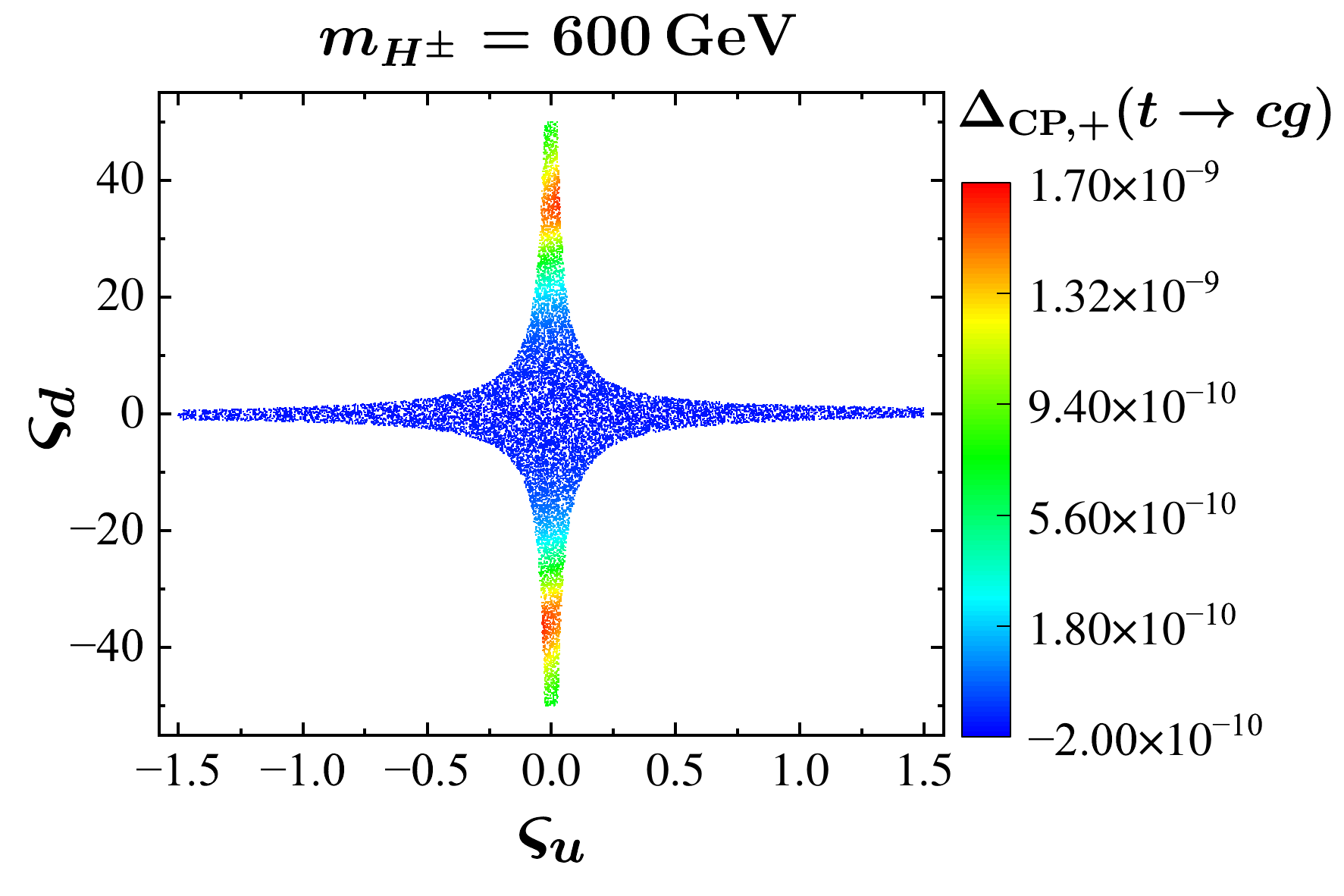}
    \includegraphics[width=0.428\textwidth]{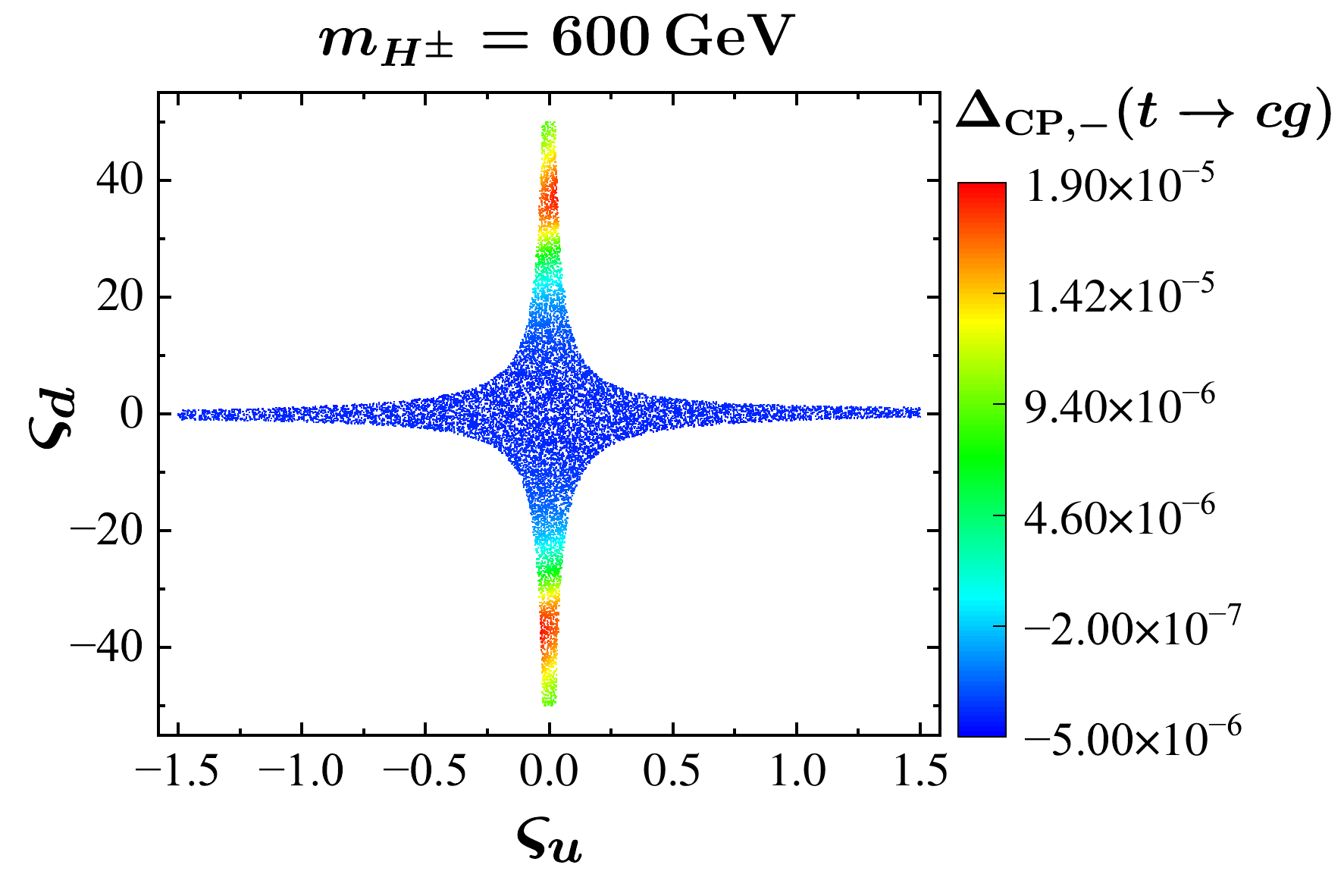}
    \caption{The polarized CP asymmetries $\Delta_{\text{CP},+}(t \to cg)$ (left column) and $\Delta_{\text{CP},-}(t \to cg)$ (right column) versus the real alignment parameters $\varsigma_u$ and $\varsigma_d$. The other captions are the same as in Fig.~\ref{Fig:tcy cpre}. \label{Fig:tcg cppmre}}
\end{figure}

\begin{figure}[htbp]
    \centering 
    \includegraphics[width=0.428\textwidth]{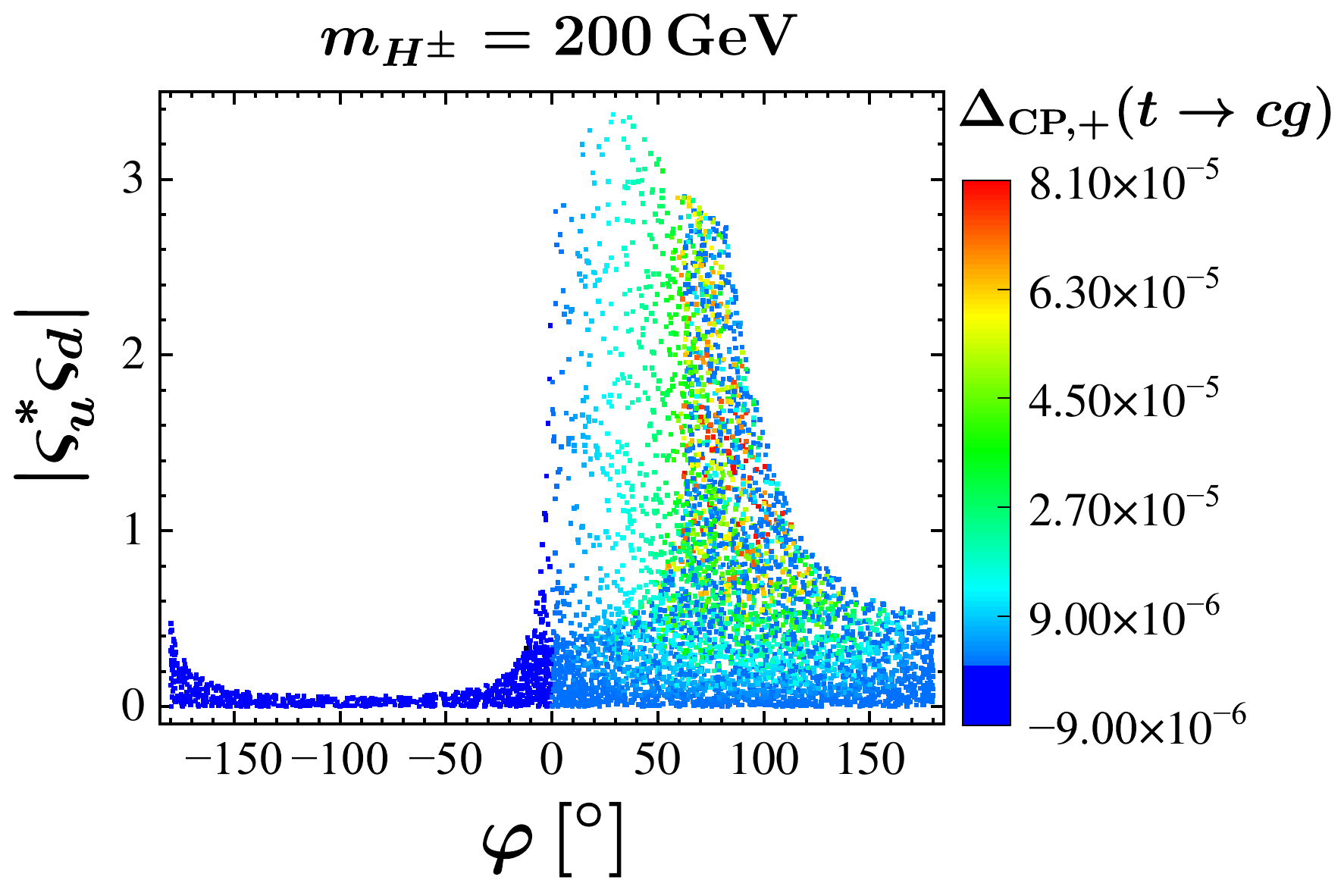}
    \includegraphics[width=0.428\textwidth]{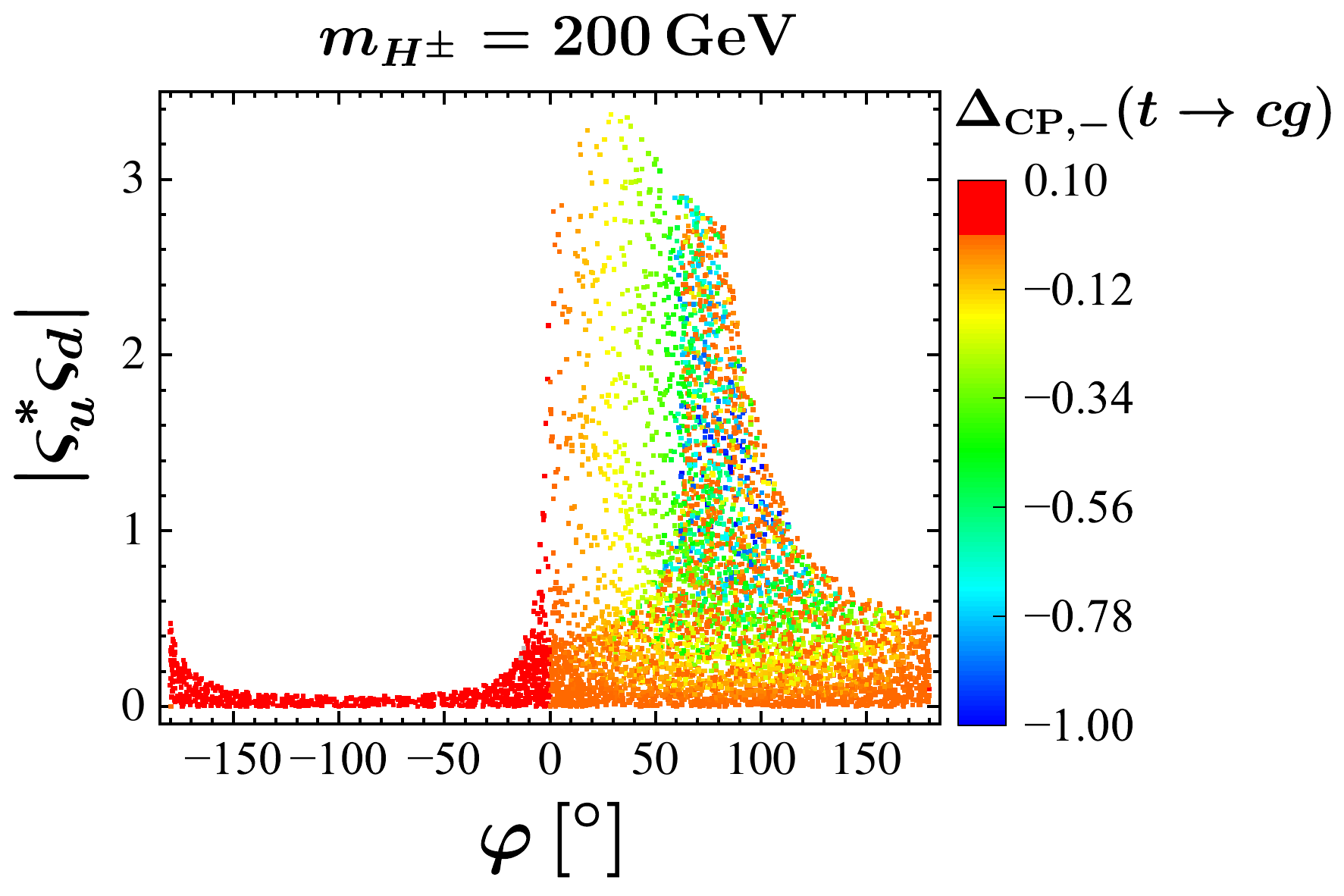}
    \includegraphics[width=0.428\textwidth]{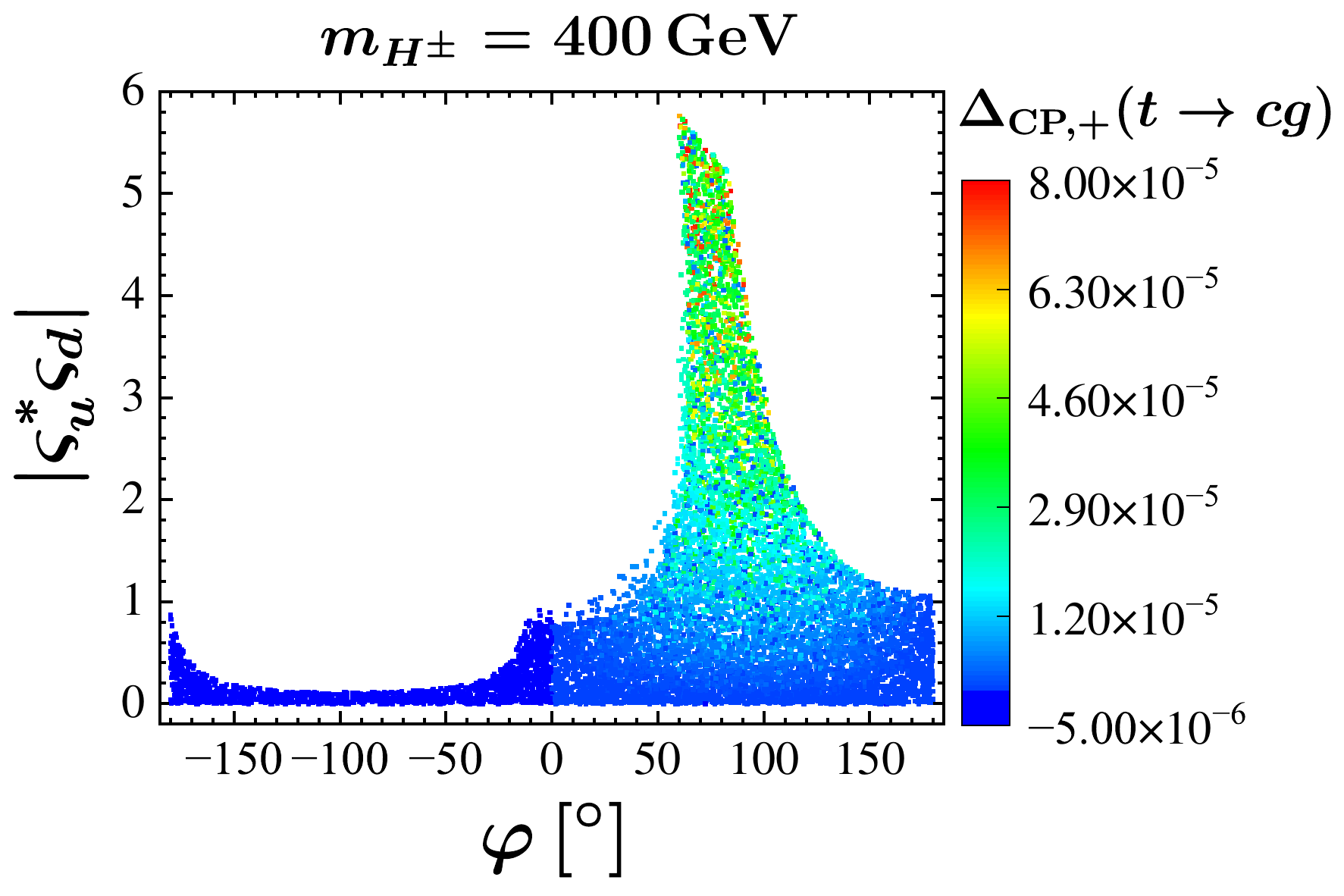}
    \includegraphics[width=0.428\textwidth]{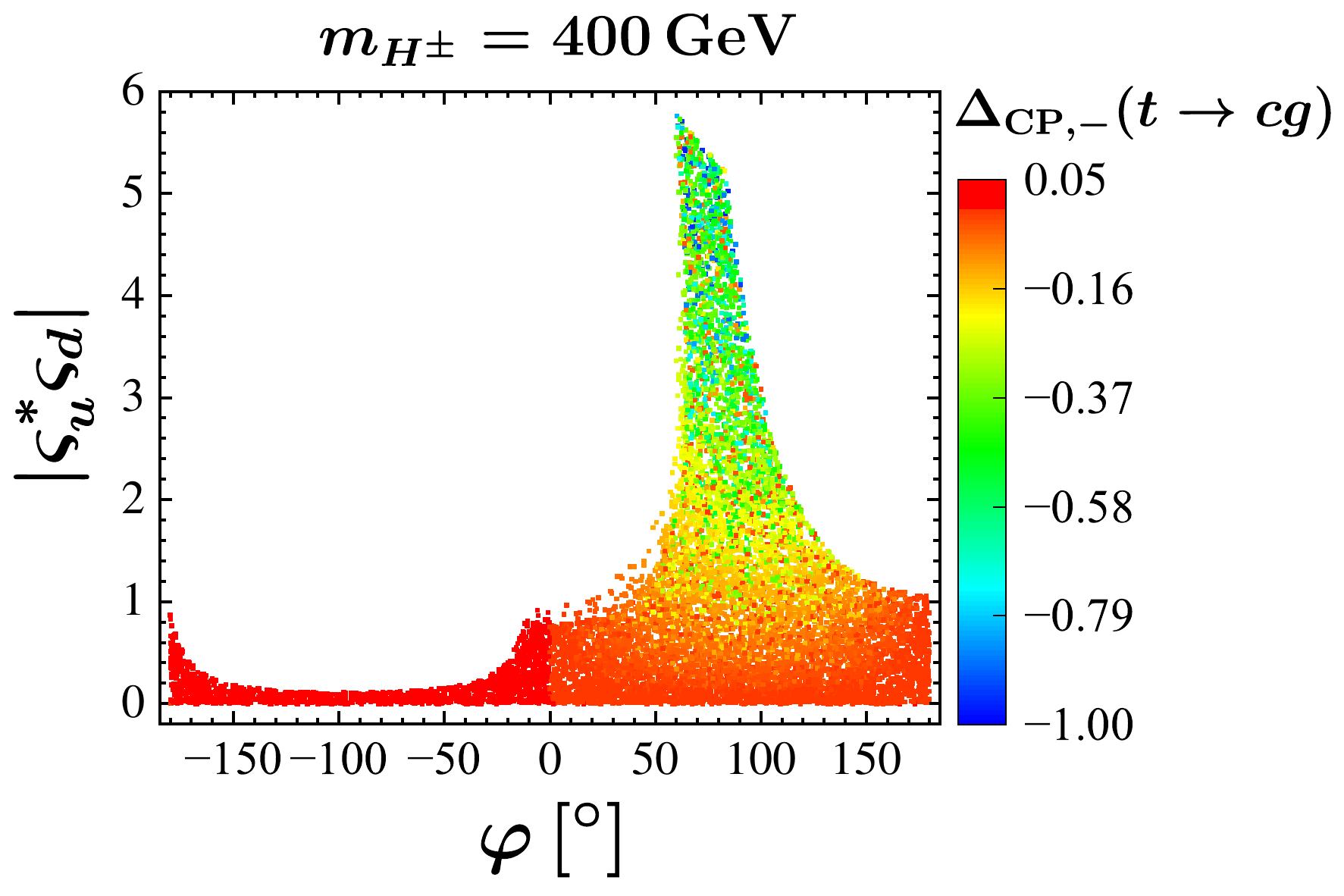}
    \includegraphics[width=0.428\textwidth]{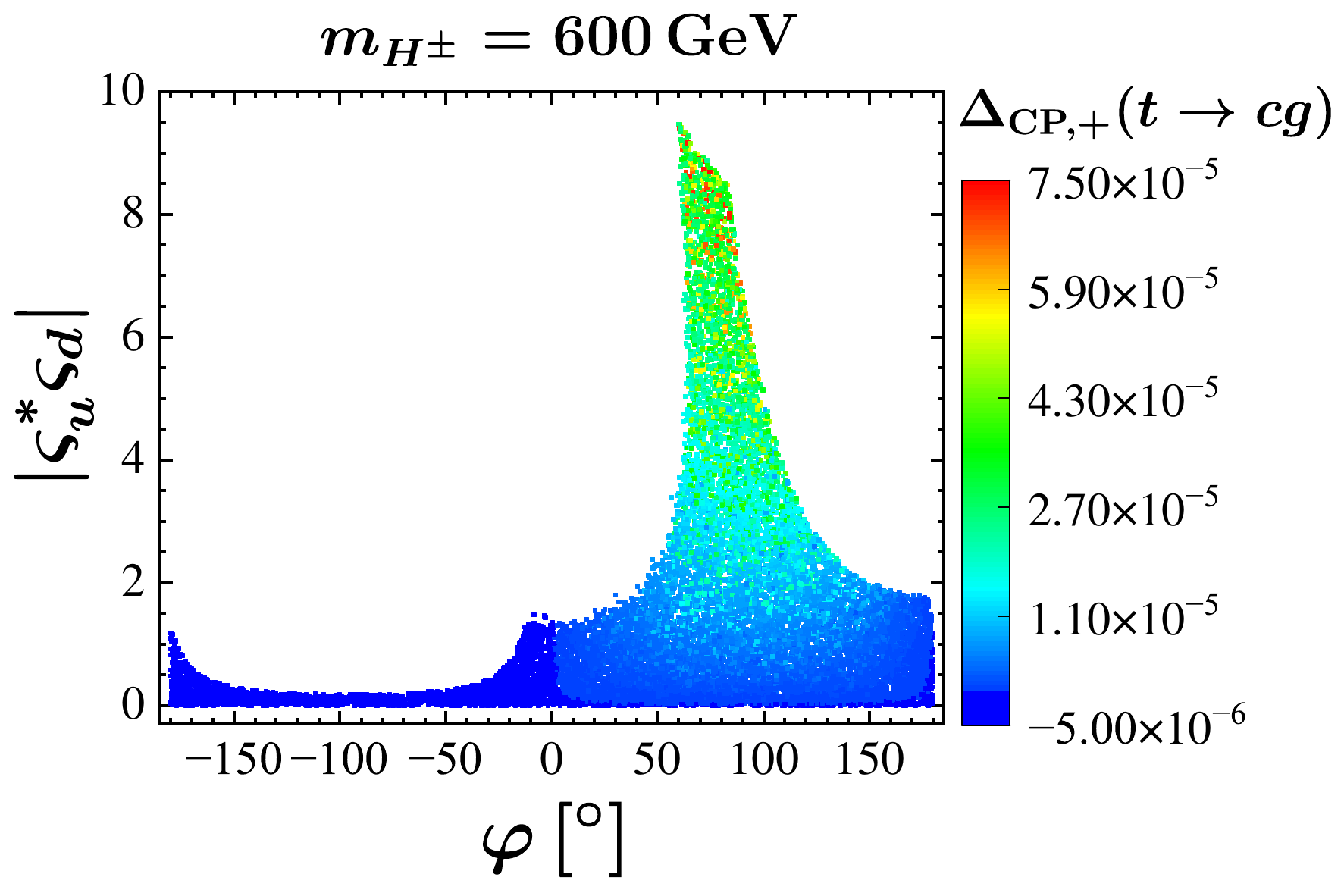}
    \includegraphics[width=0.428\textwidth]{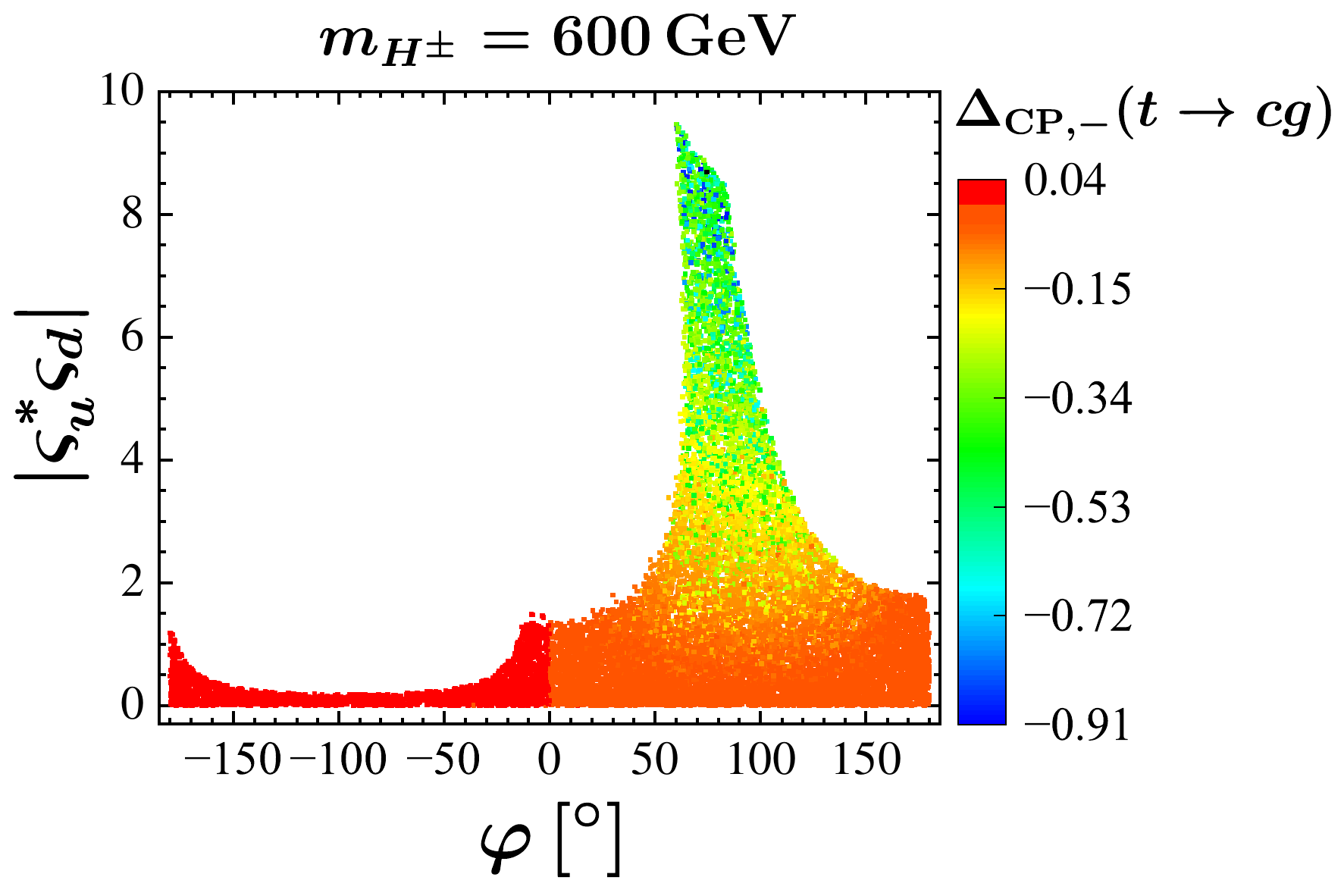}
    \caption{The polarized CP asymmetries $\Delta_{\text{CP},+}(t \to cg)$ (left column) and $\Delta_{\text{CP},-}(t \to cg)$ (right column) versus the product of the two alignment parameters $|\varsigma_{u}^{\ast}\varsigma_{d}|$ and their relative phase $\varphi$. The other captions are the same as in Fig.~\ref{Fig:tcy cpim}. \label{Fig:tcg cppmim}}
\end{figure}

In this appendix, we present the variations of the CP asymmetries of $t \to cg_{\pm}$ decays with respect to the parameters of the A2HDM, for three benchmark values of the charged-Higgs mass, $m_{H^{\pm}}=200$, $400$, and $600~\gev$. First, we show in Fig.~\ref{Fig:tcg cppmre} the dependence of the two polarized CP asymmetries $\Delta_{\text{CP},+}(t \to cg)$ (left column) and $\Delta_{\text{CP},-}(t \to cg)$ (right column) on the real alignment parameters $\varsigma_u$ and $\varsigma_d$. Considering the complex alignment parameters, we show in Fig.~\ref{Fig:tcg cppmim} the two polarized CP asymmetries $\Delta_{\text{CP},+}(t \to cg)$ (left column) and $\Delta_{\text{CP},-}(t \to cg)$ (right column) projected onto the $|\varsigma_{u}^{\ast}\varsigma_{d}|-\varphi$ plane.

\bibliographystyle{JHEP}
\bibliography{ref}

\end{document}